\documentclass[aps,twocolumn,preprintnumbers,amsmath,amssymb]{revtex4}

\usepackage{amssymb}
\usepackage{amsmath}
\usepackage{textcomp}
\usepackage{graphicx}
\usepackage{bm}
\usepackage{amsfonts}

\begin{document}

\title{Stable Mode Sorting by Two-Dimensional Parity of Photonic Transverse Spatial States}

\author{C.C. Leary}
\email{cleary@uoregon.edu}
\author{L.A. Baumgardner}
\author{M.G. Raymer}
\affiliation{Oregon Center for Optics and Department of Physics, University of Oregon, Eugene, OR USA, 97403}

\date{\today}

\begin{abstract}
We describe a mode sorter for two-dimensional parity of transverse spatial states of light based on an out-of-plane Sagnac interferometer. Both Hermite-Gauss (HG) and Laguerre-Gauss (LG) modes can be guided into one of two output ports according to the two-dimensional parity of the mode in question. Our interferometer sorts $HG_{nm}$ input modes depending upon whether they have even or odd order $n+m$; it equivalently sorts $LG^{l}_{p}$ modes depending upon whether they have an even or odd value of their orbital angular momentum $l$. It functions efficiently at the single-photon level, and therefore can be used to sort single-photon states. Due to the inherent phase stability of this type of interferometer as compared to those of the Mach-Zehnder type, it provides a promising tool for the manipulation and filtering of higher order transverse spatial modes for the purposes of quantum information processing. For example, several similar  Sagnacs cascaded together may allow, for the first time, a stable measurement of the orbital angular momentum of a true single-photon state. Furthermore, as an alternative to well-known holographic techniques, one can use the Sagnac in conjunction with a multi-mode fiber as a spatial mode filter, which can be used to produce spatial-mode entangled Bell states and heralded single photons in arbitrary first-order ($n+m=1$) spatial states, covering the entire Poincar\'{e} sphere of first-order transverse modes. 

\end{abstract}

\maketitle

\section{\label{sec:Intro}Introduction\protect}

 The manipulation of the transverse spatial degrees of freedom of photons offers avenues for quantum information studies using single photons as qubits. Several recent studies have focused on higher order transverse spatial modes, of which the Hermite-Gauss (HG) and Laguerre-Gauss (LG) modes are examples \cite{Allen}. In particular, several different interferometers of the Mach-Zehnder type have been experimentally employed for sorting single photons by their transverse mode properties. For the LG modes, an orbital angular momentum (OAM) sorter was demonstrated at the single-photon level which sorted modes with even OAM values, which exited one port of the interferometer, from those with odd OAM values, which exited the other port \cite{Padgett, Wei}. For the HG modes, a mode index sorter was theoretically proposed to measure the value of a single HG index $n$ or $m$ \cite{Xue}, while a different setup experimentally distinguished the $HG_{10}$ and $HG_{01}$ modes, thereby acting as a transverse-mode beam splitter for first-order HG modes, in analogy with a polarizing beam splitter \cite{Sasada}. A similar setup was later used to perform various manipulations on photonic qubits using a one-dimensional transverse spatial degree of freedom \cite{Yarnall}.

\begin{figure}[!t]
\includegraphics[width=0.5\textwidth]{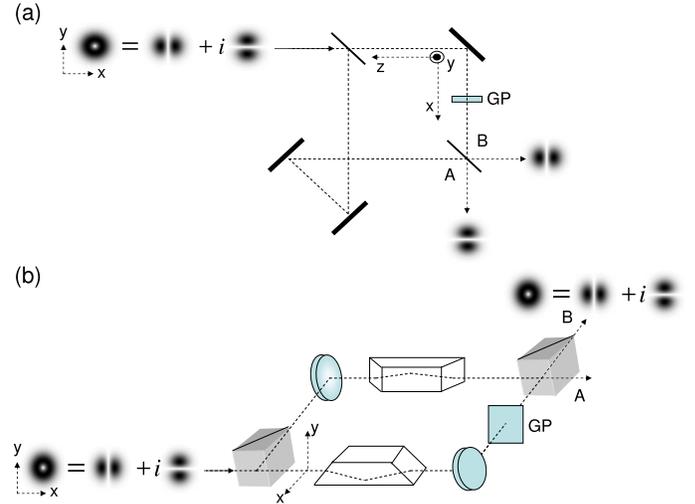}
\caption{\label{fig:Sorters} Previously realized 1-D and 2-D parity sorters of the Mach-Zehnder type, each with a tiltable phase-shifting glass plate (GP) in one arm. (a) The 1-D parity sorter is distinguished by having an extra mirror in one arm. Upon entering this interferometer, an $LG_{01}$ mode is sorted into its constituent $HG_{10}$  and $HG_{01}$ components due to complete constructive and destructive interference at the output ports. (b) In contrast, the 2-D parity sorter has a Dove prism in each arm. One of the Dove prisms is rotated 90\textdegree $ $ with respect to the other, which causes a 180\textdegree $ $ relative rotation of the two interfering beams in the transverse plane. In this case, both $HG_{10}$ and $HG_{01}$ modes exit the same port, so that an incident $LG_{01}$ mode is not decomposed.}
\end{figure}

  The two experimentally implemented sorter designs \cite{Wei} and \cite{Sasada}, shown in Fig. \ref{fig:Sorters}, share several similarities. Among these are their Mach-Zehnder character with two independent arms, and their reliance on complete constructive and destructive interference at the output beam splitter in order to sort certain transverse spatial modes into one of two distinct output ports. Also, they both have the advantage of not relying on computer-generated holograms, which are lossy in practice, so that the sorting efficiency was limited only by the transmission and reflection coefficients of the constituent mirrors, glass plates, and dove prisms. However, though they are similar, the two aforementioned sorters actually treat transverse spatial modes in fundamentally different ways. The first order HG mode beam splitter in Fig. \ref{fig:Sorters} (a) distinguishes modes on the basis of one-dimensional (1-D) parity about the axis normal to the plane of propagation, due to the extra mirror reflection in one of the interferometer arms. This means in particular that $HG_{10}$ and $HG_{01}$ input modes propagating in the \textit{x}-\textit{z} plane exit different ports as shown since they respectively have odd and even parity with respect to reflections about the \textit{y}-axis, described by the transformation $E\left(x,y\right)\to E\left(-x,y\right)$. On the other hand, the OAM sorter in Fig. \ref{fig:Sorters} (b) distinguishes modes on the basis of the \textit{two}-dimensional (2-D) parity of the transverse (\textit{x}-\textit{y}) plane--that is, whether the mode is even or odd with respect to the transformation $E\left(x,y\right)\to E\left(-x,-y\right)$. Therefore, conversely to the prior case, the $HG_{10}$ and $HG_{01}$ input modes will be sorted into the \textit{same} port by the OAM sorter. Since, to our knowledge, no experiments have been reported in which a sorter of the OAM type was employed to sort HG modes, this fact seems to have been overlooked and the difference between the two sorters has not been emphasized until now. We will henceforth refer to sorters of the aforementioned types as 1-D and 2-D parity sorters, respectively.

\begin{figure}
\includegraphics[width=0.5\textwidth]{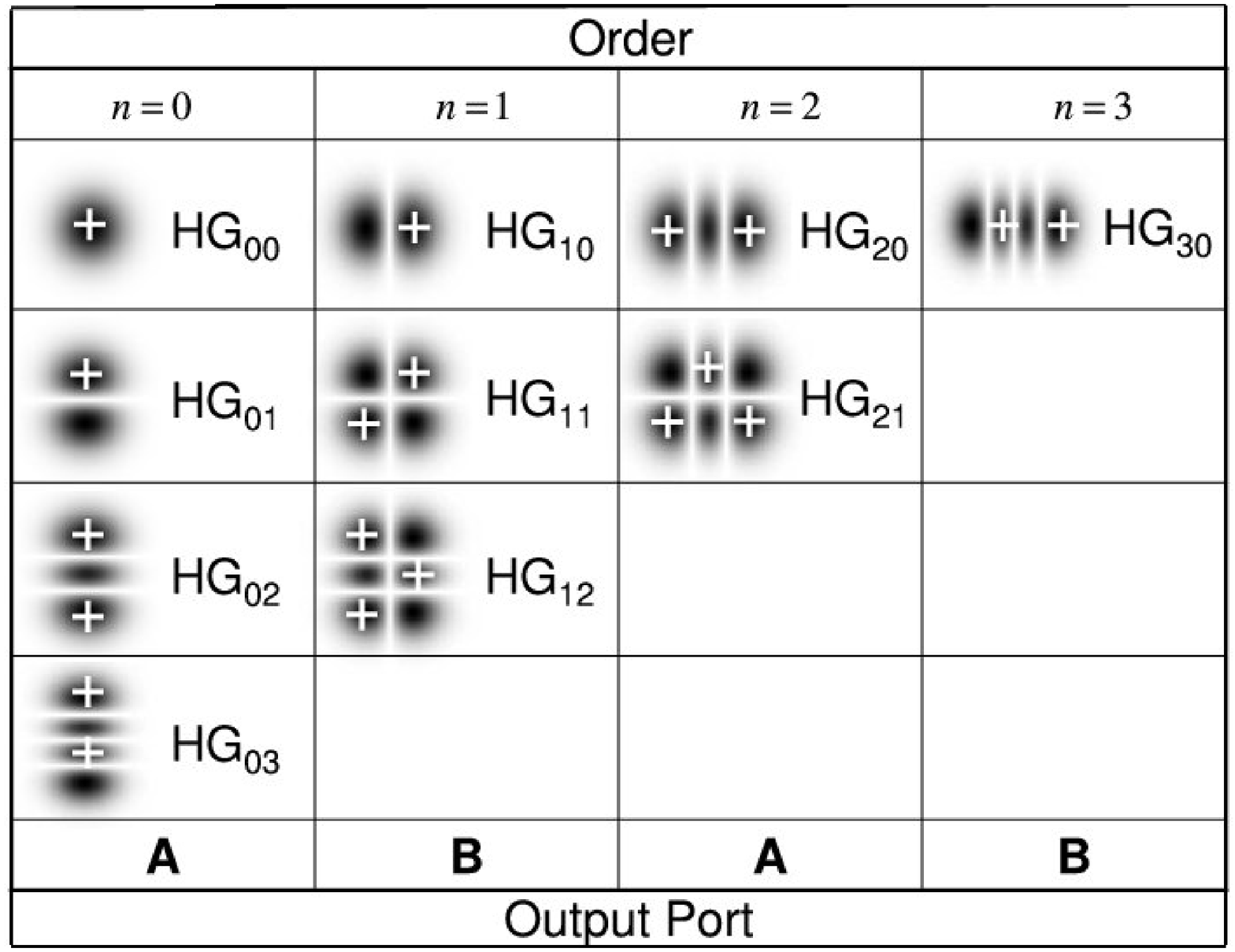}
\caption{\label{fig:1DParityTable} A tabulation of the action of the the 1-D sorter upon the HG modes of order $n+m\leq3$. The plus signs designate modal lobes which are in phase with each other and 180\textdegree $ $ out of phase with the unmarked lobes. Modes with an even value for $n$ exit port A, while modes with odd $n$-value exit port B (see Fig. \ref{fig:Sorters} (a)).}
\end{figure}

\begin{figure}
\includegraphics[width=0.5\textwidth]{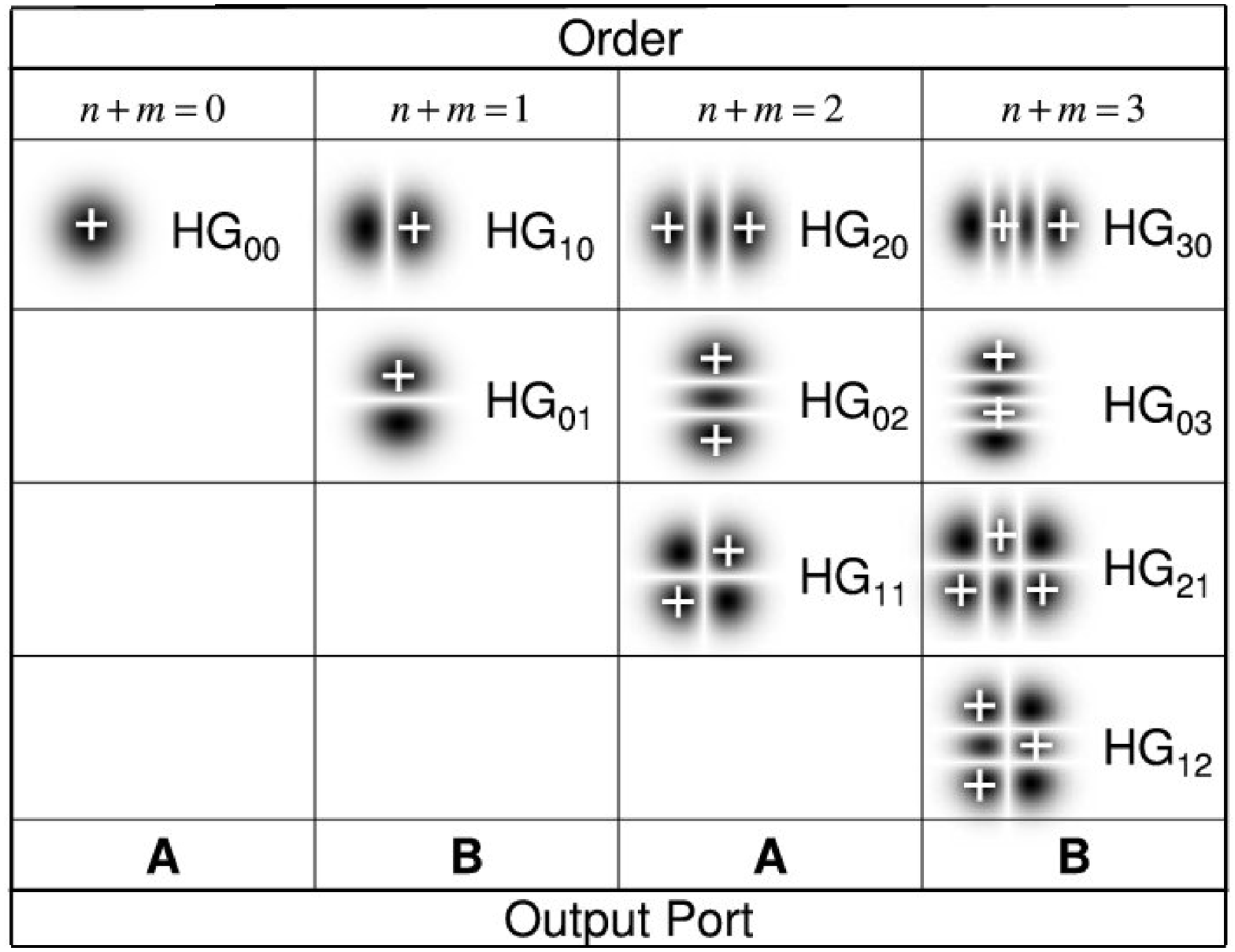}
\caption{\label{fig:2DParityTable} A tabulation of the action of the the 2-D sorter upon the HG modes of order $n+m\leq3$. The plus signs designate the relative phases of the lobes as in Fig. \ref{fig:1DParityTable}. Modes with an even value for $n+m$ exit port A, while modes with odd $n+m$-value exit port B (see Fig. \ref{fig:Sorters} (b)).}
\end{figure}

Figs. \ref{fig:1DParityTable} and \ref{fig:2DParityTable} give a general outline of how $HG_{nm}$ modes are sorted by the 1-D and 2-D parity sorters. Since the reflection from the extra mirror of the 1-D sorter inverts the sign of the \textit{x}-variable only, it sorts according to whether the single index $n$ (which corresponds to the \textit{x} spatial variable) is even (output port A) or odd (output port B).  For the 2-D sorter, the two interfering beams undergo a 180\textdegree $ $ relative rotation in the transverse plane (equivalent to sign inversion about \textit{both} the \textit{x} and \textit{y} axes) so that it sorts according to whether the \textit{sum} of the two indices $n+m$ is even (port A) or odd (port B).

 A major disadvantage of both the 1-D and 2-D parity sorters mentioned above is their Mach-Zehnder character, or more specifically the existence of two independent beam paths in each interferometer. This makes them subject to phase noise and drift so that in practice it is difficult to keep them aligned. In this work, we overcome this difficulty for the latter design by demonstrating experimentally a novel type of 2-D parity sorter in the form of an out-of-plane Sagnac interferometer. Since the two interfering paths of the Sagnac are counter-propagating (see Fig. \ref{fig:Sagnac}), the interferometer is automatically phase-stable. As a result, experiments involving multiple cascading interferometers are considerably more feasible when employing a Sagnac then with a Mach-Zehnder interferometer. Such cascading allows full 2-D sorting of arbitrary superpositions of modes, including the OAM sorting schemes described in \cite{Padgett, Wei}.

 In what follows we discuss the design of our out-of-plane Sagnac interferometer. We present, to our knowledge, the first direct 2-D parity measurements of HG modes. We also propose several applications of this interferometer, including its use as an alternative to holograms in spatial mode filtering, the production of Bell states entangled in first-order transverse spatial modes, and the production of heralded single photons in first-order transverse spatial states corresponding to an arbitrary point of the first-order spatial mode Poincar\'{e} sphere \cite{van Enk}.

\section{\label{sec:Sagnac}Phase-Stable Parity Sorting Sagnac Interferometer\protect}

\begin{figure}
\includegraphics[width=0.5\textwidth]{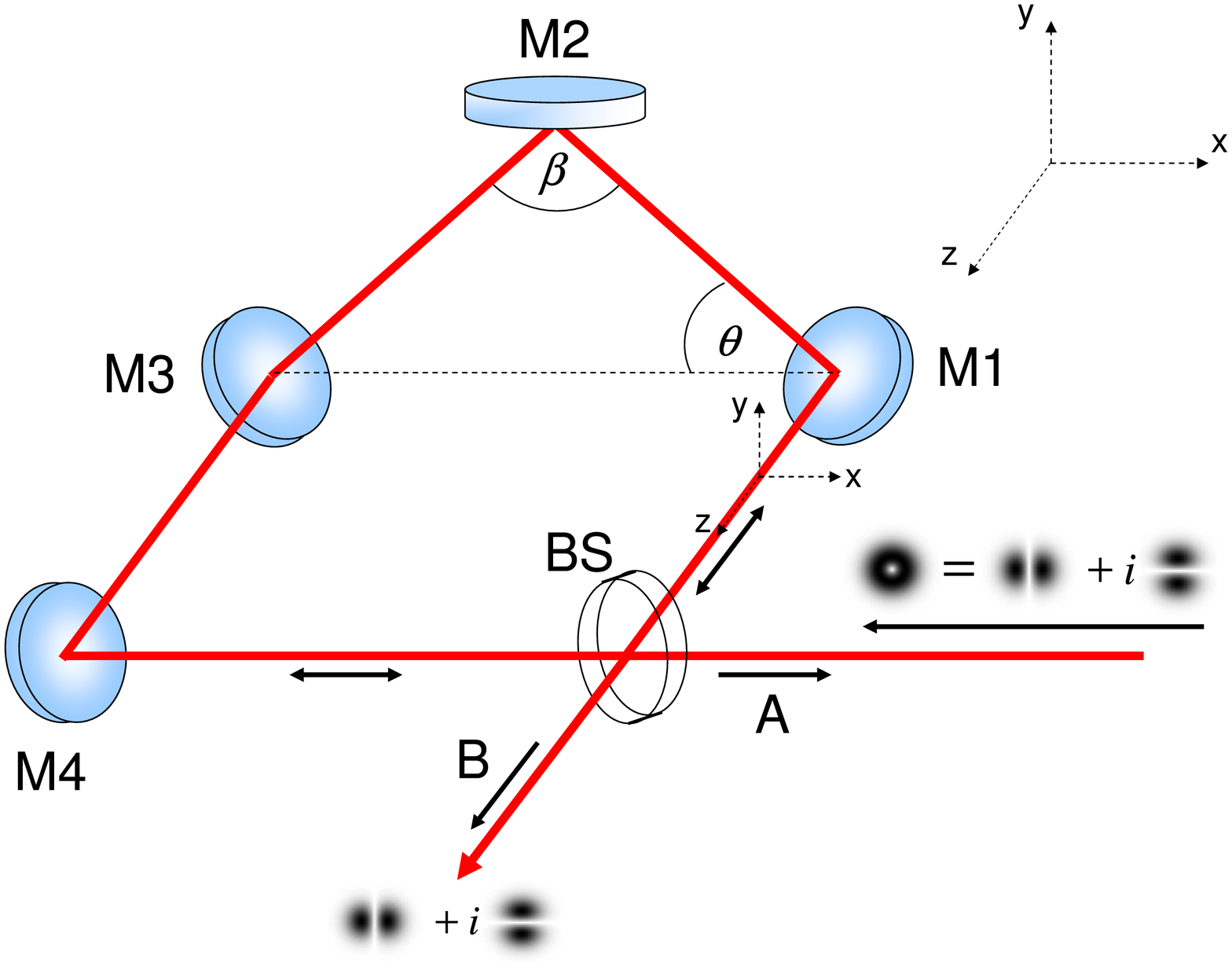}
\caption{\label{fig:Sagnac} The 2-D parity sorting Sagnac interferometer. Mirrors M1-M3 cause the beam path to travel out of the \textit{x-z} plane, tracing out two congruent sides of an isosceles triangle in the \textit{x-y} plane, with congruent base angles $\theta$ and apex angle $\beta=\pi-2\theta$. Incident $HG_{10}$ and $HG_{01}$ modes exit the same port of this interferometer, as do $LG_{01}$ modes, as is the case for the sorter in Fig. \ref{fig:Sorters} (b). See text for further discussion.}
\end{figure}

The Sagnac interferometer can be made to operate as a mode sorter because of a geometric phase accumulation implemented by the out-of-plane reflections at three of the interferometer mirrors, which results in a rotation of the transverse spatial profile of the beam through an angle $\Omega$ about the beam axis \cite{Galvez, Smith}. To see this, consider a simplified version of our 2-D parity-sorting interferometer, shown in Fig. \ref{fig:Sagnac}. The mirrors labeled M1-M3 in the figure are oriented such that the reflected beam path travels out of the \textit{x-z} plane, and traces out the two congruent sides of an isosceles triangle which lies completely in the \textit{x-y} plane. The action of the 50:50 beam splitter (BS in the figure) gives rise to two counter-propagating paths of equal path length, which is the reason for the phase-stability of the interferometer. We denote the path involving reflections off of the mirrors M1-M4 in the successive order M4-M3-M2-M1 as the clockwise path, and that involving reflections of order M1-M2-M3-M4 as the anti-clockwise path.

\begin{figure}
\includegraphics[width=0.5\textwidth]{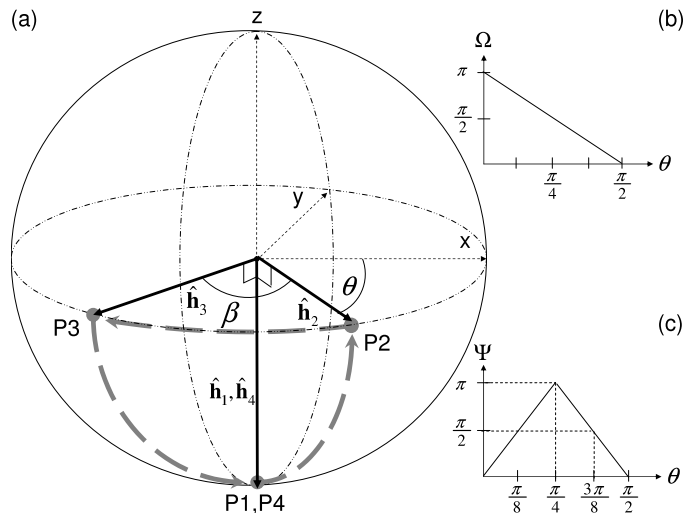}
\caption{\label{fig:Poincare} (a) The unit sphere construction corresponding to the beam path in Fig. \ref{fig:Sagnac} as discussed in the text. The beam path angles $\theta$ and $\beta$ from Fig. \ref{fig:Sagnac} define the helicity vectors $\hat{\mathbf{h}}_{1}-\hat{\mathbf{h}}_{4}$, which in turn define the points P1-P4 on the spherical surface as shown. The dashed lines denote the triangular closed loop that results from successively connecting these points with geodesic curves. The figure is drawn for the special case where $\theta=45$\textdegree $ $, so that $\beta=90$\textdegree $ $. In this case, the angles between the distinct helicity vectors are all 90\textdegree $ $, so that the area enclosed by the loop is exactly one eighth of the area of the entire sphere, which corresponds to a transverse rotation of $\Omega=\frac{\pi}{2}$ rads, or 90\textdegree $ $. (b) Plot of the beam rotation angle $\Omega$ vs. the parameter $\theta$. (c) Plot of the relative rotation angle $\Psi$ (modulo $\pi$) of the two counter-propagating beams after passing through the Sagnac interferometer vs. $\theta$, emphasizing the two special cases where $\Psi=180$\textdegree $ $ and $\Psi=90$\textdegree $ $. These two cases correspond to the first and second cascaded interferometer stages for the OAM sorting scheme in \cite{Padgett}.} 
\end{figure}

A convenient parameter for discussing the state of the beam is the unit helicity vector $\hat{\mathbf{h}}$, defined as $\hat{\mathbf{h}}\equiv\frac{\left(-1\right)^{n}\mathbf{k}}{\left|\mathbf{k}\right|}$ where $n$ is the number of reflections, and $\mathbf{k}$ is the beam wave vector \cite{Galvez}. Employing this, we now consider the four vectors $\hat{\mathbf{h}}_{1}$ through $\hat{\mathbf{h}}_{4}$ associated with the anti-clockwise-traveling beam, defined as the unit helicity vectors prior to the respective reflections off of M1-M4 as shown in the figure. These vectors define four points P1-P4 on the unit sphere (See Fig. \ref{fig:Poincare} (a)); however, due to the odd number of out-of-plane reflections, we have $\hat{\mathbf{h}}_{1}=\hat{\mathbf{h}}_{4}$, which implies that $P1=P4$, so that only three of these points are unique. Therefore, connecting in order the points P1-P4 on the spherical surface with geodesic curves (great circles) results in a triangular closed loop as shown in the figure. 

The spherical surface area enclosed by this loop is equal to the azimuthal rotation angle $\Omega$ of the transverse beam, as discussed in \cite{Galvez}. In the present special case of three mirror reflections resulting in a closed triangular geodesic surface, one may apply the following relation (due to Euler) involving the surface area $\Omega$ of the sphere \cite{Koch}:

\begin{equation} \label{Euler}
\cos{\frac{\Omega}{2}}=\frac{1+\cos{\alpha}+\cos{\beta}+\cos{\gamma}}{4\cos{\frac{\alpha}{2}}\cos{\frac{\beta}{2}}\cos{\frac{\gamma}{2}}}
\end{equation}

\noindent In \eqref{Euler}, $\alpha\equiv\hat{\mathbf{h}}_{1}\cdot\hat{\mathbf{h}}_{2}$, $\beta\equiv\hat{\mathbf{h}}_{2}\cdot\hat{\mathbf{h}}_{3}$, and $\gamma\equiv\hat{\mathbf{h}}_{3}\cdot\hat{\mathbf{h}}_{4}$ are the angles between the respective helicity unit vector pairs. In our present case (see Figs. \ref{fig:Sagnac}-\ref{fig:Poincare}), $\alpha=\gamma=\frac{\pi}{2}$, and $\beta=\pi-2\theta$, so that the right hand side of \eqref{Euler} simplifies to $\frac{1+\cos{\left(\pi-2\theta\right)}}{2\cos{\left(\frac{\pi}{2}-\theta\right)}}$, which is equivalent to $\sin{\theta}$.  Therefore, we find for our design that $\Omega$ depends on $\theta$ through the simple relationship 

\begin{equation} \label{Simplified Euler}
\cos{\frac{\Omega}{2}}=\sin{\theta}
\end{equation}

\noindent The plot of $\Omega$ vs. $\theta$ in Fig. \ref{fig:Poincare}(b), which has been composed from \eqref{Simplified Euler}, shows that $\Omega$ can be tuned to any angle between 0\textdegree $ $ to 180\textdegree $ $ by choosing the appropriate value of the parameter $\theta$, which ranges from $0\leq\theta\leq\pi$. Now, for the clockwise traveling beam, the direction of the geodesic path of Fig. \ref{fig:Poincare}(a) is reversed, so that the transverse rotation is through angle $-\Omega$, or opposite that of the anti-clockwise traveling beam. We therefore find that the absolute relative rotation $\Psi$ between the two counter-propagating beams can be expressed as $\Psi=\pi-\left|2\Omega-\pi\right|$, as given through the plot of $\Psi$ vs. $\theta$ in Fig. \ref{fig:Poincare}(c). 

 By orienting M1 and M3 in order to select the successive values $\theta=\frac{\pi}{4},\frac{3\pi}{8},\frac{7\pi}{16},...$ one can achieve the relative rotations $\Psi=\pi,\frac{\pi}{2},\frac{\pi}{4},...$ required for each of the interferometer stages of the OAM sorter scheme in \cite{Padgett, Wei}. As discussed in \cite{Wei}, a tiltable phase-shifting glass plate in one of the Mach-Zehnder arms introduces an adjustable relative phase difference between the two interferometer paths, in addition to the phase difference already introduced by the transverse spatial rotation. This additional relative phase is needed in order to distinguish between modes possessing odd OAM values without the use of holograms, which are lossy in practice. Since our device is a common-path interferometer, a glass plate cannot introduce the required relative phase shift between the counter-propagating interferometer paths. However, one may surmount this difficulty by employing a device consisting of a tiltable birefringent waveplate surrounded on both sides by Faraday glass which produces the requisite phase shift as discussed in Appendix \ref{app:OAM}. Therefore, like the device of Fig. \ref{fig:Sorters} (b), our interferometer is able to sort and measure photons possessing arbitrary absolute OAM values with 100 \% efficiency in principle. Furthermore, as our Sagnac interferometer is phase-stable, it provides, for the first time, a practical way to realize a multiple-stage cascaded sorting scheme for single photon states. 

 In the special case where $\theta=45$\textdegree $ $ is chosen, the relative rotation $\Psi=180$\textdegree $ $ and our Sagnac interferometer becomes a 2-D Parity sorter. We note that this type of Sagnac interferometer (with $\theta=45$\textdegree $ $) was used in \cite{Smith} to measure the transverse spatial Wigner function for light at the single-photon level, while a similar interferometer was previously employed for similar purposes in \cite{Mukamel}. The fact that the transverse Wigner function is the expectation value of the 2-D parity operator displaced in phase space \cite{Royer, Smith} provides another way to look at our sorter, which simply measures the \textit{non}-displaced parity operator (corresponding to the transverse Wigner function evaluated at zero transverse position and momentum) for the special case of $\theta=45$\textdegree $ $. Following \cite{Mukamel}, one could in principle use a single rotatable Dove prism in the place of mirrors M1-M3, however there are various disadvantages to this approach including a significantly lower numerical aperture and the introduction of astigmatism \cite{Smith}. One may avoid these difficulties by substituting the out-of-plane mirror configuration described in \cite{van Exter}. Furthermore, since an \textquotedblleft all mirror\textquotedblright $ $ Sagnac has already been successfully employed (for other purposes than the sorting of HG or LG modes) at the single-photon level in \cite{Smith}, we conclude that a Sagnac with $\theta=45$\textdegree $ $ is capable of measuring the 2-D parity of single photon states (or equivalently, whether a single photon's OAM is even or odd).

 Finally, we note that our 2-D parity sorter can also be used to sort the output of a three-mode fiber into its constituent zero- and first-order mode families with 100\% efficiency in principle. In addition to presenting direct 2-D parity measurements, we carry out this fiber filtering experiment in Section \ref{sec:Experiment}, and discuss applications of this technique to quantum information processing in Section \ref{sec:Applications}.

\section {\label{sec:Experiment}2-D Parity Sorting Experiment\protect}

\begin{figure}
\includegraphics[width=0.5\textwidth]{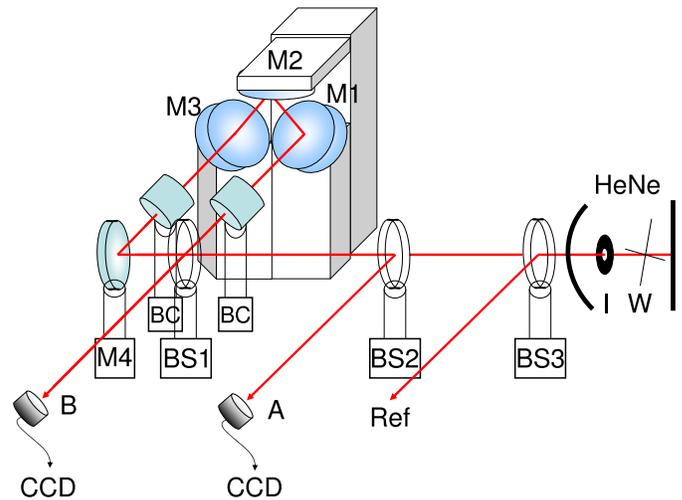}
\caption{\label{fig:Apparatus} The Experimental Apparatus. The propagating and counter-propagating beams interfere at a 50:50 beam splitter (BS1), while a second 50:50 beam splitter (BS2) separates the backward-propagating output mode (port A) from the forward-propagating input mode. An external cavity helium neon laser (HeNe) acts as the source, with thin crossed wires (W) and an iris (I) inserted into the cavity in order to select the higher order HG modes. A third beam splitter (BS3) picks off part of the source beam to use a a reference beam (Ref) for the interferometry experiments discussed in the main text. Two Berek polarization compensators (BC) are placed inside the interferometer to correct for the Fresnel polarization changes due to reflections from the out-of-plane dielectric mirrors. The output fields at both port A and port B were measured with CCD cameras.}
\end{figure}

 Our experimental apparatus was set up as shown in Fig. \ref{fig:Apparatus}, which includes two Berek polarization compensators as well as two additional 50:50 beam splitters in relation to the simplified version of the Sagnac shown in Fig. \ref{fig:Sagnac}. The Berek compensators (New Focus model number 5540) were inserted into the interferometer as shown in the figure in order to improve the interference visibility by correcting for the Fresnel polarization changes due to the reflections from the dielectric mirrors M1-M3. Without the compensators, the propagating and counter-propagating beams ended up with unequal elliptical polarization states upon interference at the beam splitter, which led to poor interference visibility and therefore inefficient sorting. Since one of the interferometer's output ports (port A in Figs. \ref{fig:Sagnac} and \ref{fig:Apparatus}) is counter-propagating with respect to the input beam, the addition of beam splitter BS2 acted to deflect half of the output beam from this path so that it could be imaged on the CCD as shown in the figure. 

\begin{figure}[!t]

$\begin{array}{c} 

{\textrm{\Large{Experiment}}} \\ 

{\begin{array}{ccccc} 

{\begin{array}{c} {\textrm{\scriptsize{Input}}} \\ {\textrm{\scriptsize{Mode}}} \end{array}}& 
{\begin{array}{c} {\textrm{\scriptsize{Input}}} \\ {\textrm{\scriptsize{Intensity}}} \\ {\textrm{\scriptsize{Profile}}} \end{array}}& 
{\textrm{\scriptsize{Port A}}}& 
{\textrm{\scriptsize{Port B}}}& 
{\begin{array}{c} {\textrm{\scriptsize{Interference}}} \\ {\textrm{\scriptsize{Intensity}}} \\ {\textrm{\scriptsize{Profile}}} \end{array}} \\ 

{\begin{array} {c} {HG_{00}} \\ { } \end{array}}&
{\includegraphics[width=000.075\textwidth]{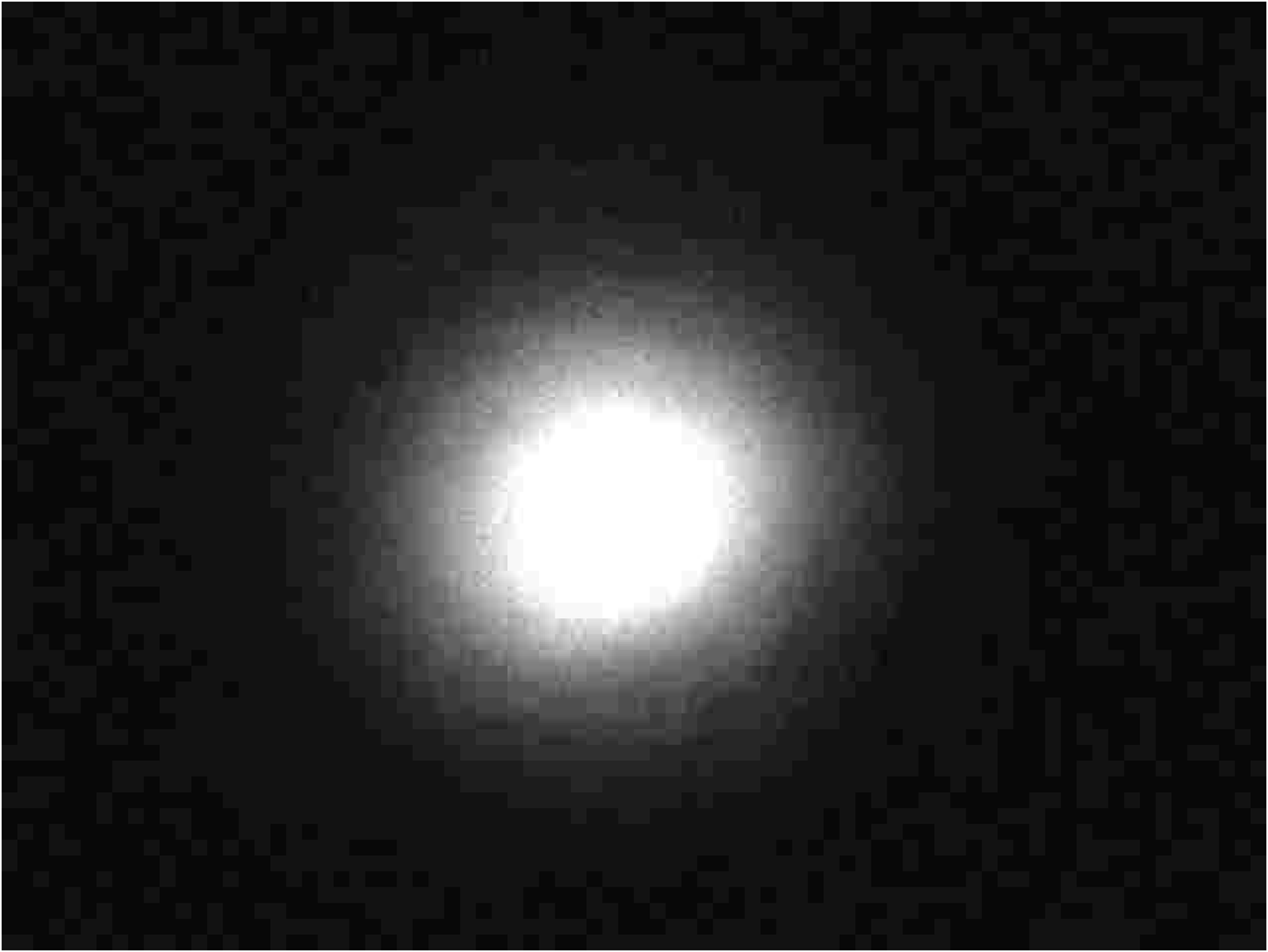}}&
{\includegraphics[width=000.075\textwidth]{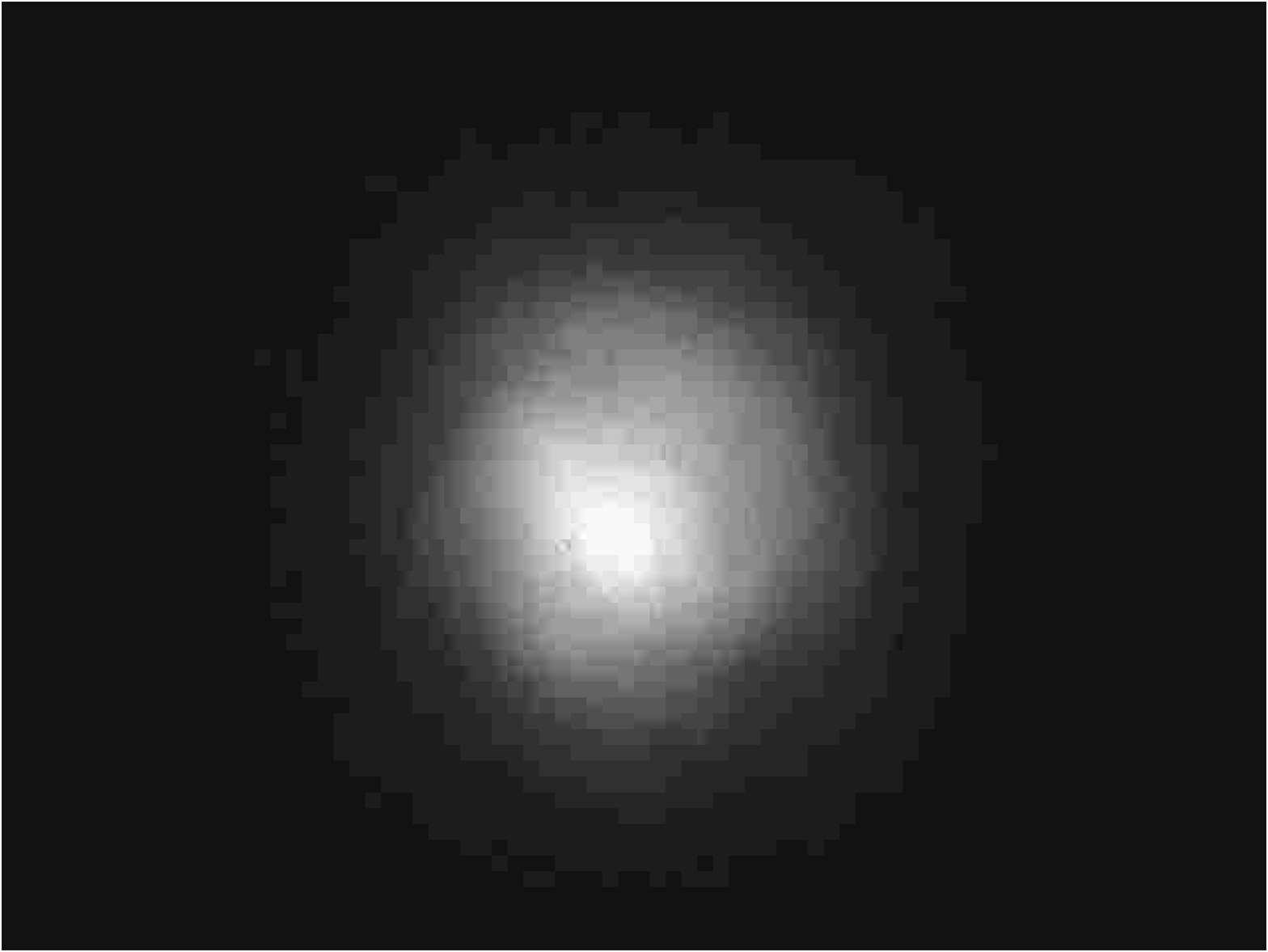}}& 
{\includegraphics[width=000.075\textwidth]{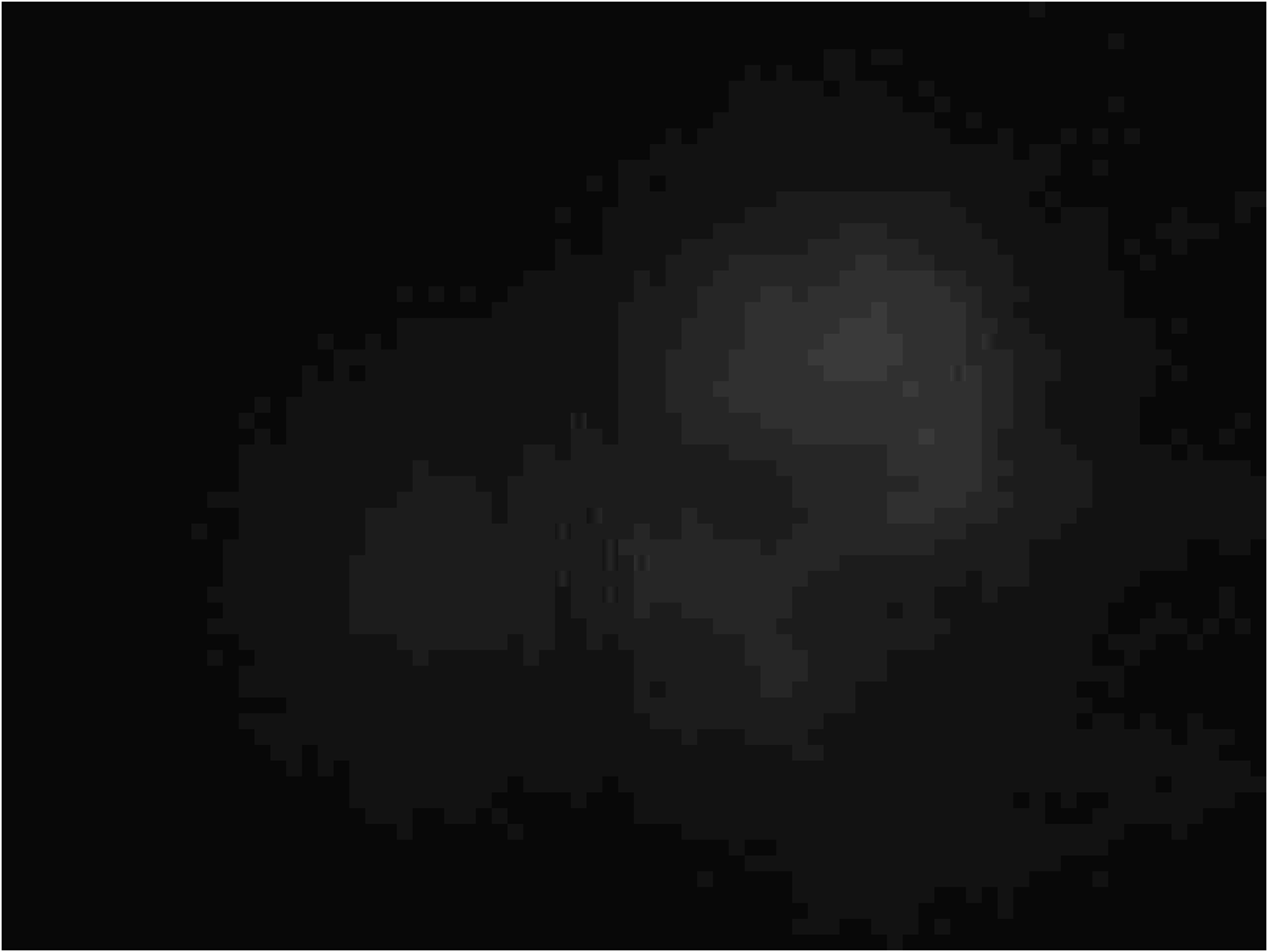}}& 
{\includegraphics[width=000.075\textwidth]{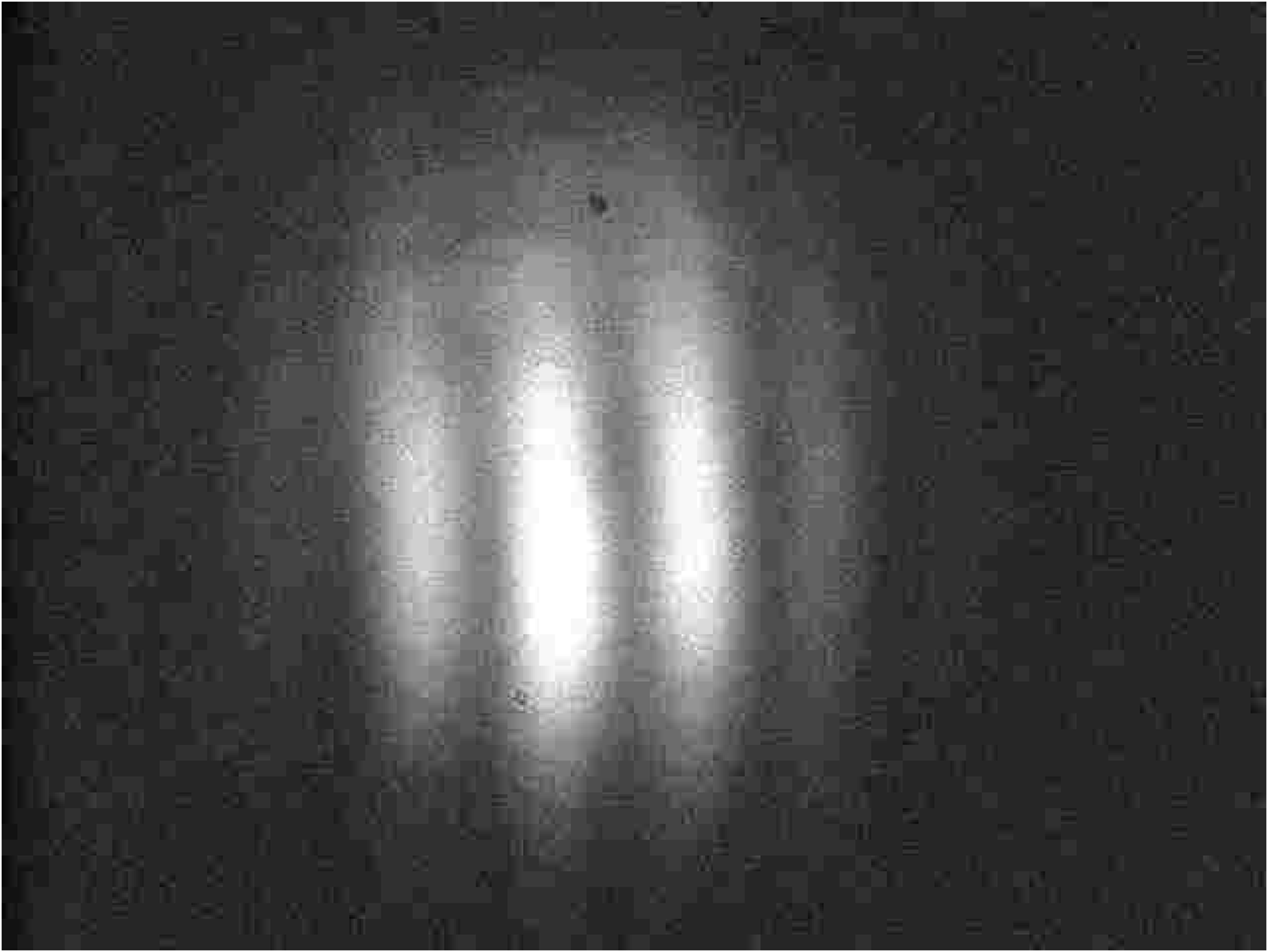}}\\ 

{\begin{array} {c} {HG_{01}} \\ { } \end{array}}& 
{\includegraphics[width=000.075\textwidth]{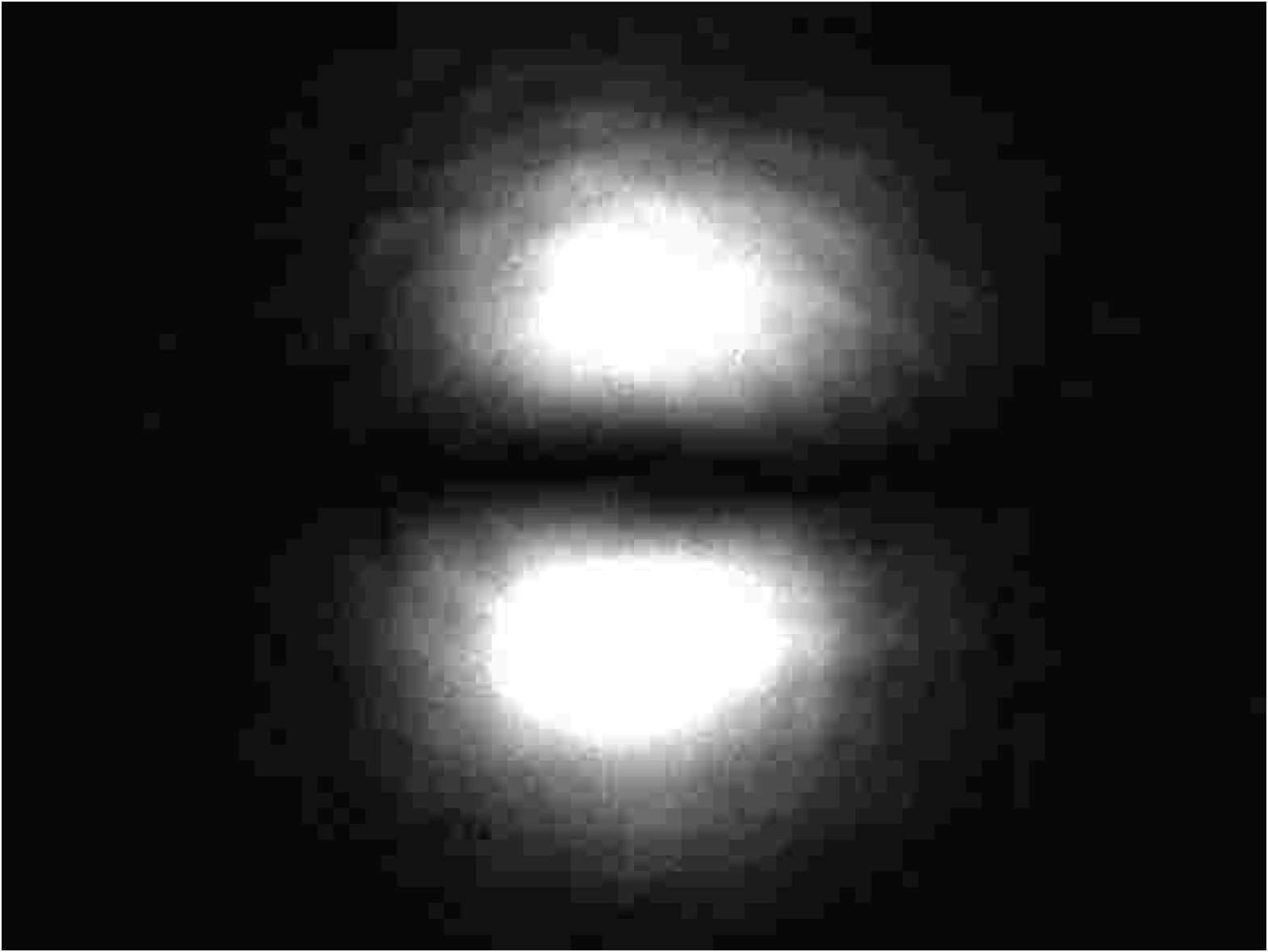}}& 
{\includegraphics[width=000.075\textwidth]{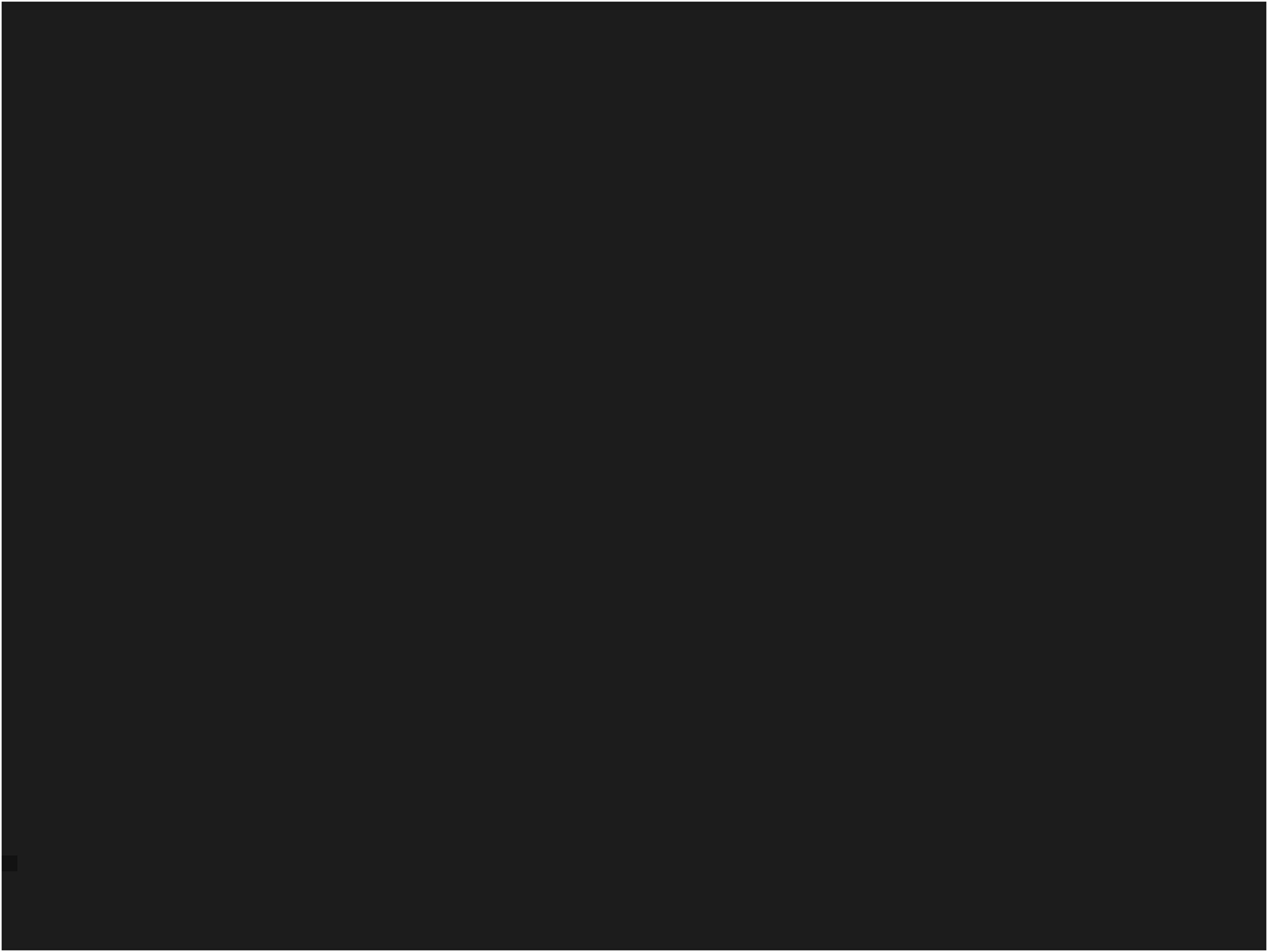}}& 
{\includegraphics[width=000.075\textwidth]{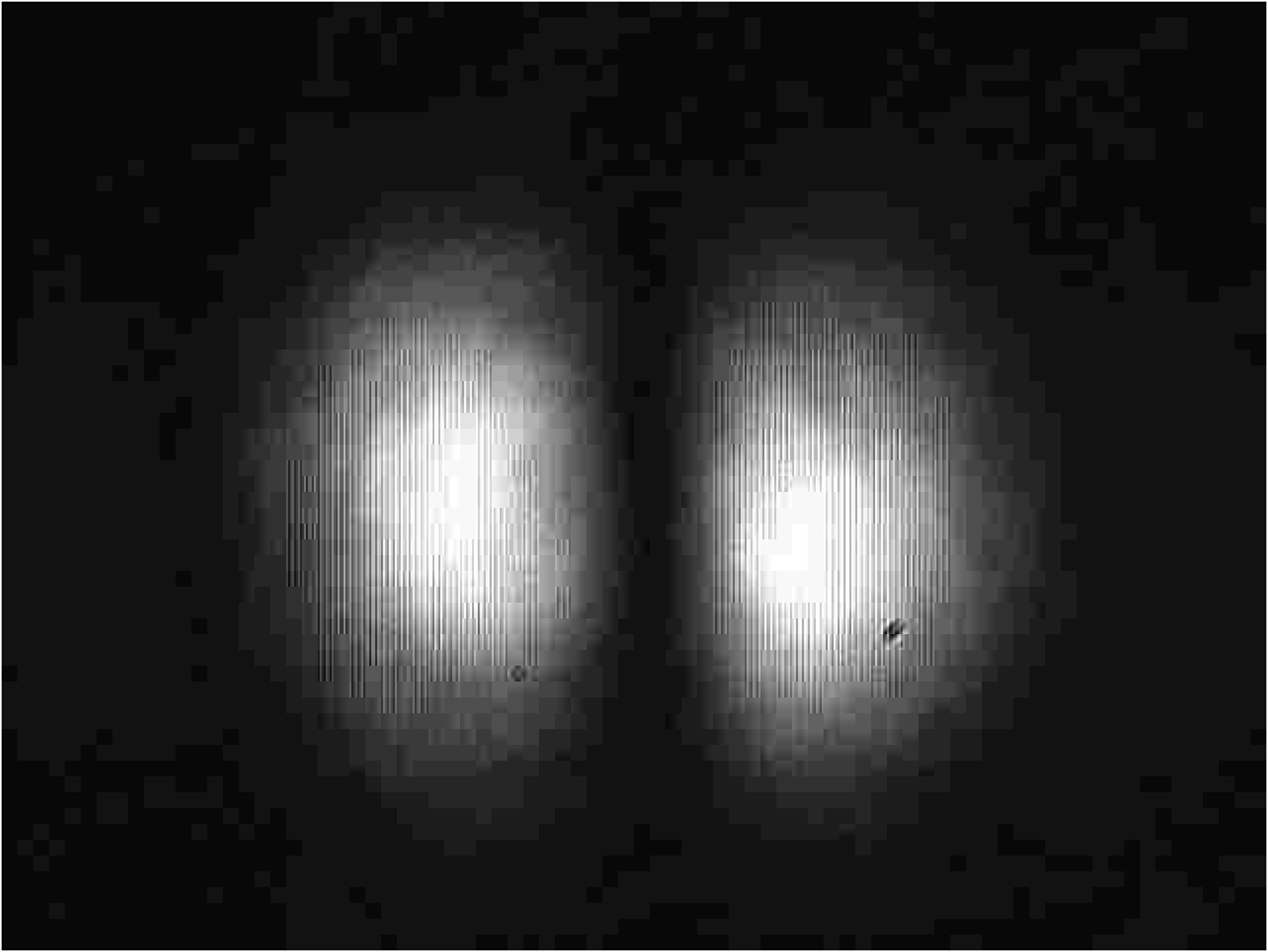}}& 
{\includegraphics[width=000.075\textwidth]{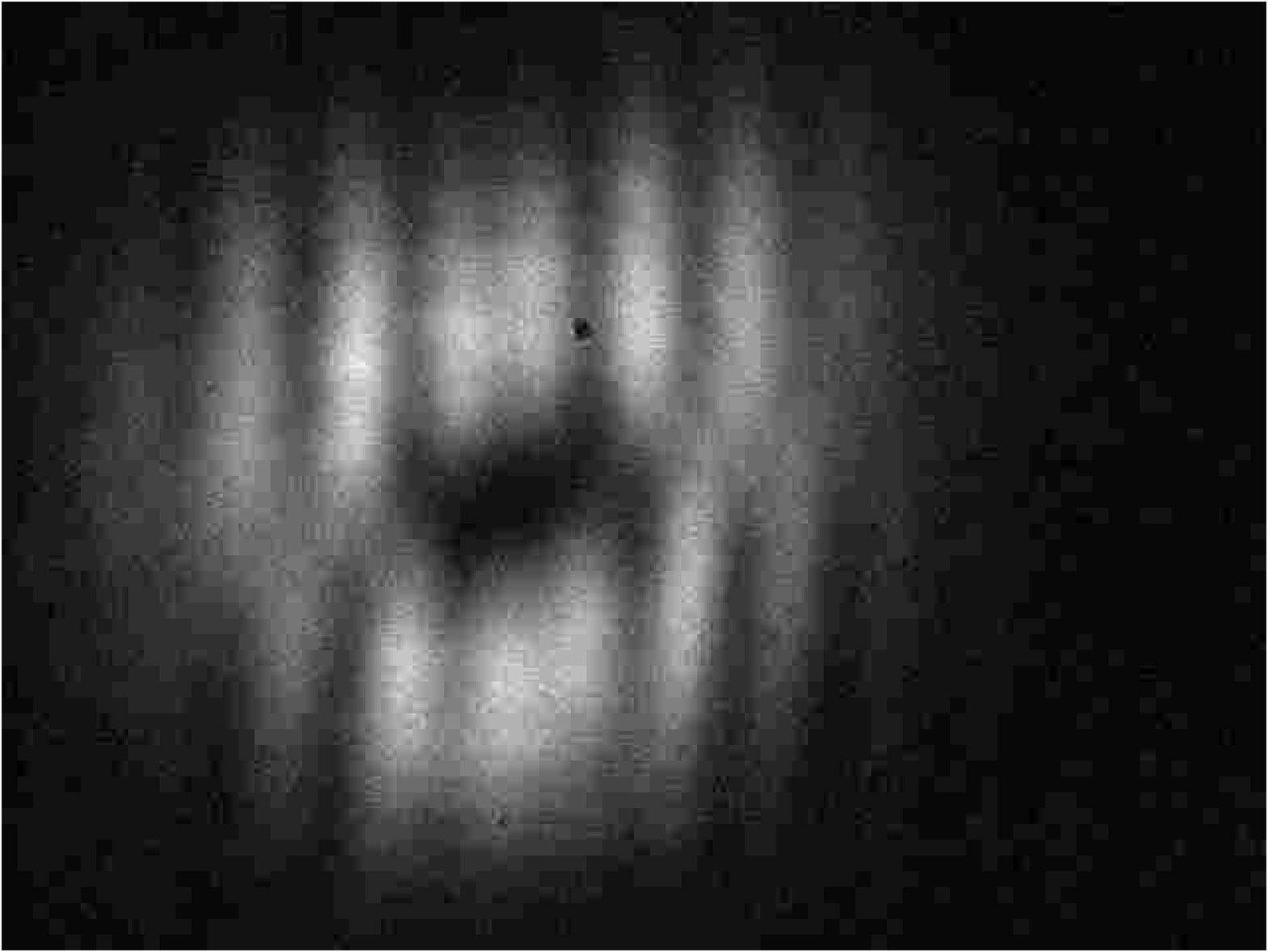}}\\ 

{\begin{array} {c} {HG_{10}} \\ { } \end{array}}& 
{\includegraphics[width=000.075\textwidth]{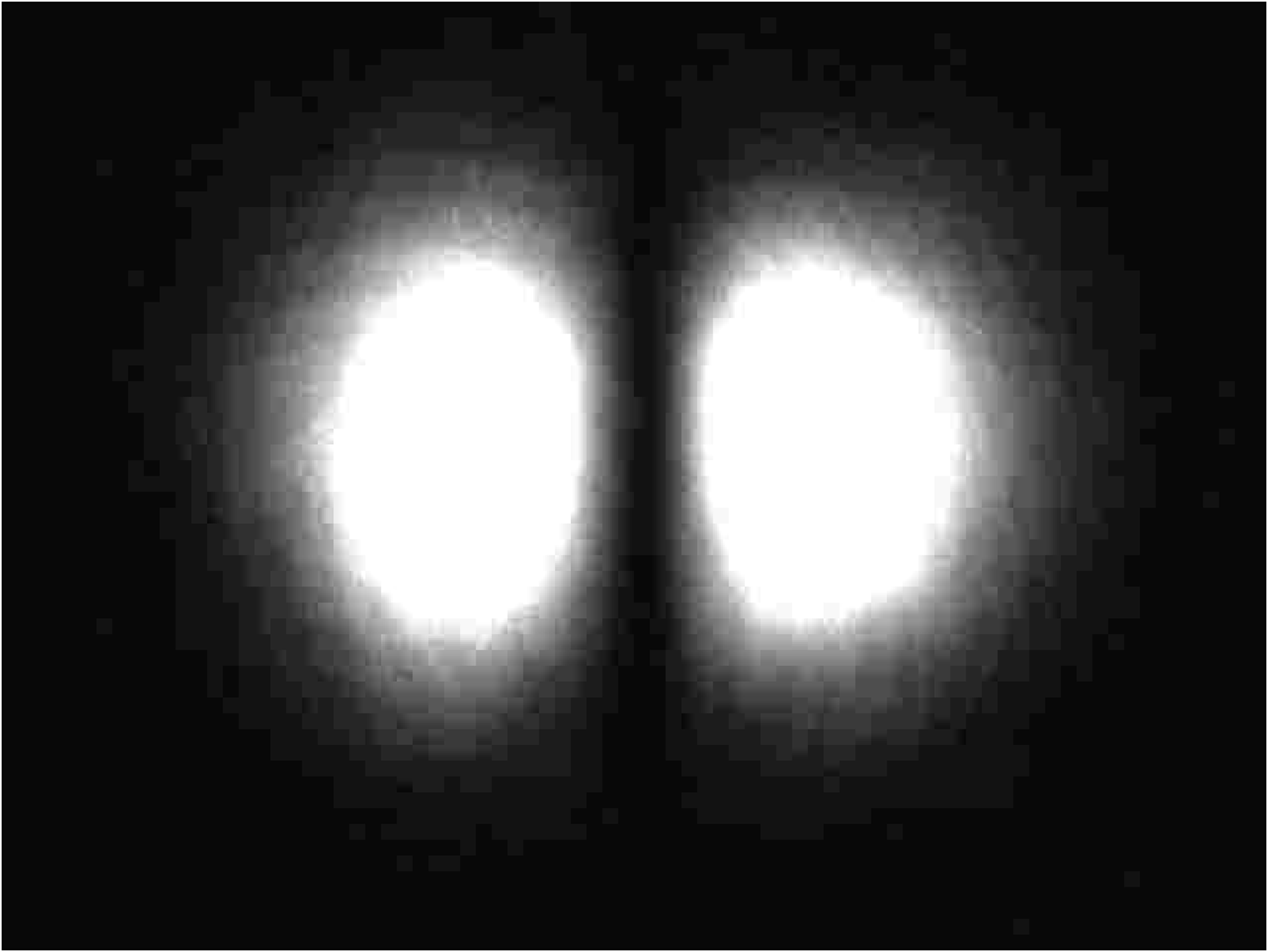}}& 
{\includegraphics[width=000.075\textwidth]{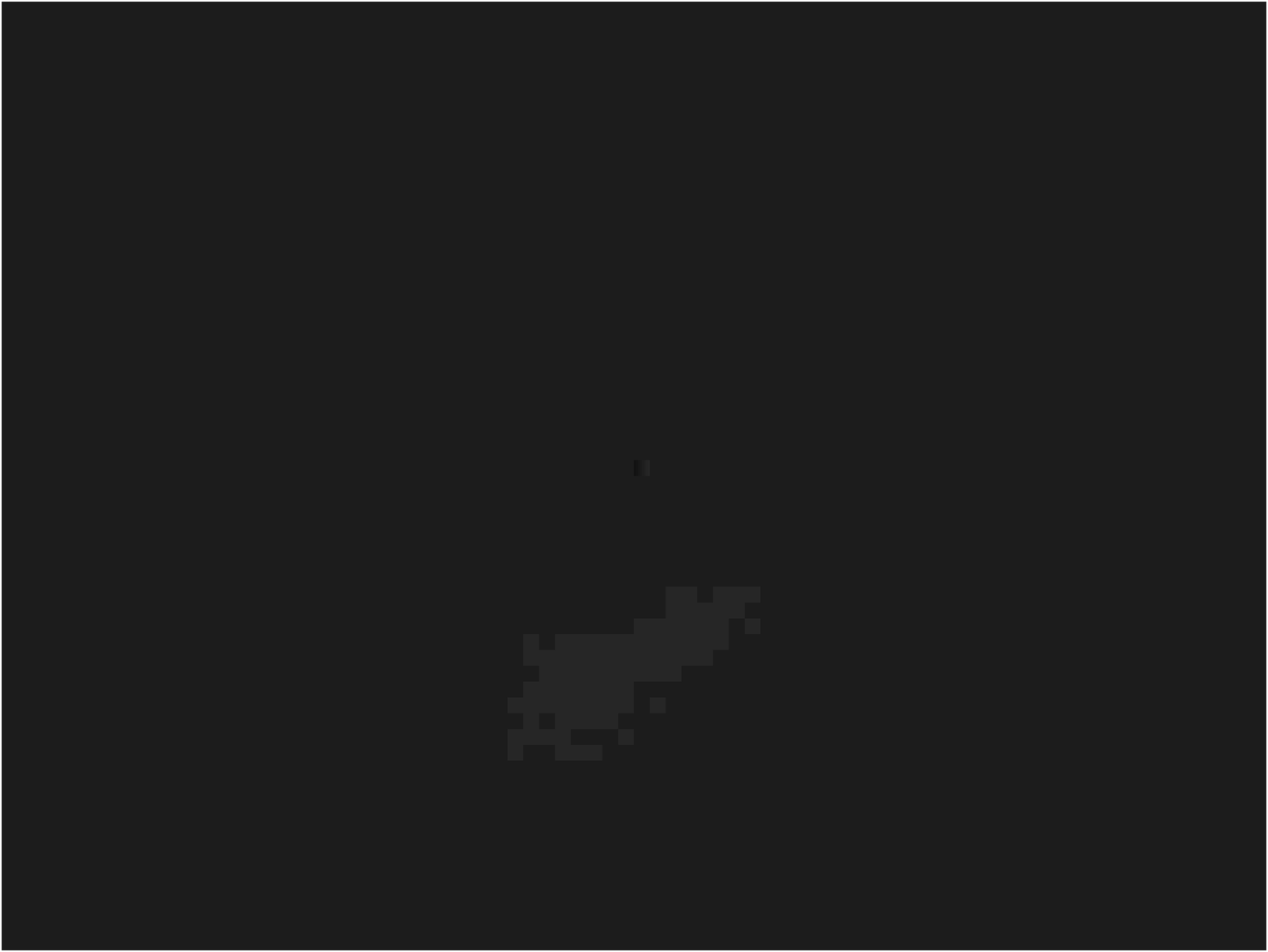}}&
{\includegraphics[width=000.075\textwidth]{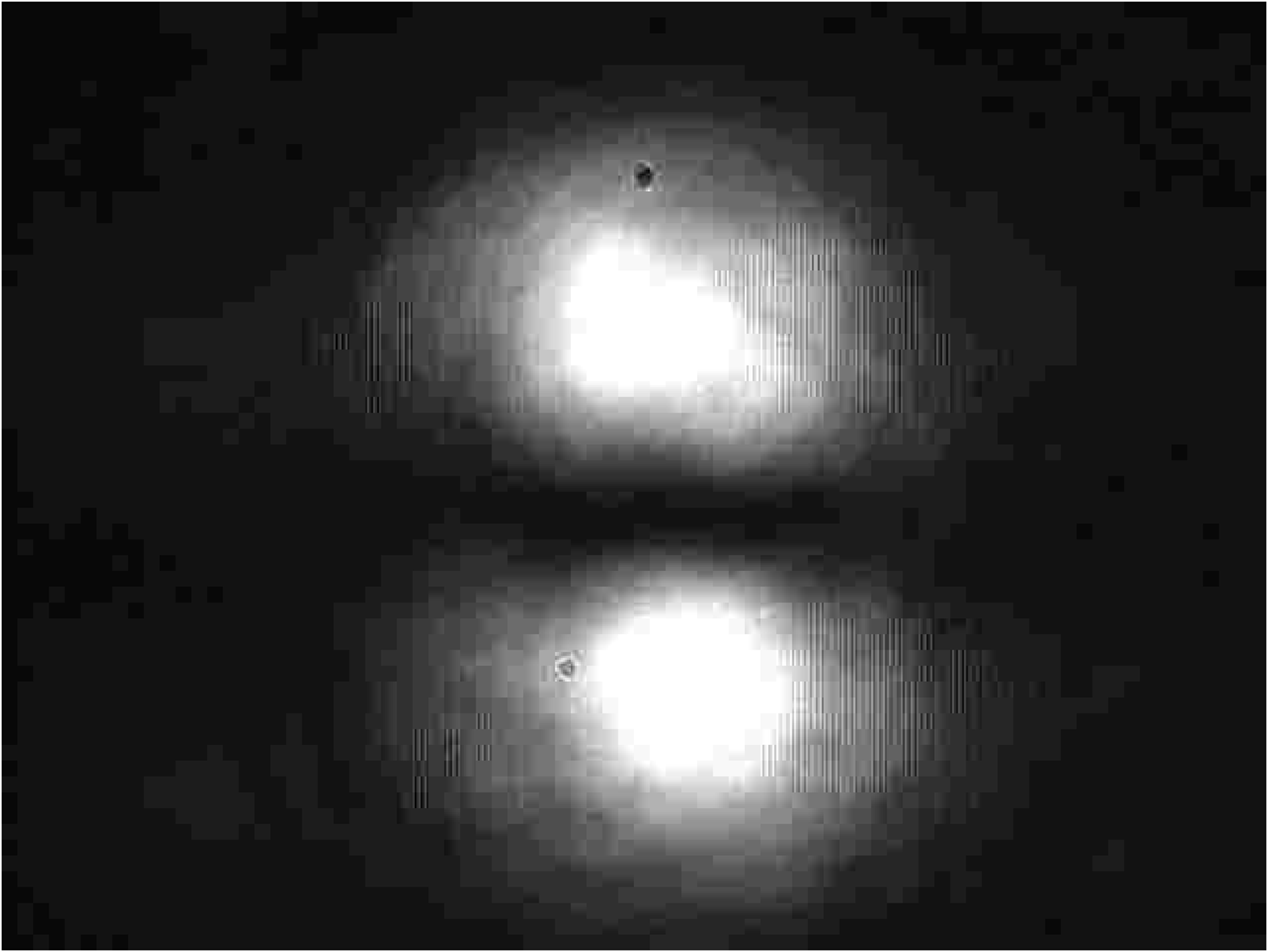}}&
{\includegraphics[width=000.075\textwidth]{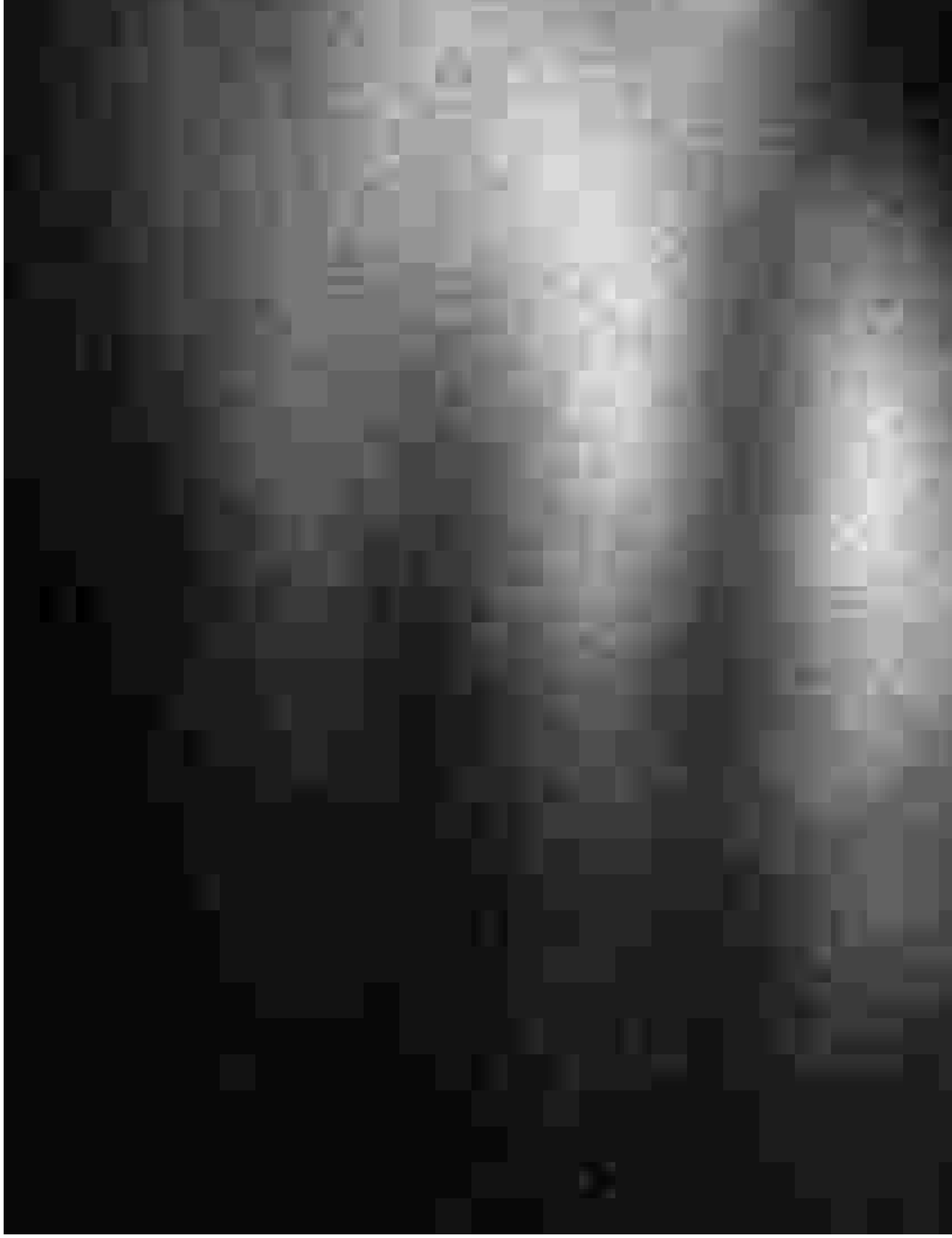}} \\ 

{\begin{array} {c} {HG_{45^{\circ}}} \\ { } \end{array}}&
{\includegraphics[width=000.075\textwidth]{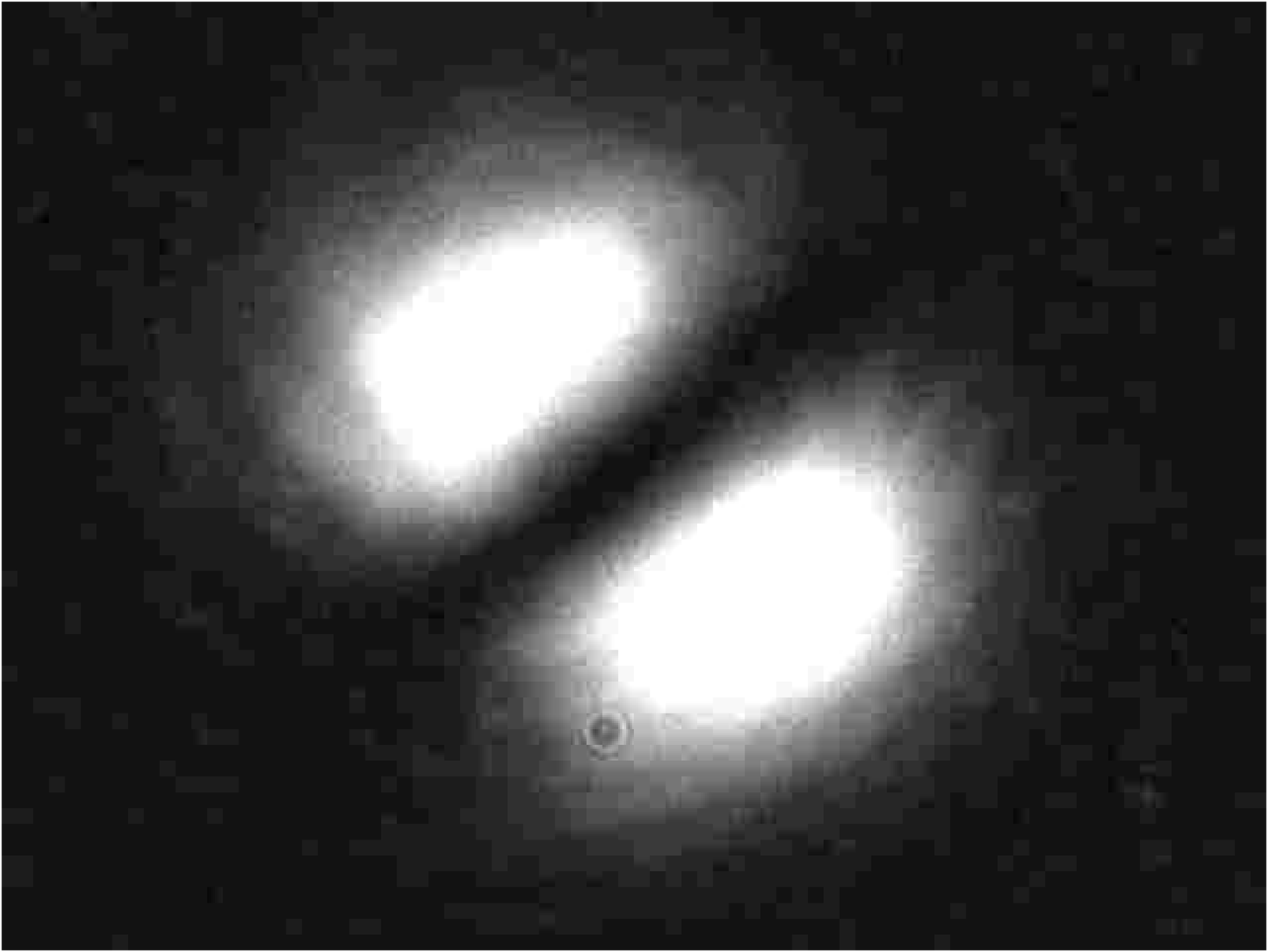}}&
{\includegraphics[width=000.075\textwidth]{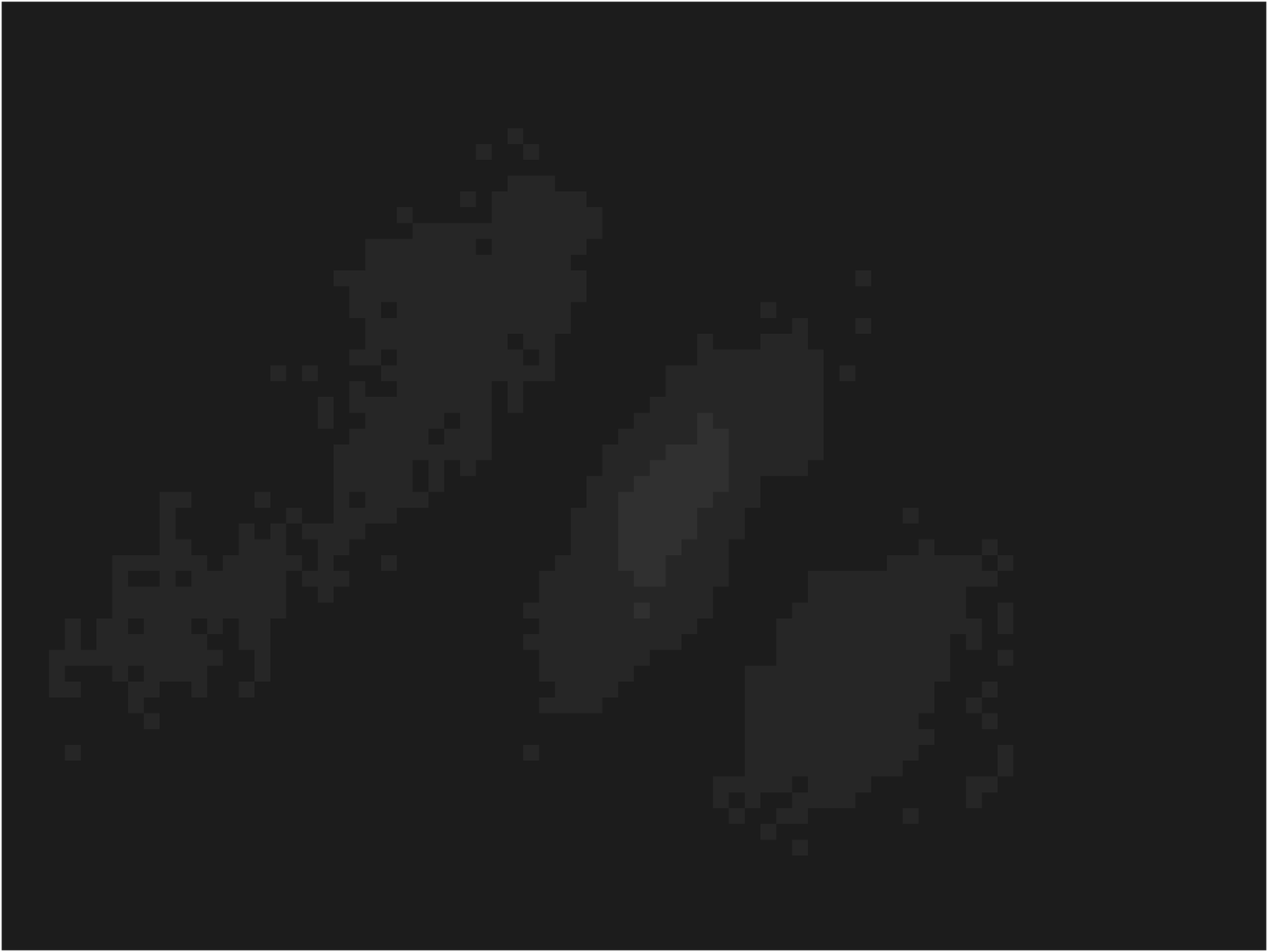}}& 
{\includegraphics[width=000.075\textwidth]{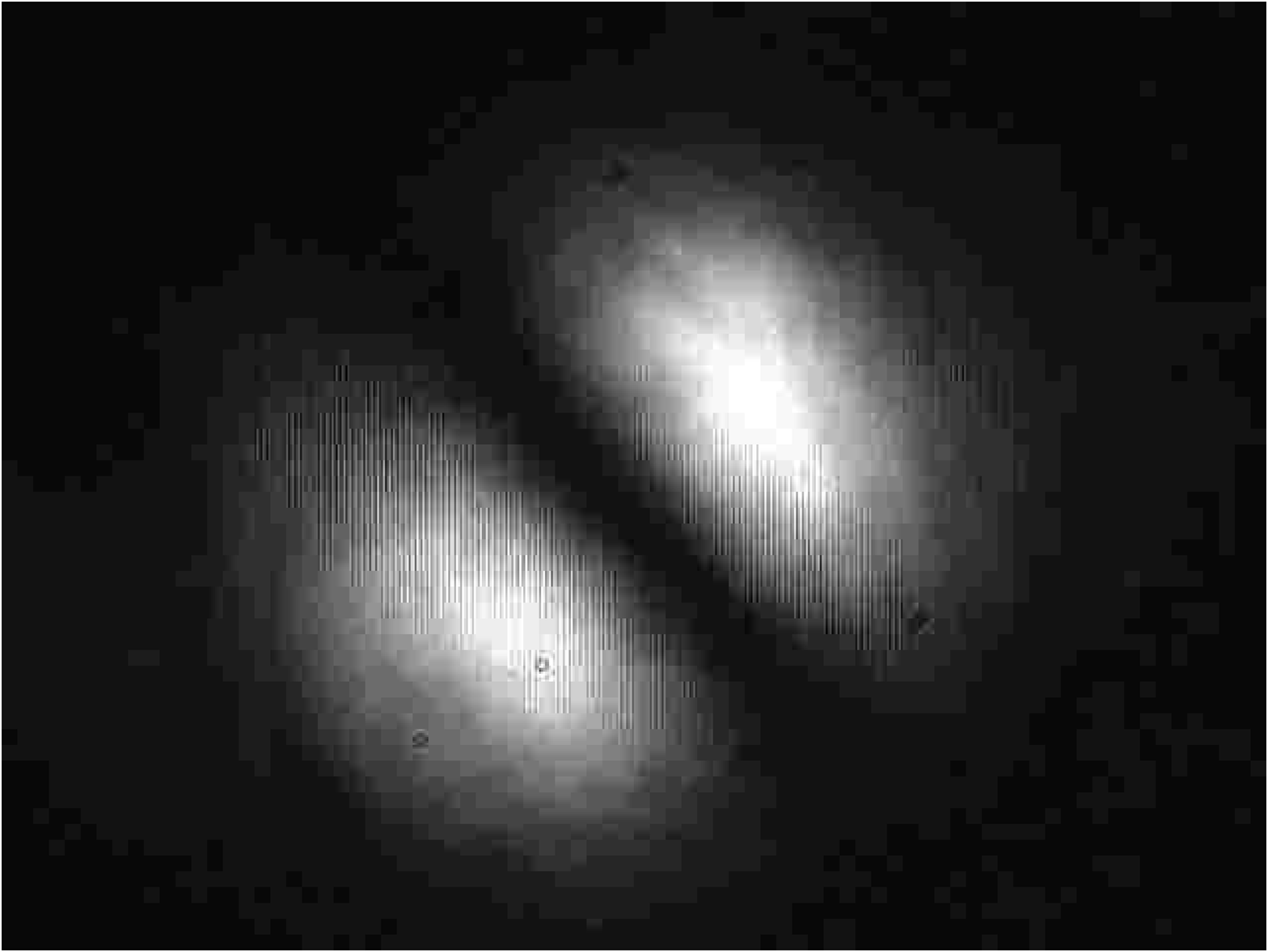}}& 
{\includegraphics[width=000.075\textwidth]{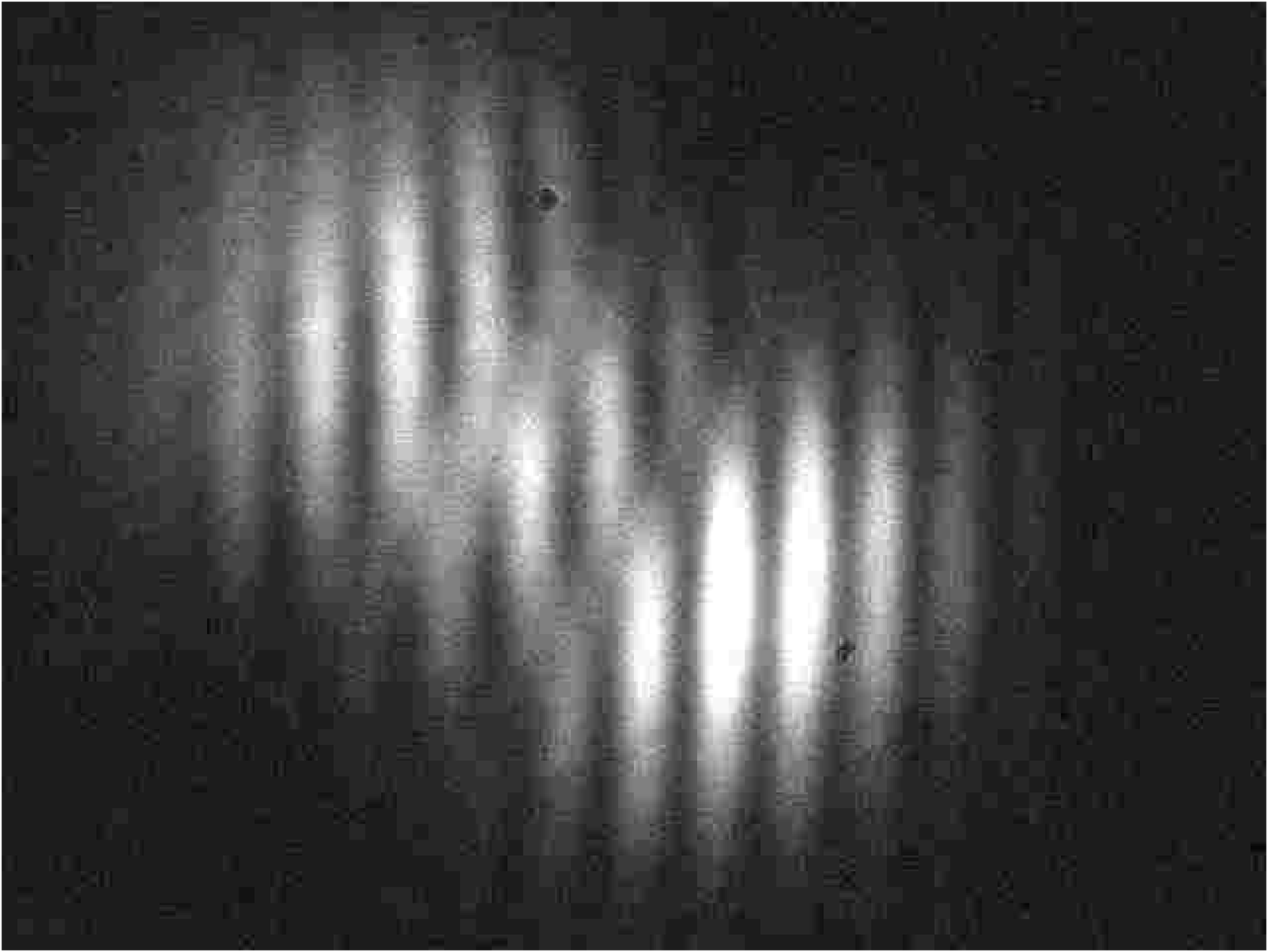}}\\ 

{\begin{array} {c} {HG_{11}} \\ { } \end{array}}&
{\includegraphics[width=000.075\textwidth]{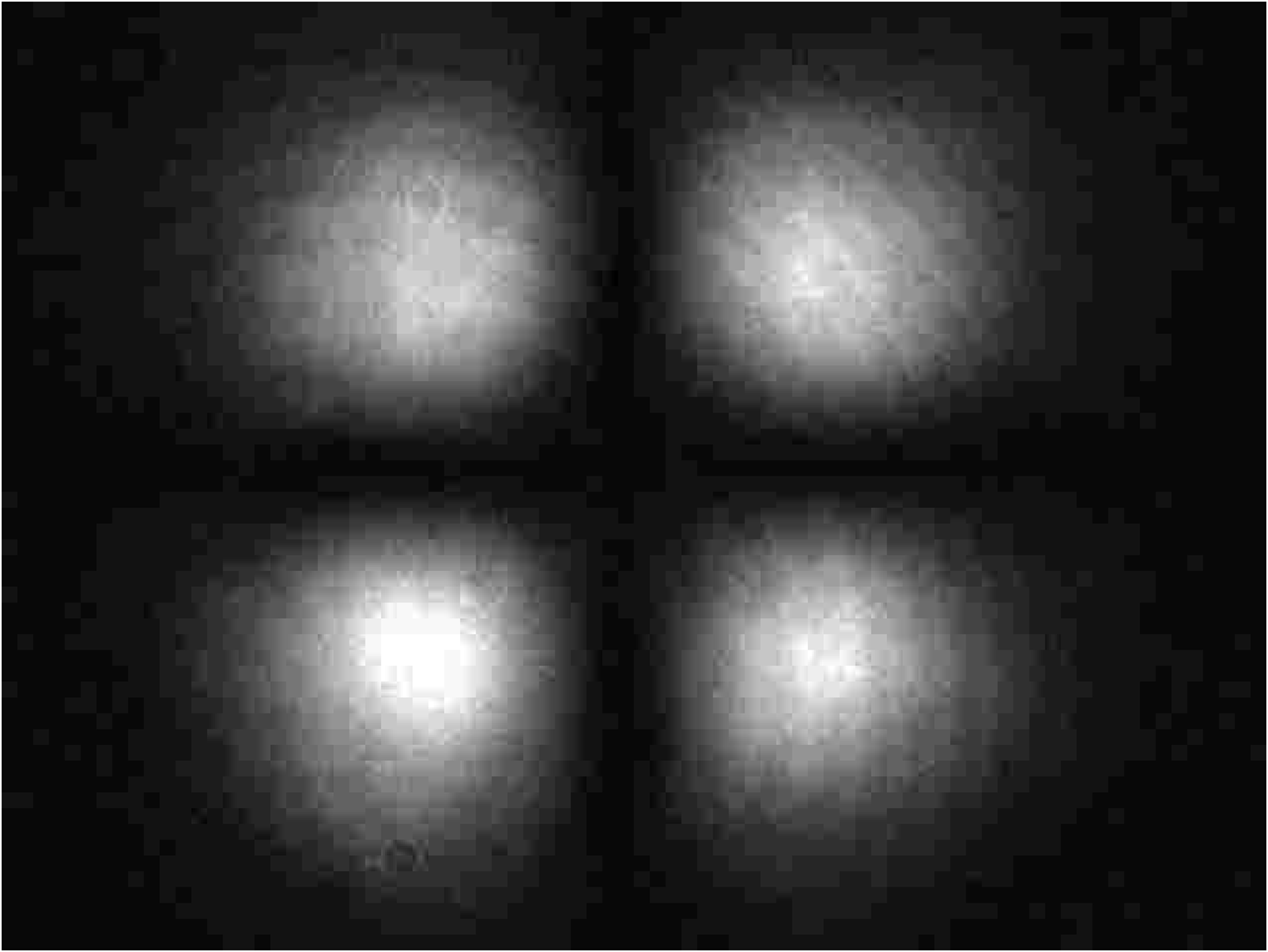}}&
{\includegraphics[width=000.075\textwidth]{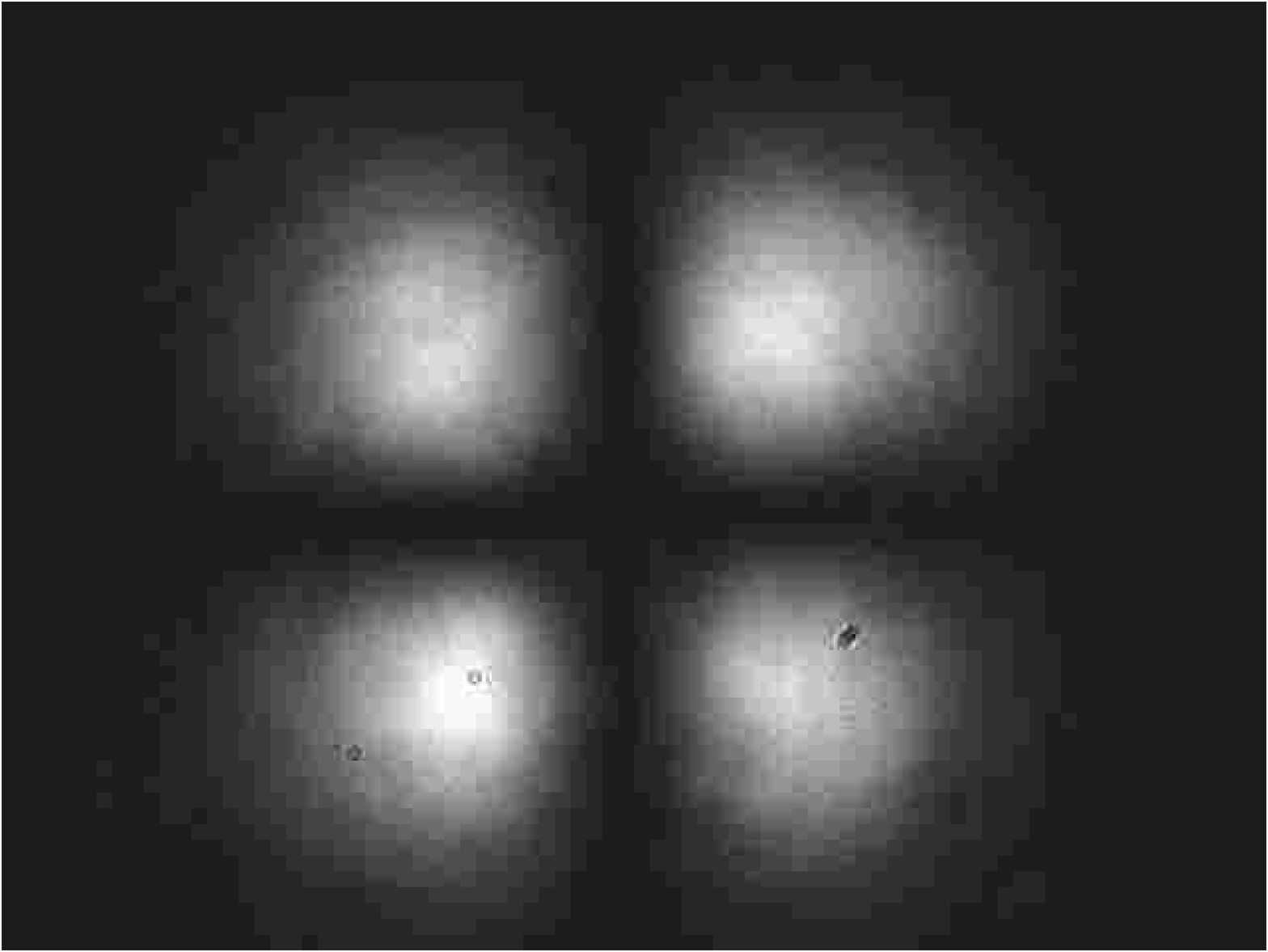}}& 
{\includegraphics[width=000.075\textwidth]{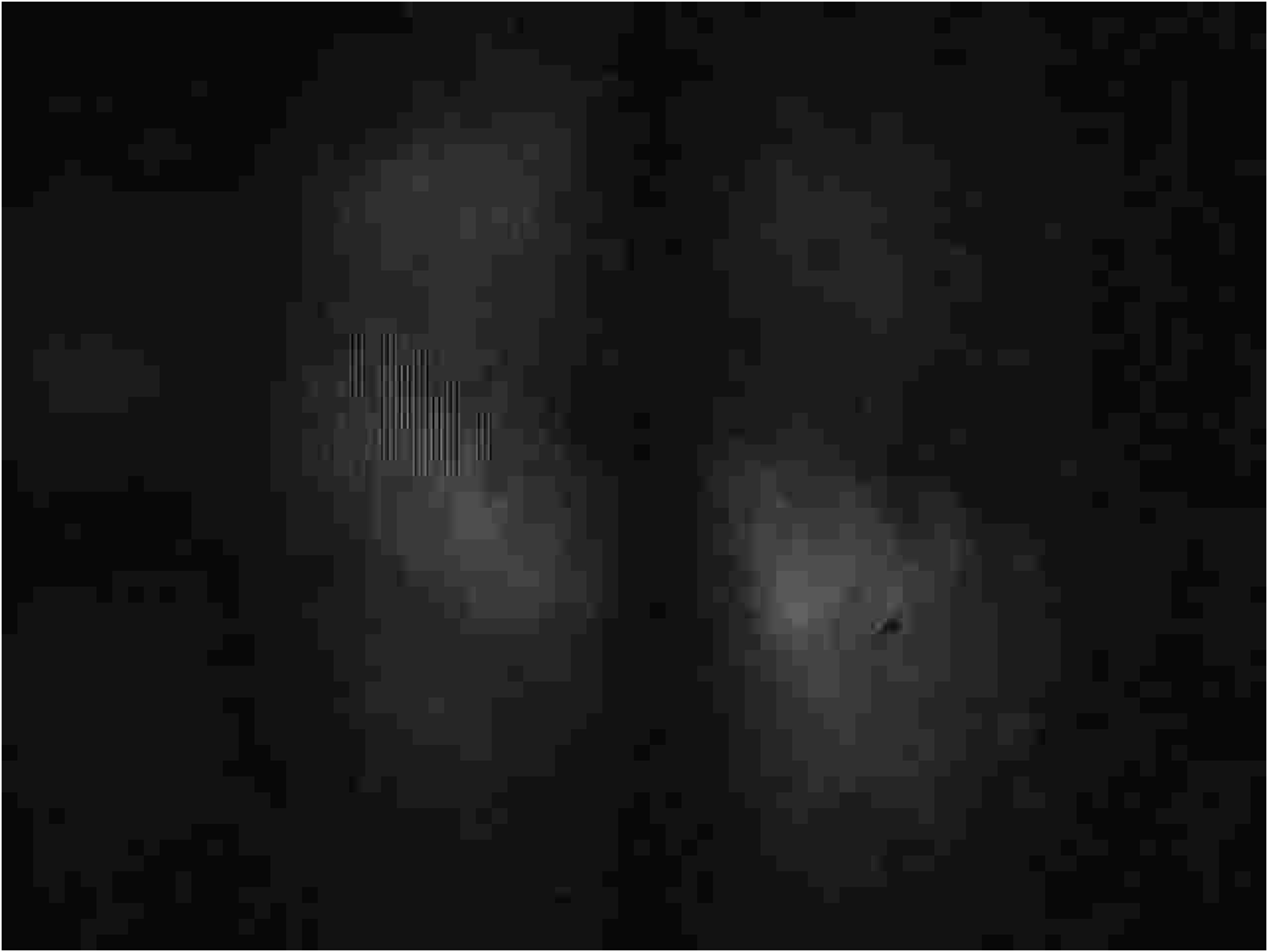}}& 
{\includegraphics[width=000.075\textwidth]{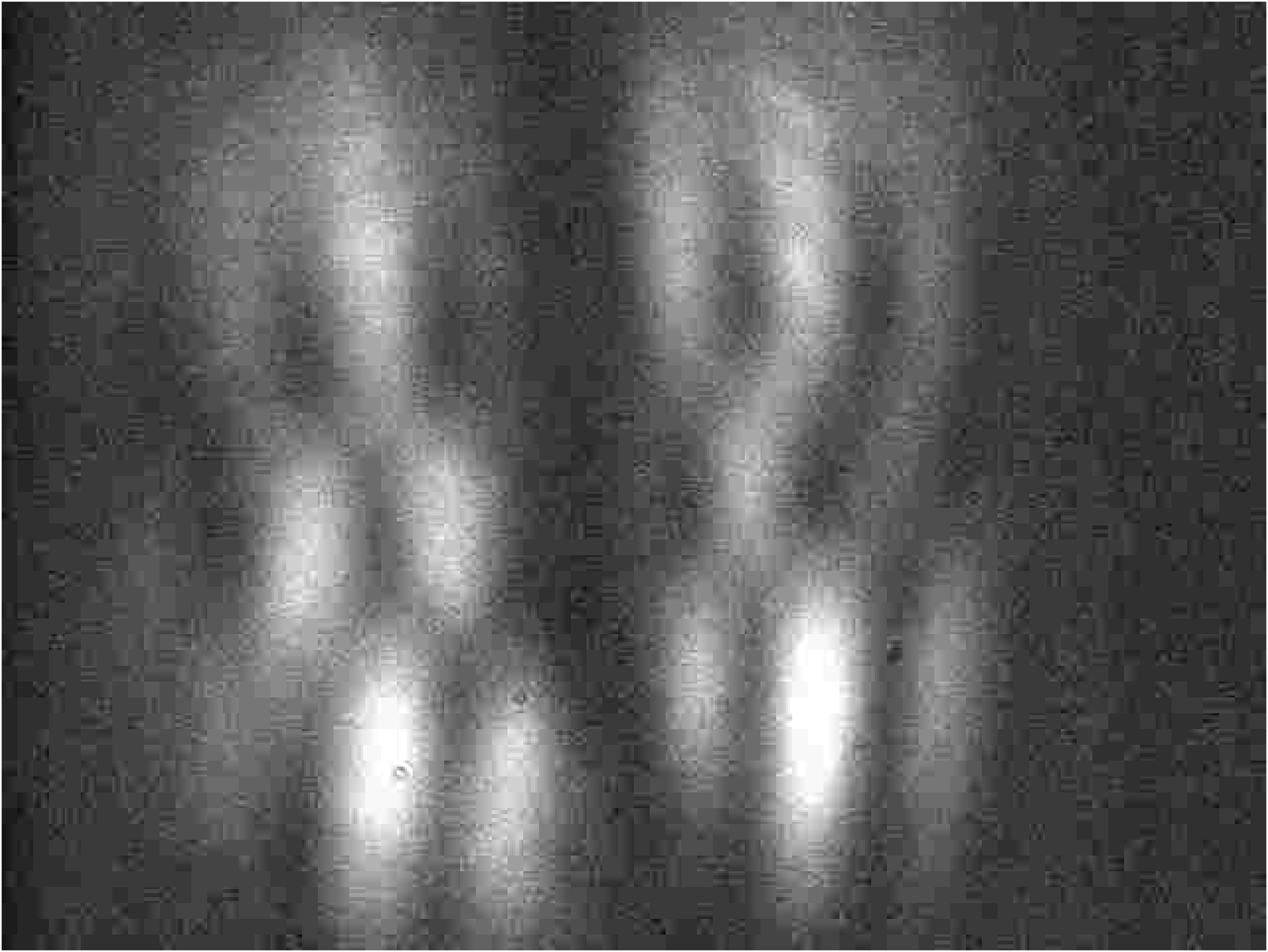}}\\ 

{\begin{array} {c} {HG_{15}} \\ { } \end{array}}& 
{\includegraphics[width=000.075\textwidth]{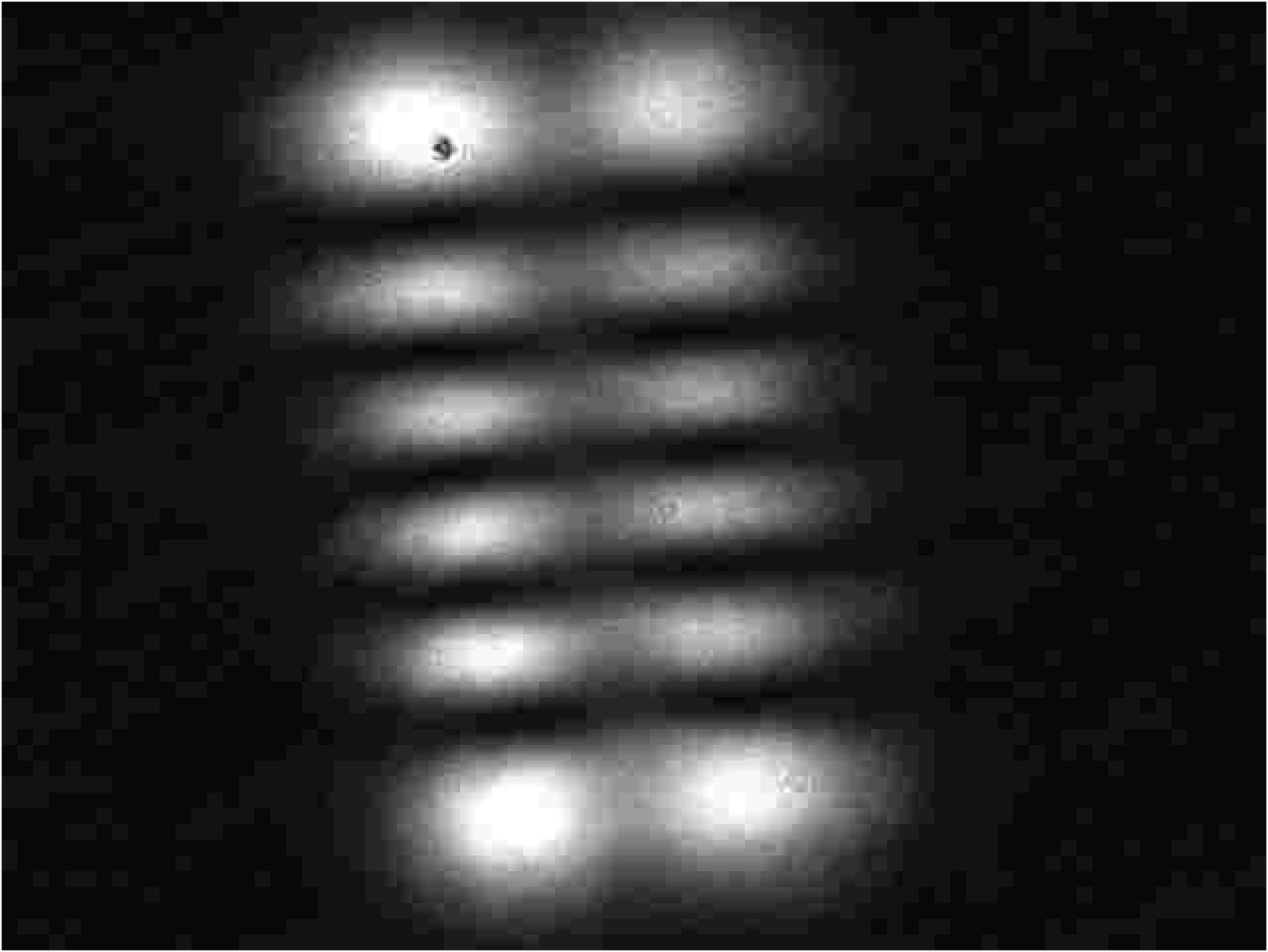}}& 
{\includegraphics[width=000.075\textwidth]{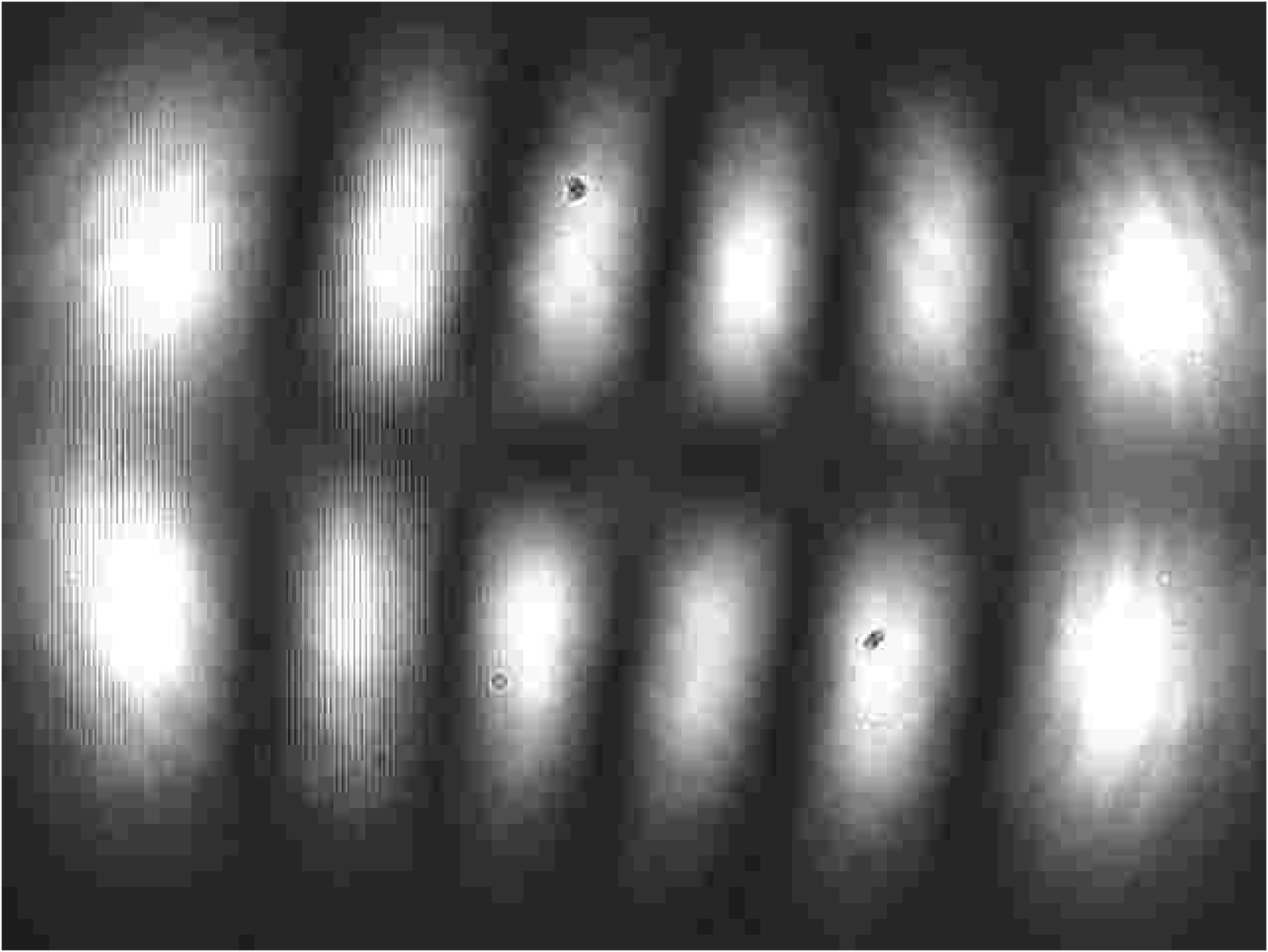}}& 
{\includegraphics[width=000.075\textwidth]{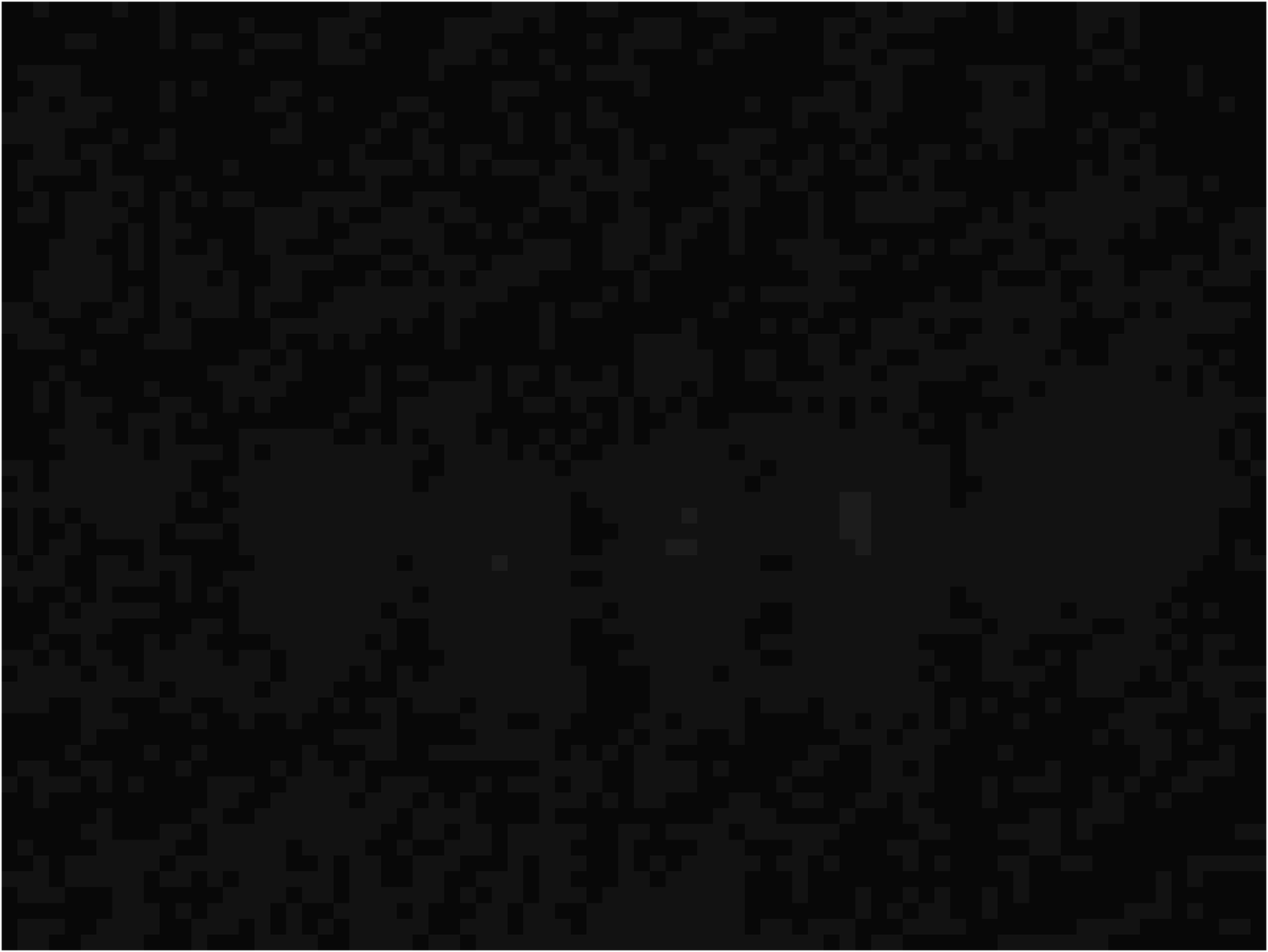}}& 
{ }\\ 

{\begin{array} {c} {HG_{32}} \\ { } \end{array}}& 
{\includegraphics[width=000.075\textwidth]{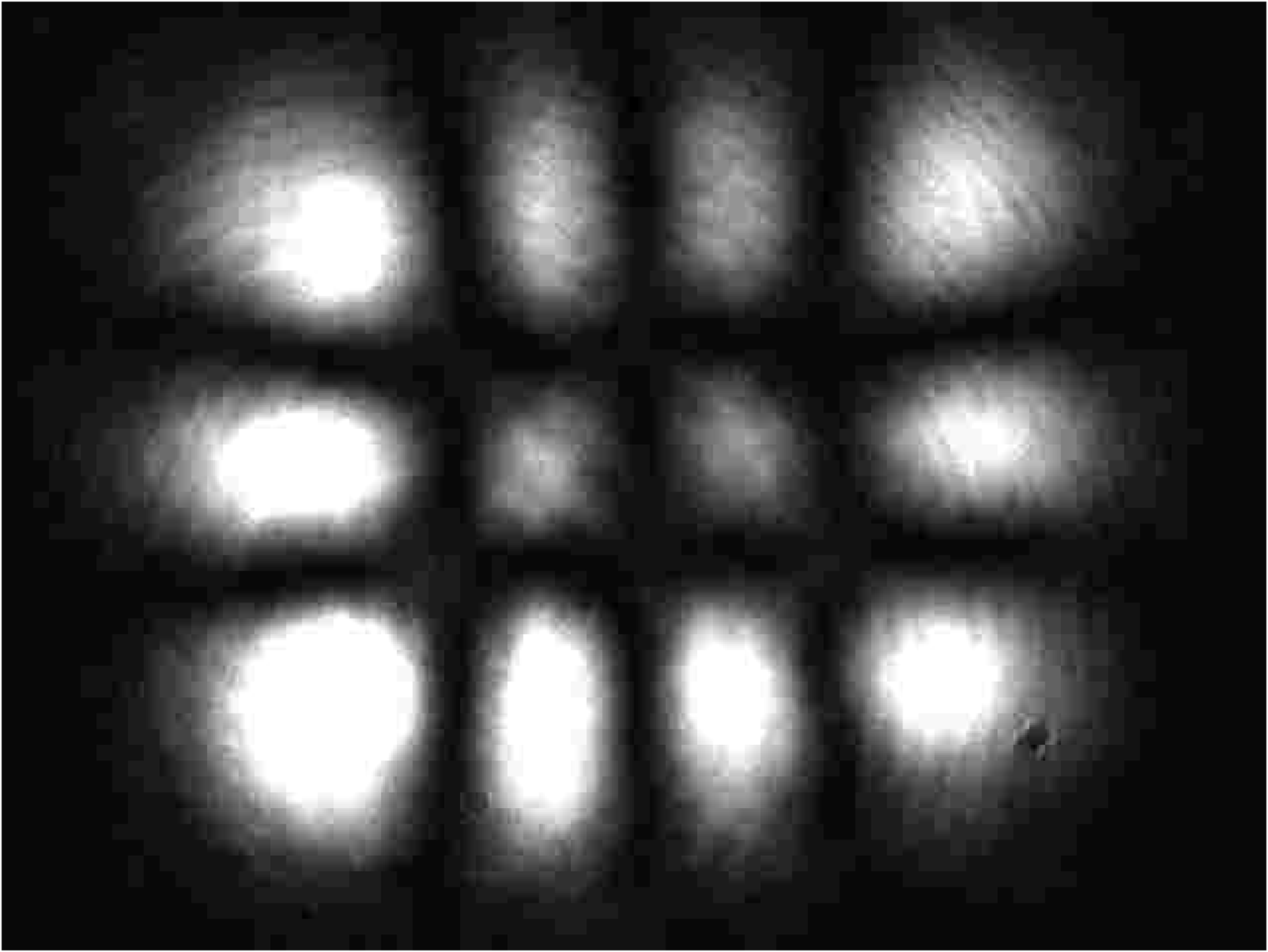}}& 
{\includegraphics[width=000.075\textwidth]{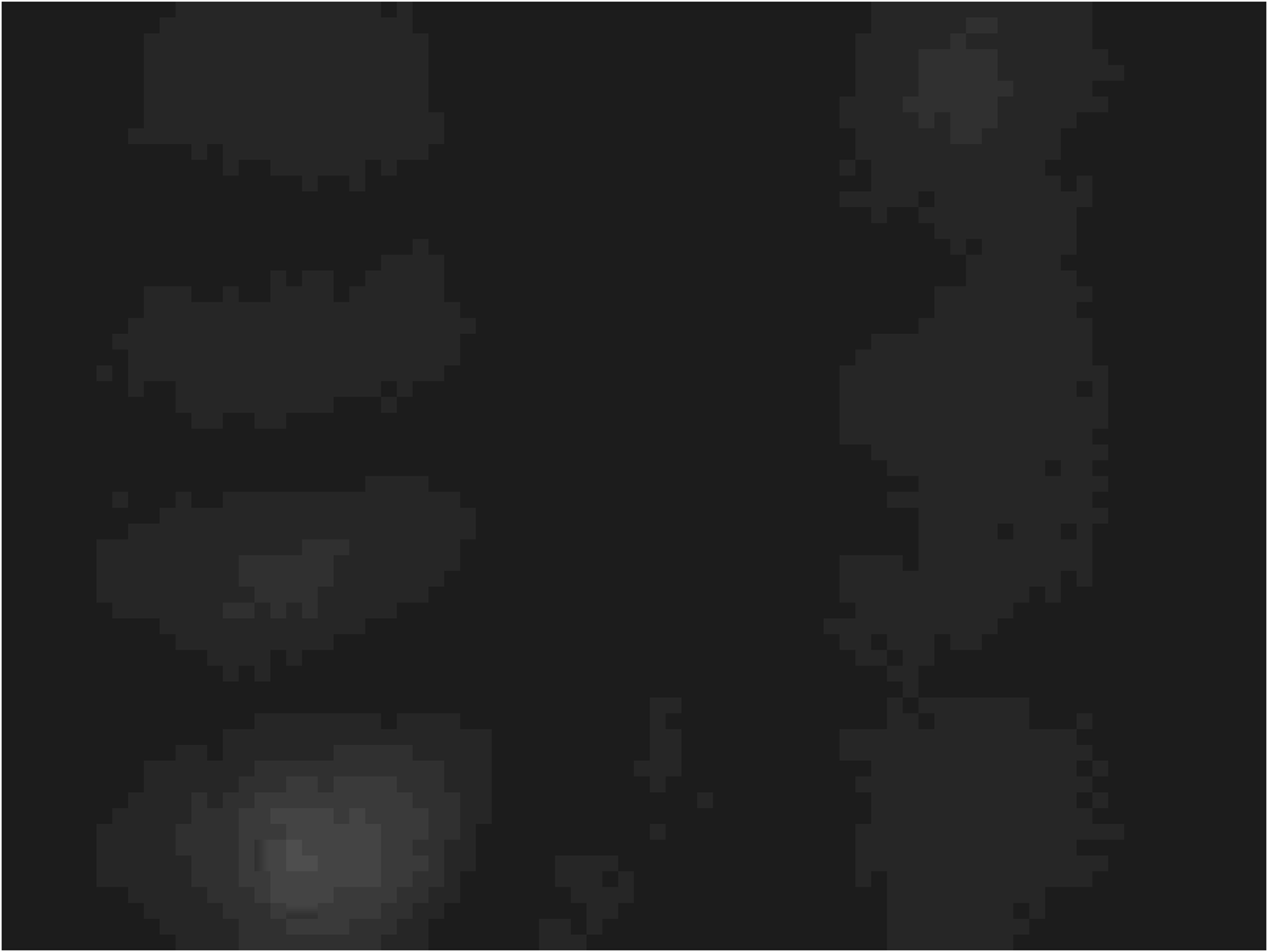}}&
{\includegraphics[width=000.075\textwidth]{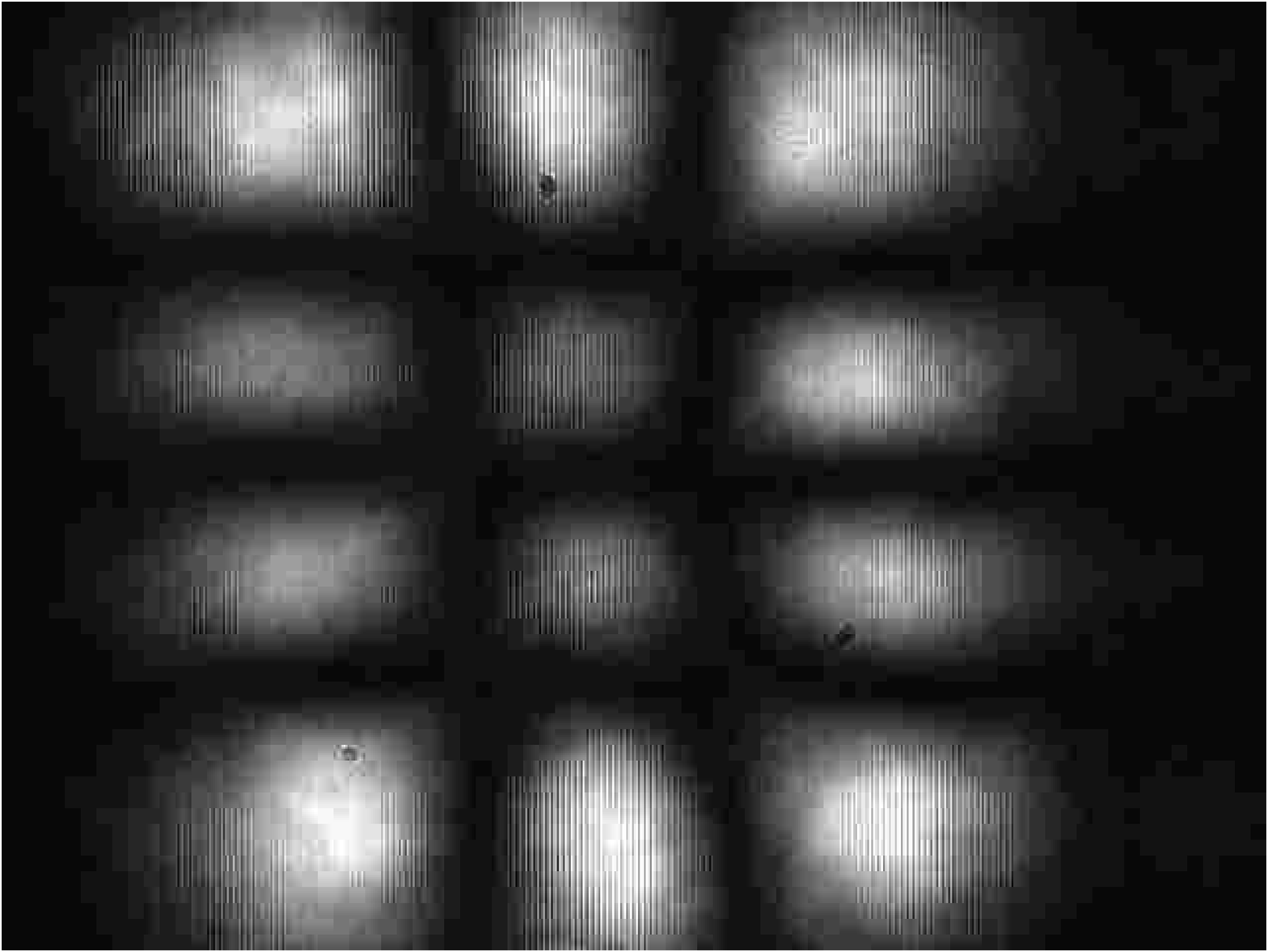}}&
{ } \\ 

{\begin{array}{c} {\textrm{\scriptsize{3 Mode}}} \\ {\textrm{\scriptsize{Fiber}}} \\ { } \\ { }\end{array}}&
{\includegraphics[width=000.075\textwidth]{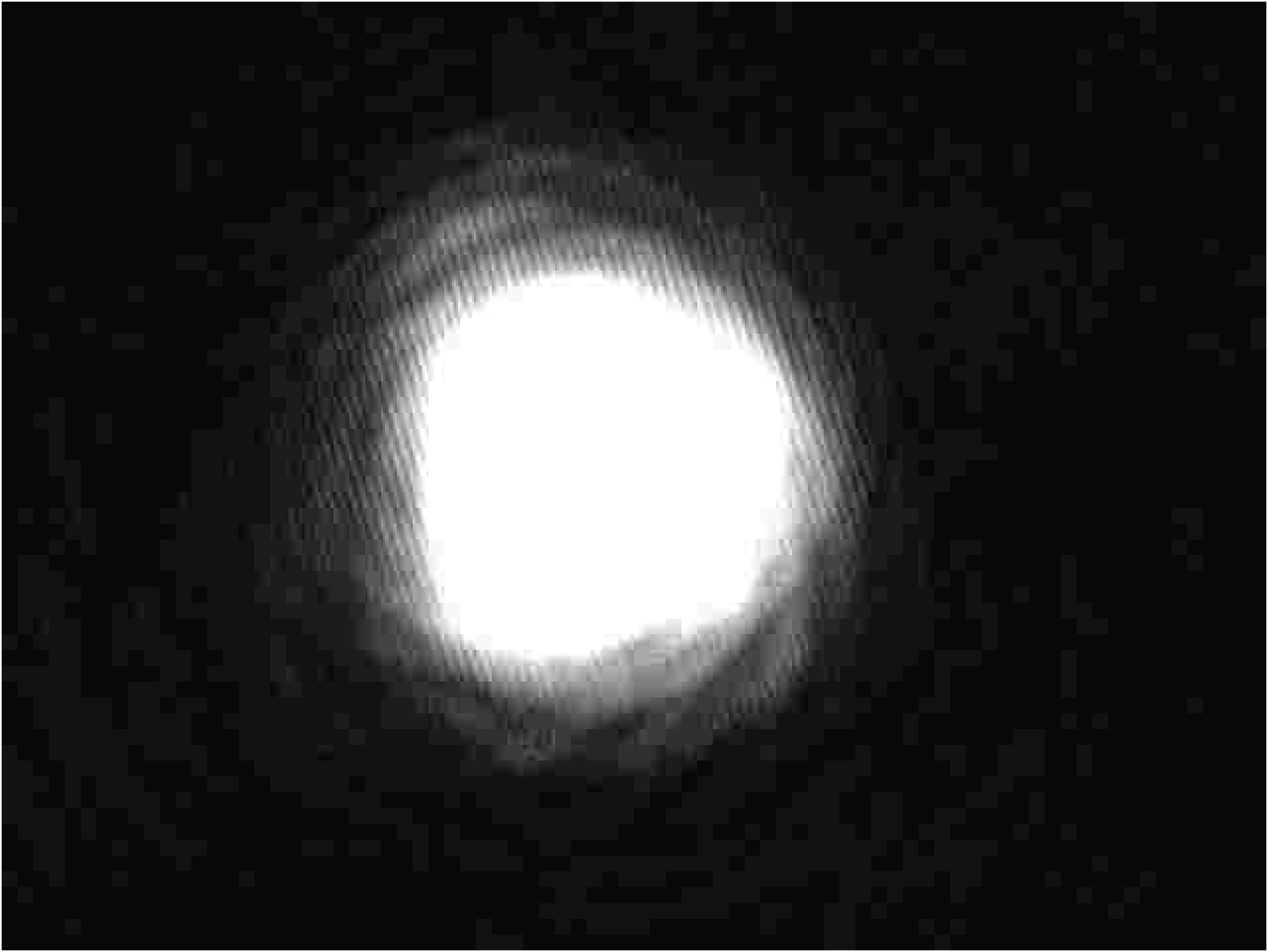}}&
{\includegraphics[width=000.075\textwidth]{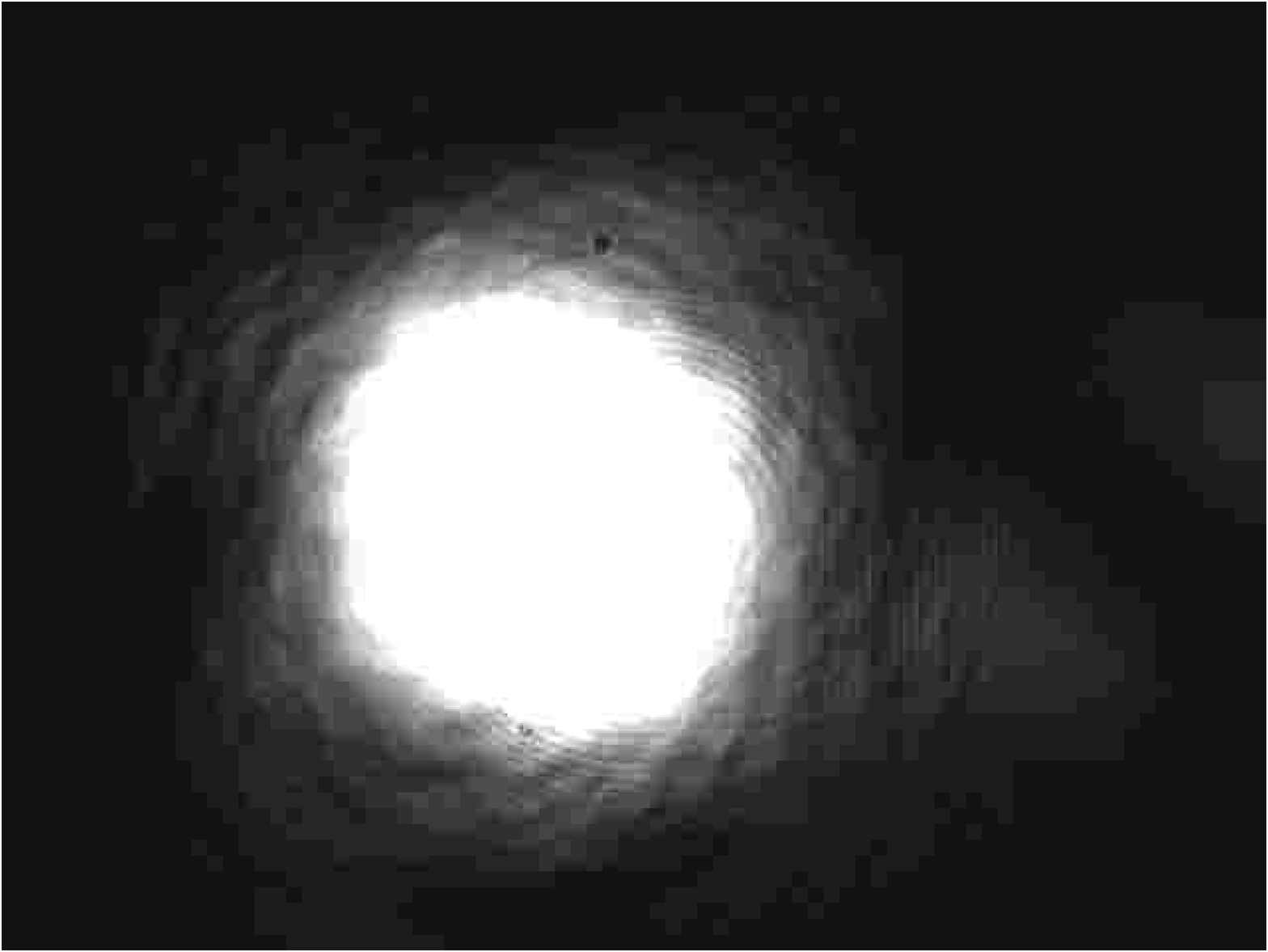}}&
{\includegraphics[width=000.075\textwidth]{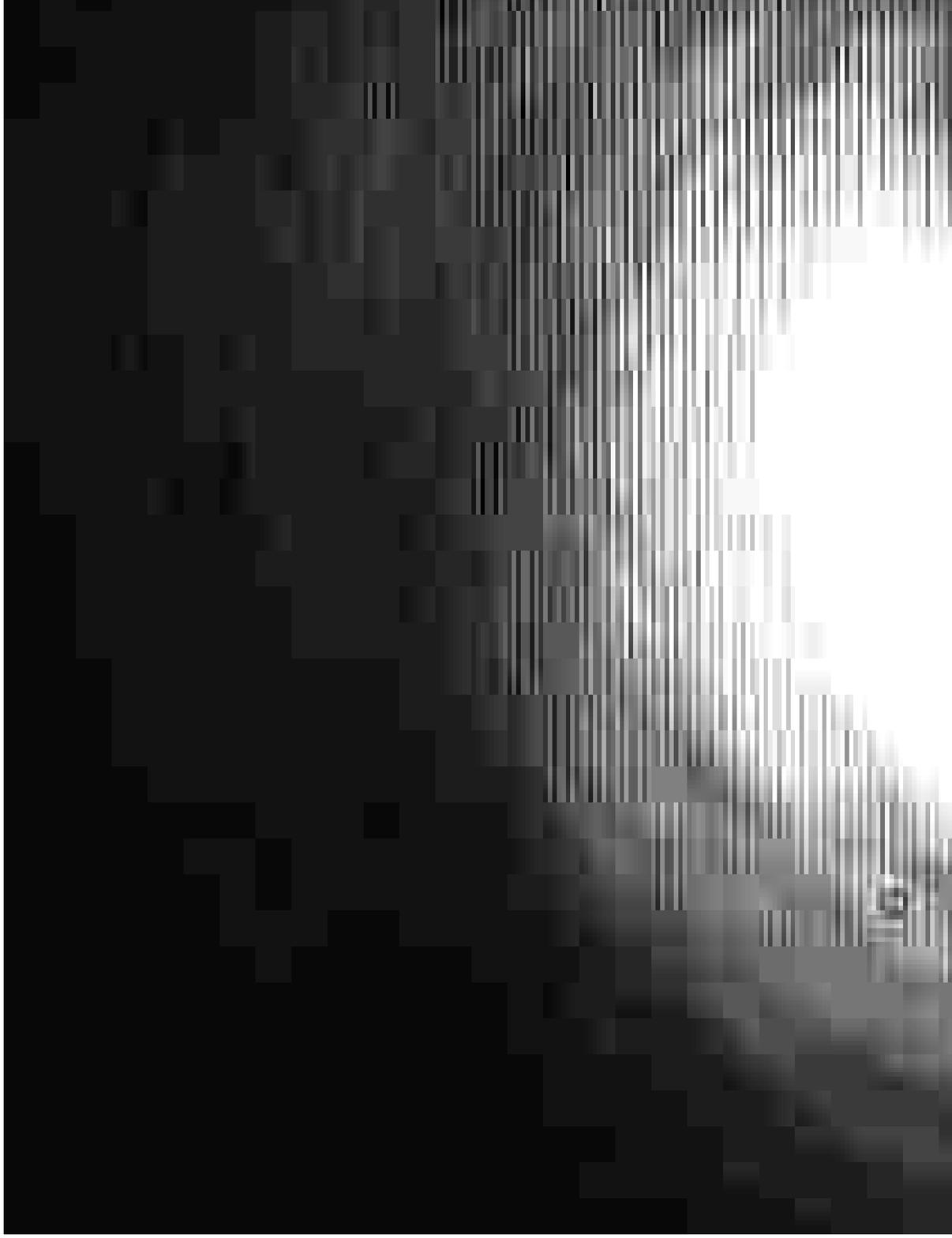}}&
{ }

\end{array}} 

\end{array}$

\caption{\label{fig:Experiment} Observed output intensity profiles and interference patterns for a given input field.} 

\end{figure}

\begin{figure}[!t] 

$\begin{array}{c} 

{\textrm{\Large{Theory}}} \\ 

{\begin{array}{ccccc} 

{\begin{array}{c} {\textrm{\scriptsize{Input}}} \\ {\textrm{\scriptsize{Mode}}} \end{array}}& 
{\begin{array}{c} {\textrm{\scriptsize{Input}}} \\ {\textrm{\scriptsize{Intensity}}} \\ {\textrm{\scriptsize{Profile}}} \end{array}}&
{\textrm{\scriptsize{Port A}}}& 
{\textrm{\scriptsize{Port B}}}& 
{\begin{array}{c} {\textrm{\scriptsize{Interference}}} \\ {\textrm{\scriptsize{Intensity}}} \\ {\textrm{\scriptsize{Profile}}} \end{array}} \\ 

{\begin{array} {c} {HG_{00}} \\ { } \end{array}}&
{\includegraphics[width=000.075\textwidth]{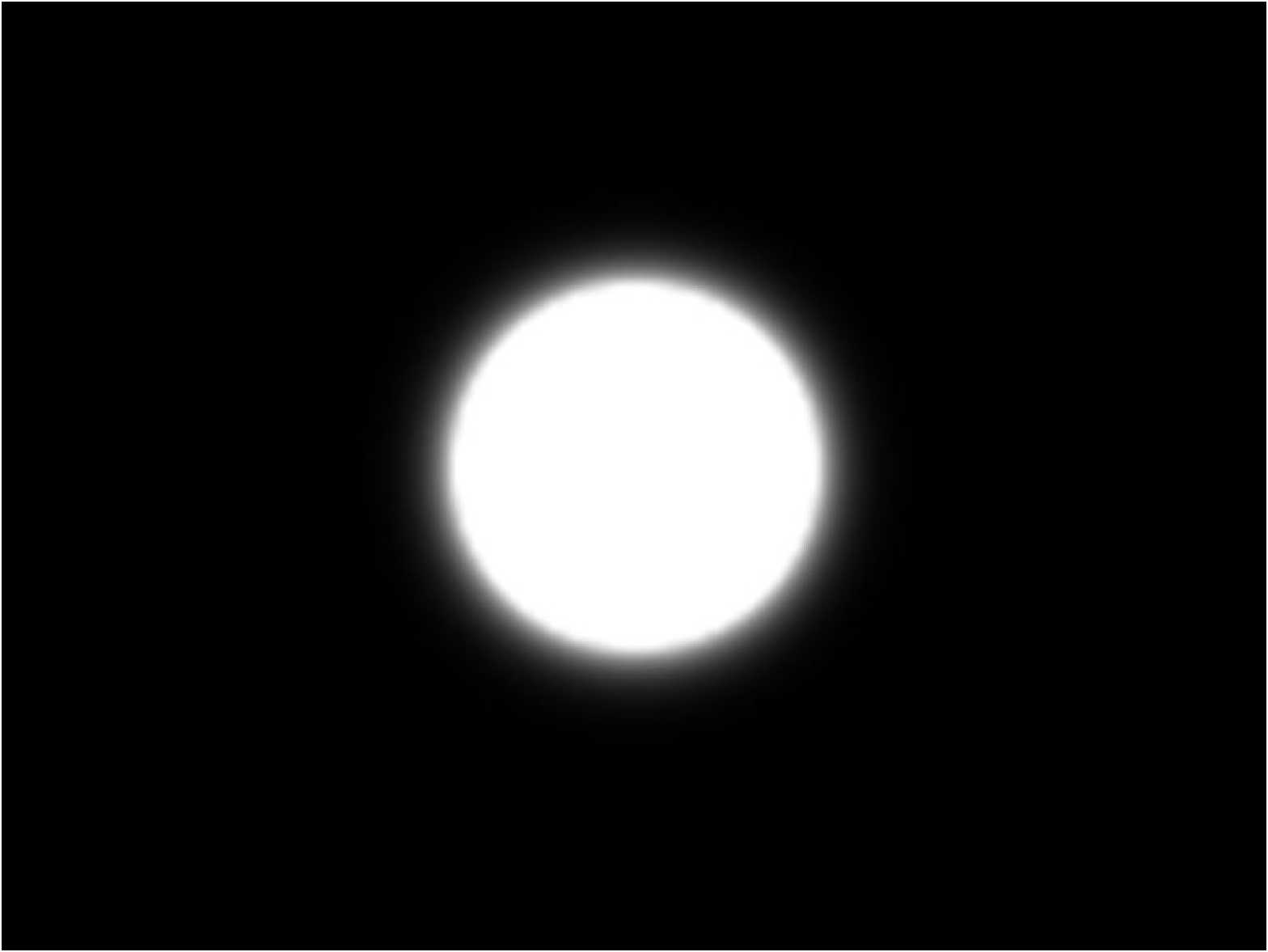}}&
{\includegraphics[width=000.075\textwidth]{00t.eps}}& 
{\includegraphics[width=000.075\textwidth]{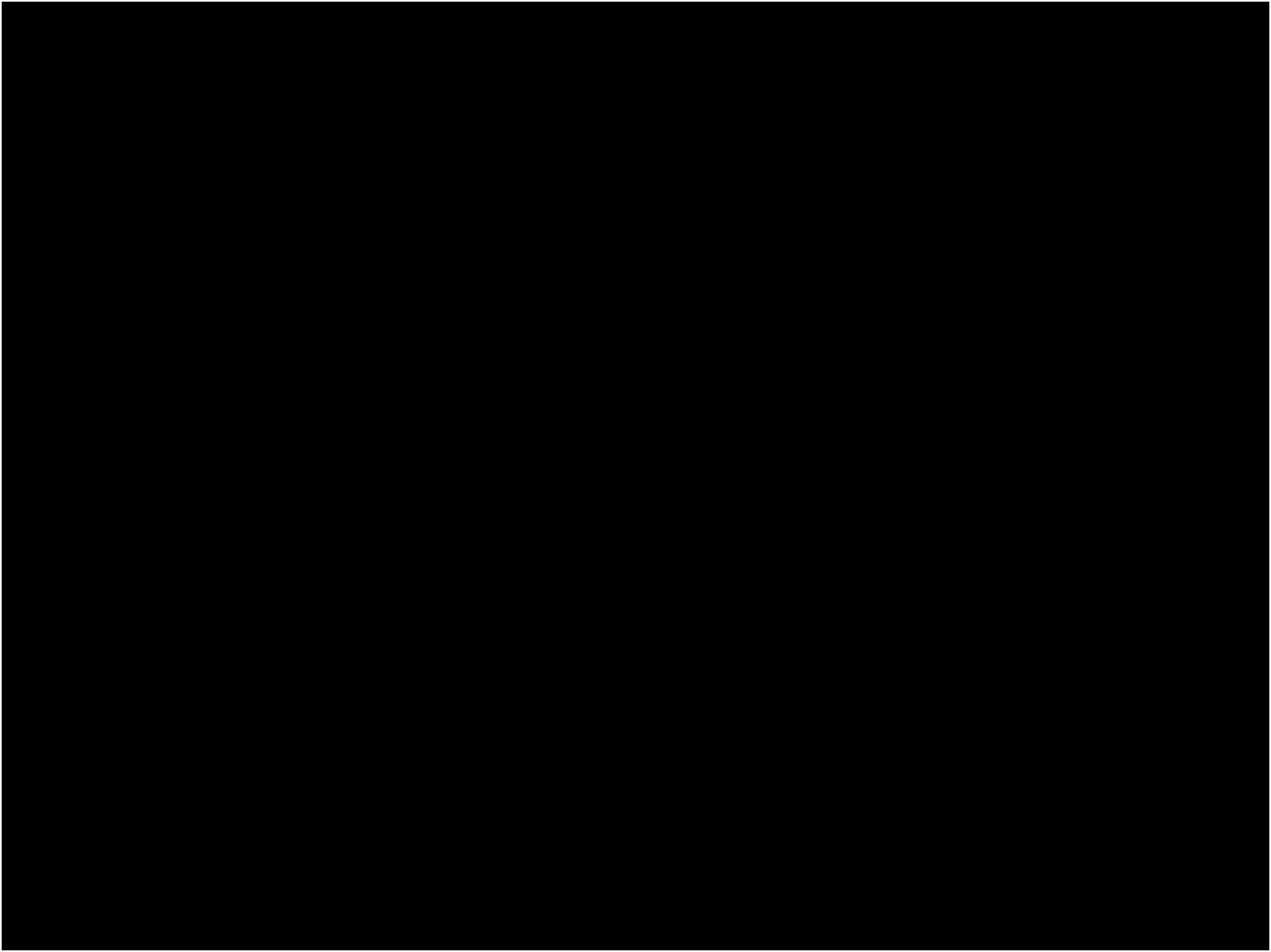}}& 
{\includegraphics[width=000.075\textwidth]{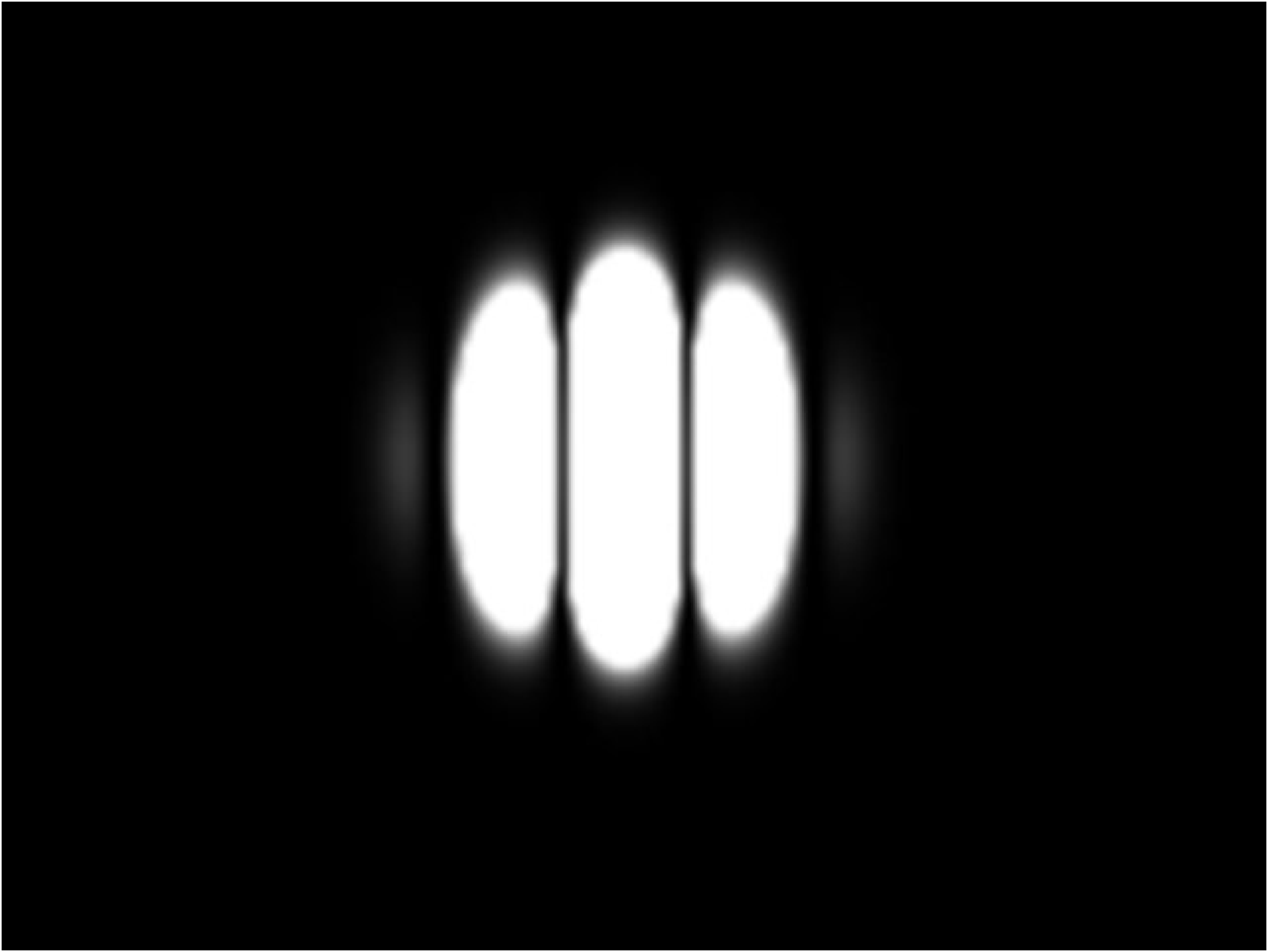}}\\ 

{\begin{array} {c} {HG_{01}} \\ { } \end{array}}& 
{\includegraphics[width=000.075\textwidth]{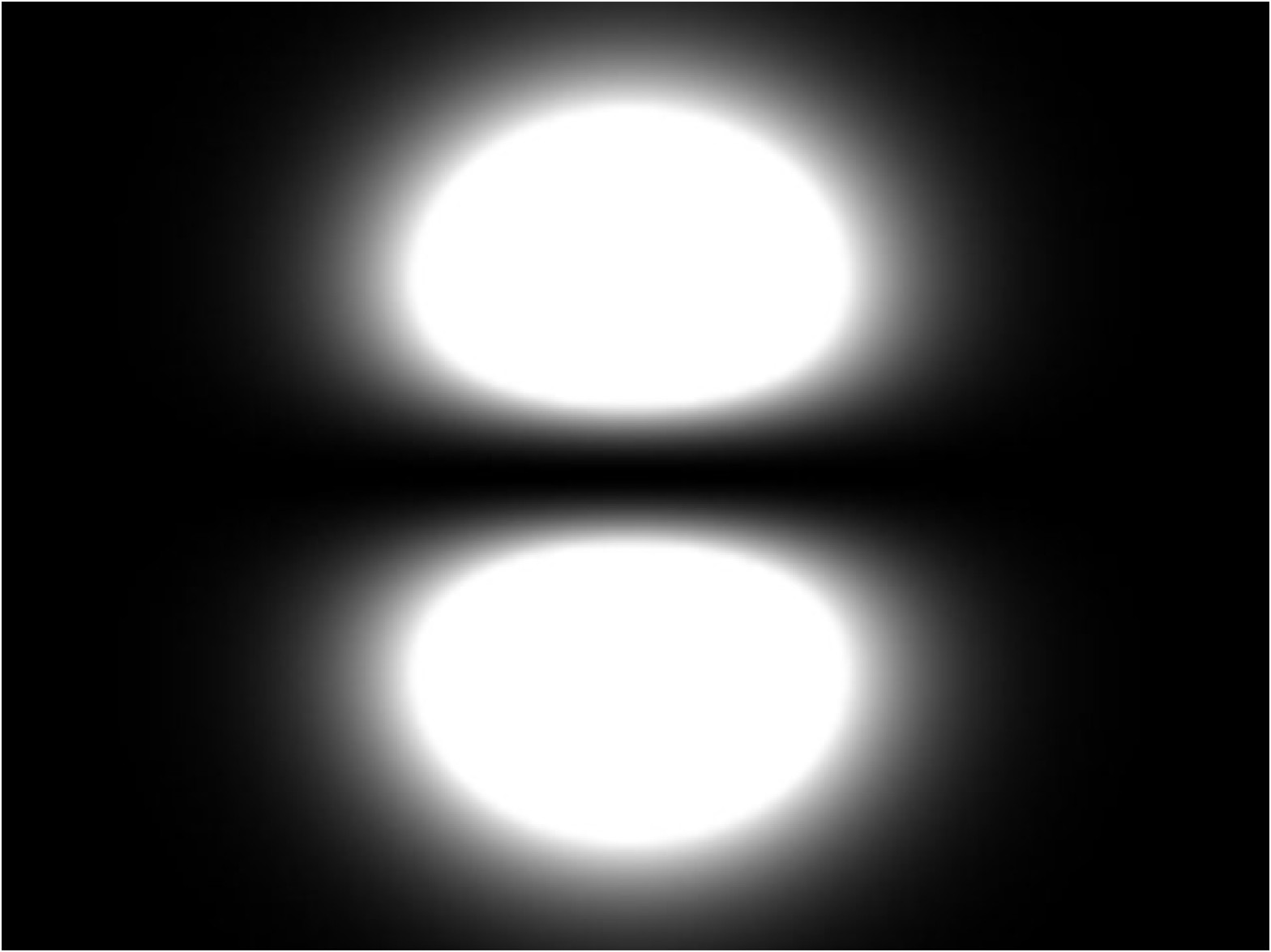}} & 
{\includegraphics[width=000.075\textwidth]{Blank.eps}}& 
{\includegraphics[width=000.075\textwidth]{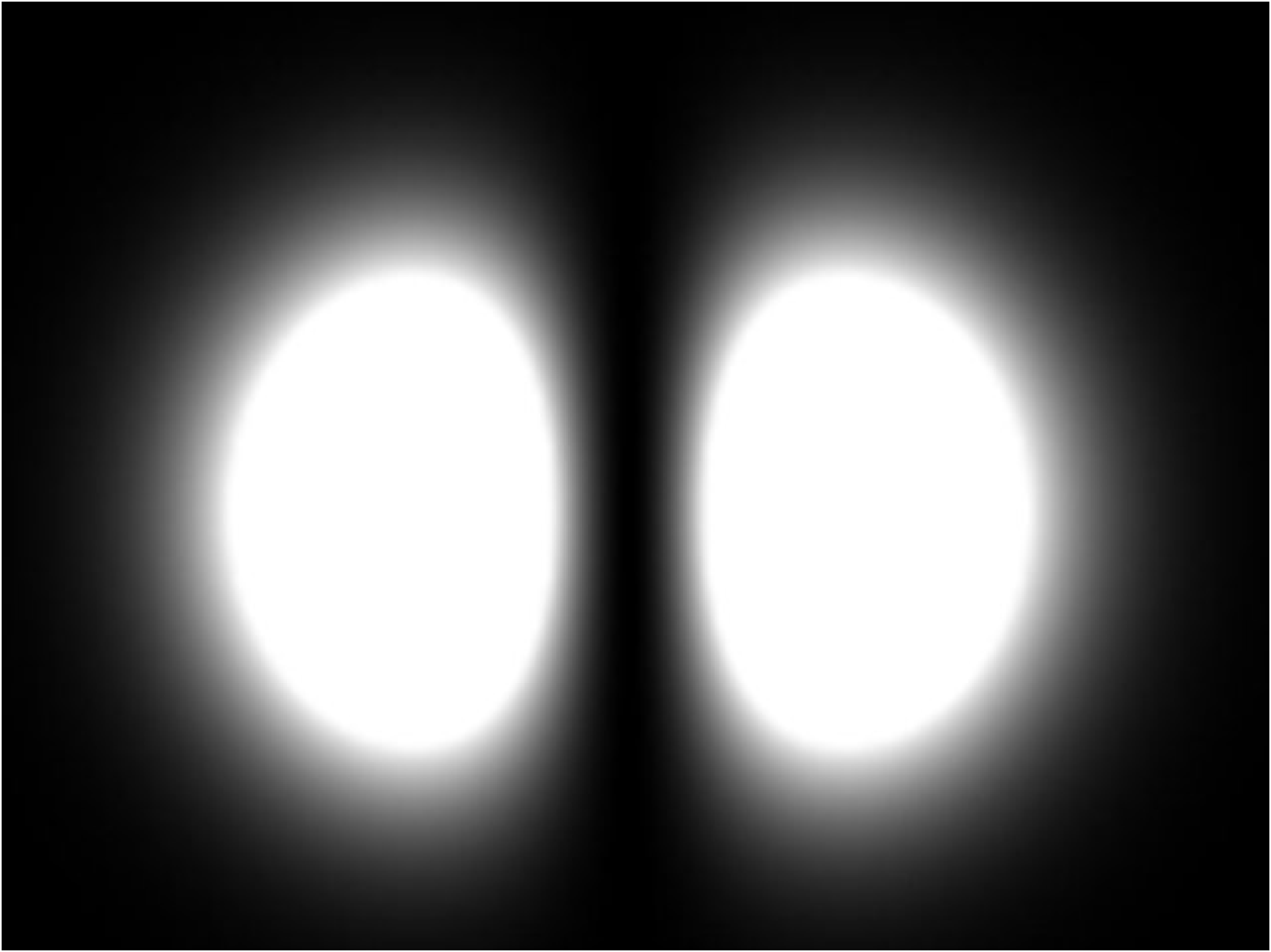}}& 
{\includegraphics[width=000.075\textwidth]{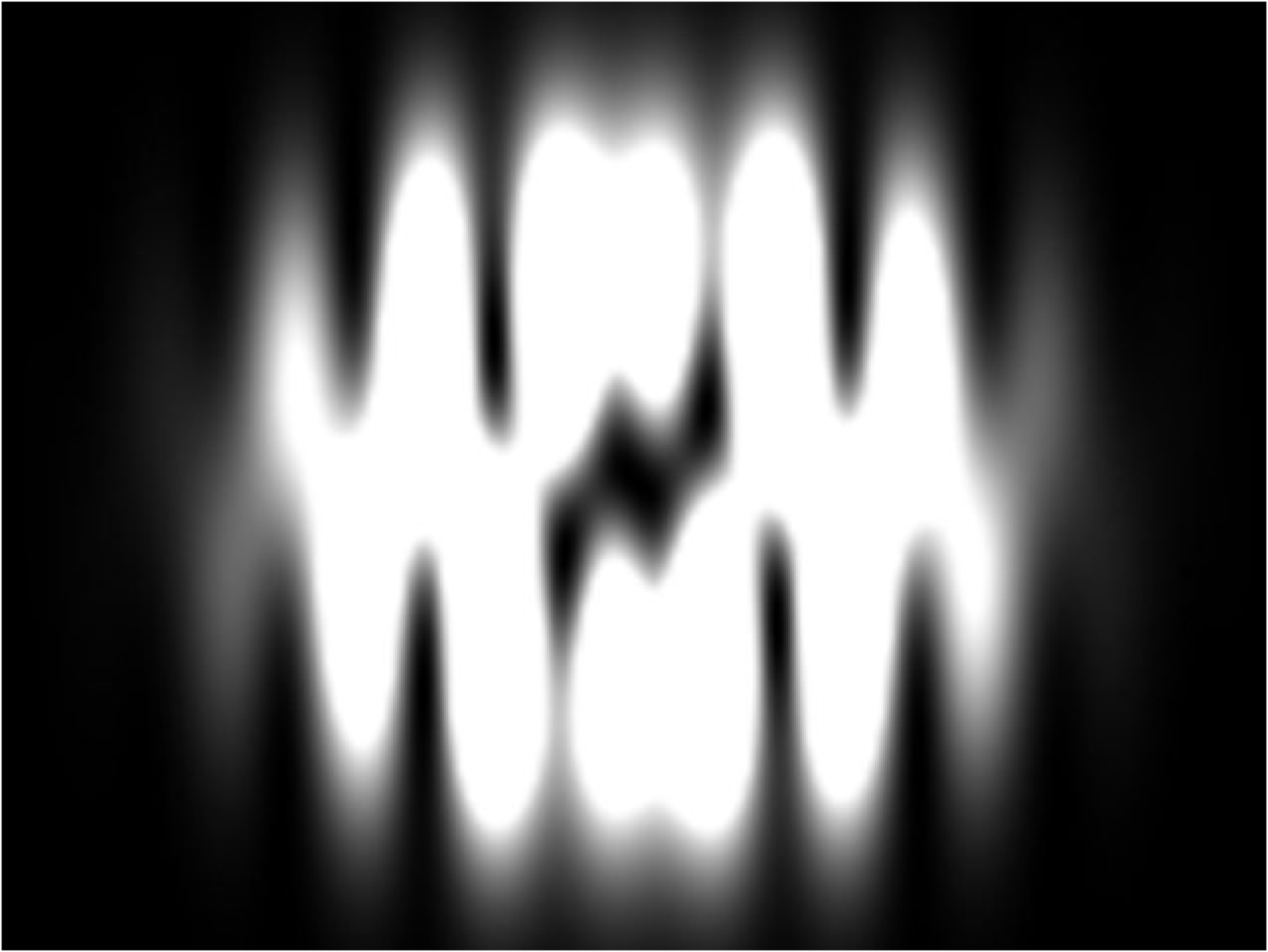}}\\ 

{\begin{array} {c} {HG_{10}} \\ { } \end{array}}& 
{\includegraphics[width=000.075\textwidth]{10t.eps}}& 
{\includegraphics[width=000.075\textwidth]{Blank.eps}}&
{\includegraphics[width=000.075\textwidth]{01t.eps}}&
{\includegraphics[width=000.075\textwidth]{10it.eps}} \\ 

{\begin{array} {c} {HG_{45^{\circ}}} \\ { } \end{array}}&
{\includegraphics[width=000.075\textwidth]{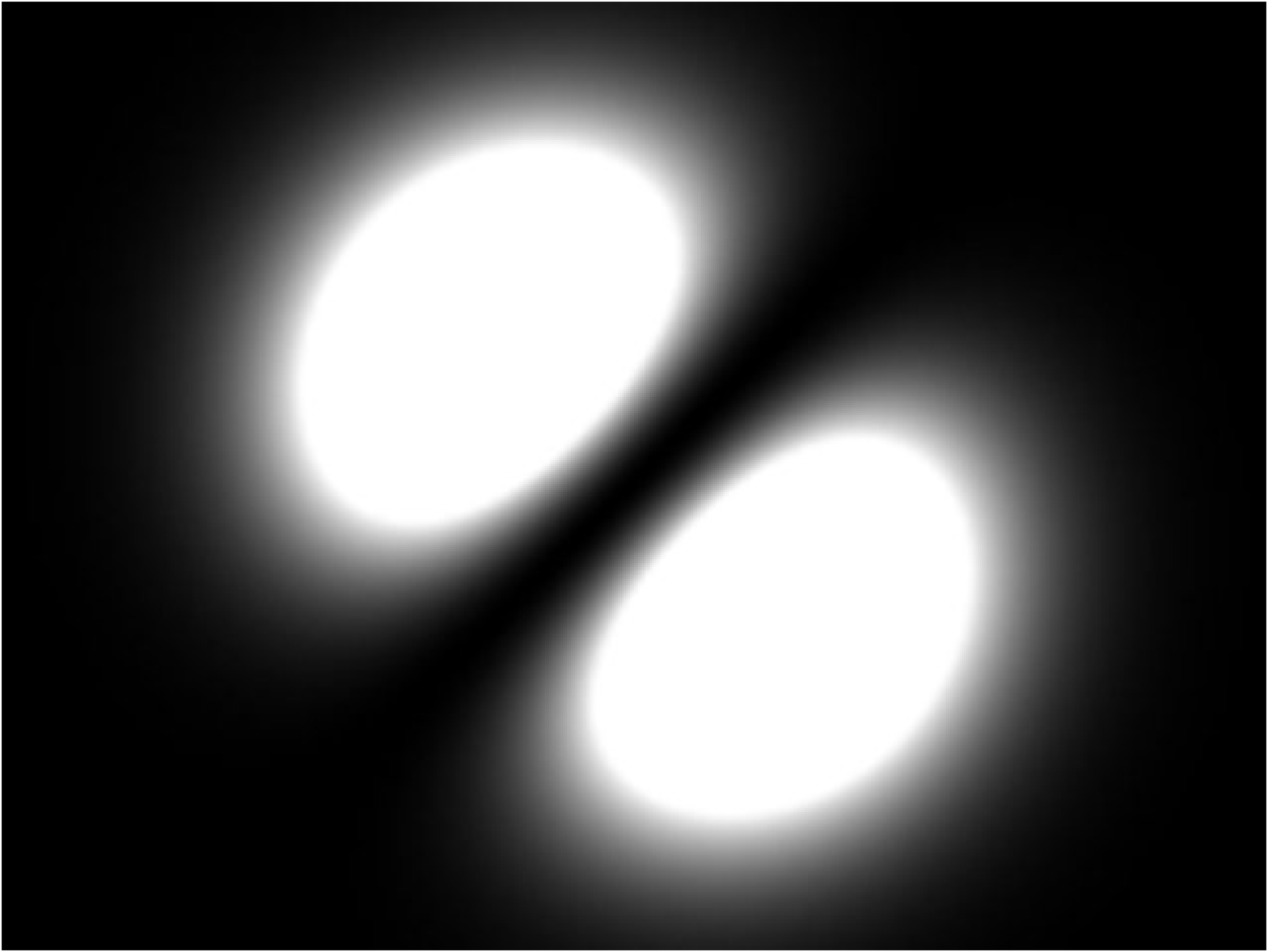}}&
{\includegraphics[width=000.075\textwidth]{Blank.eps}}& 
{\includegraphics[width=000.075\textwidth]{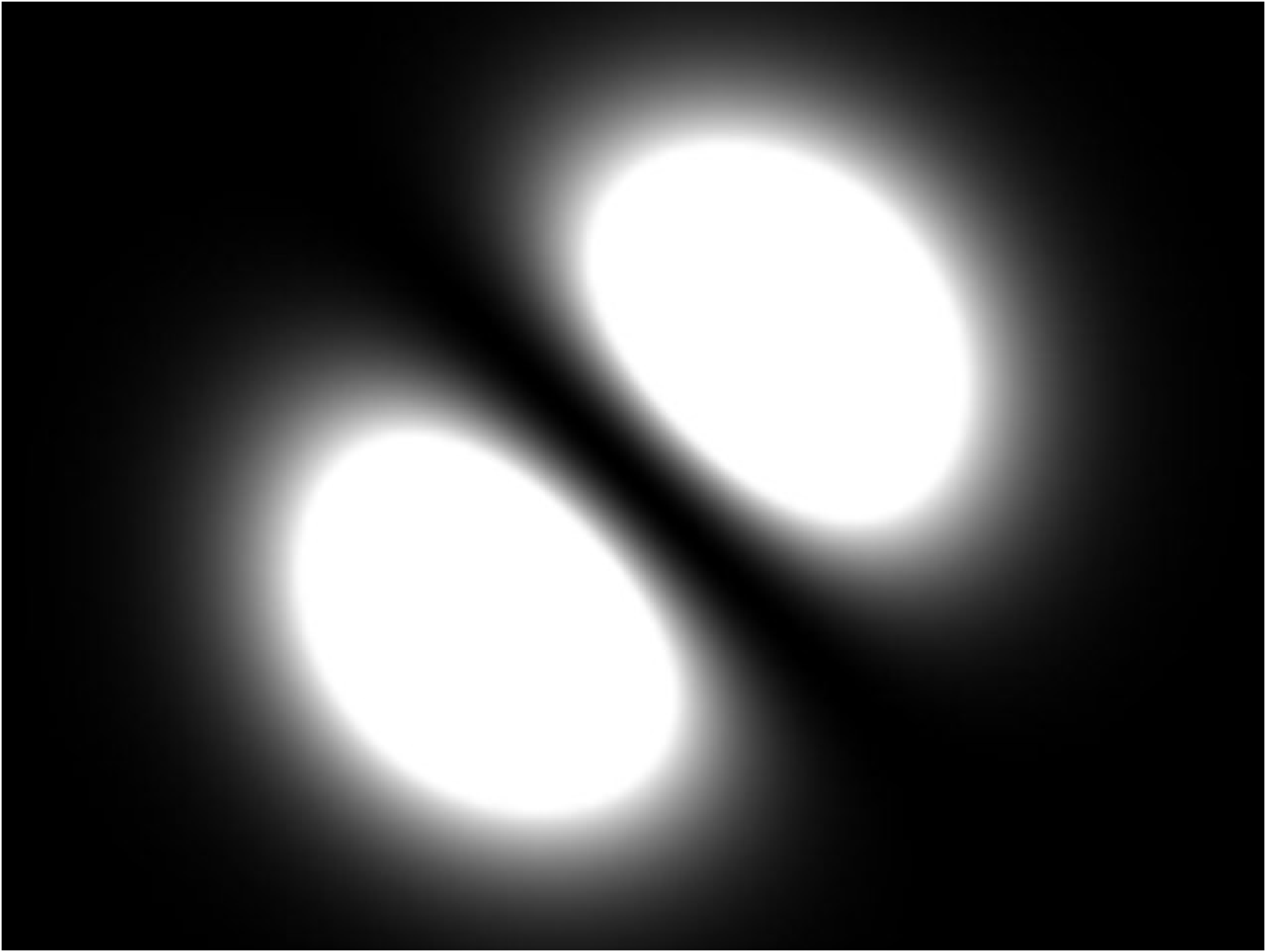}}& 
{\includegraphics[width=000.075\textwidth]{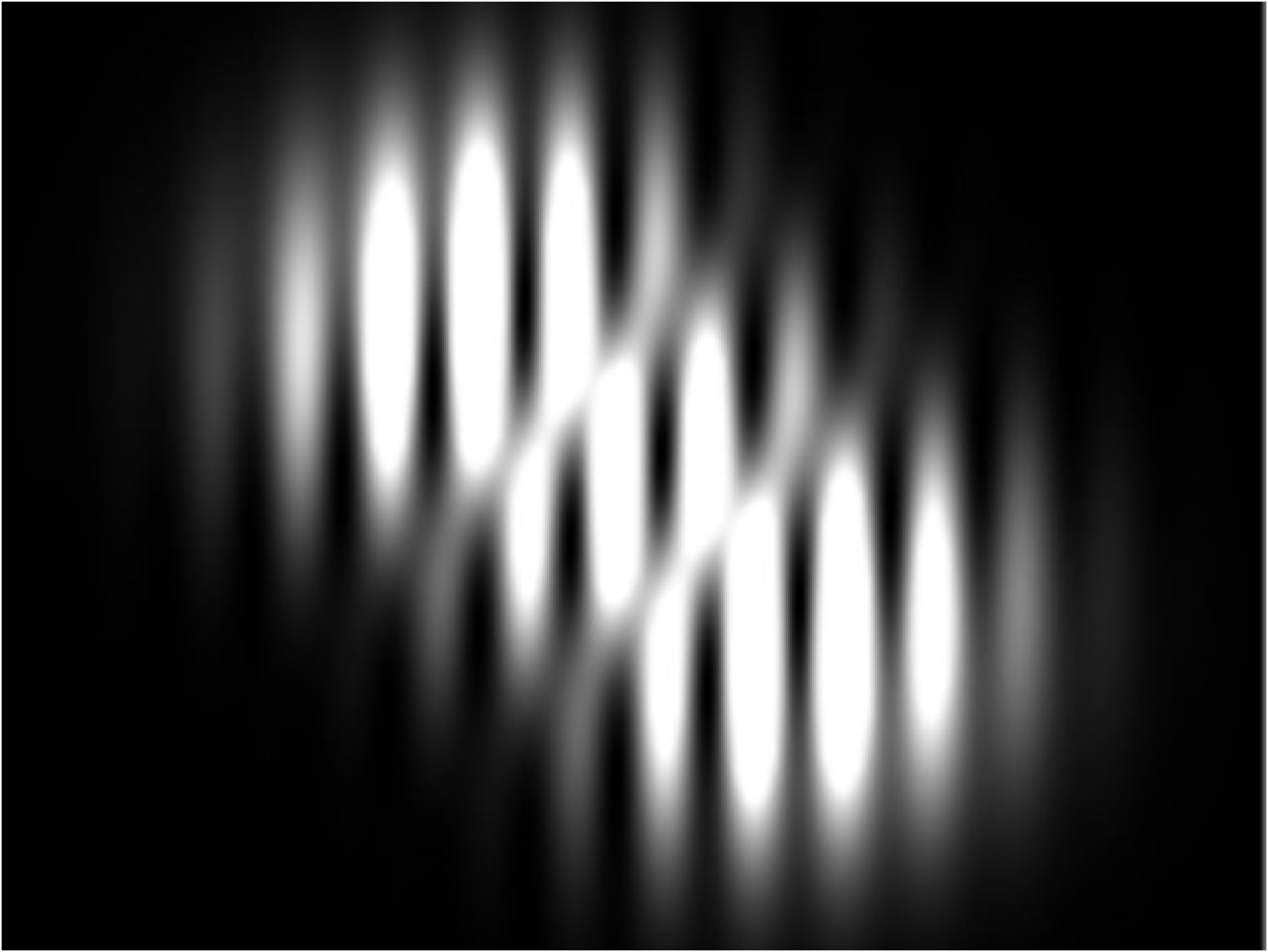}}\\ 
 
{\begin{array} {c} {HG_{11}} \\ { } \end{array}}&
{\includegraphics[width=000.075\textwidth]{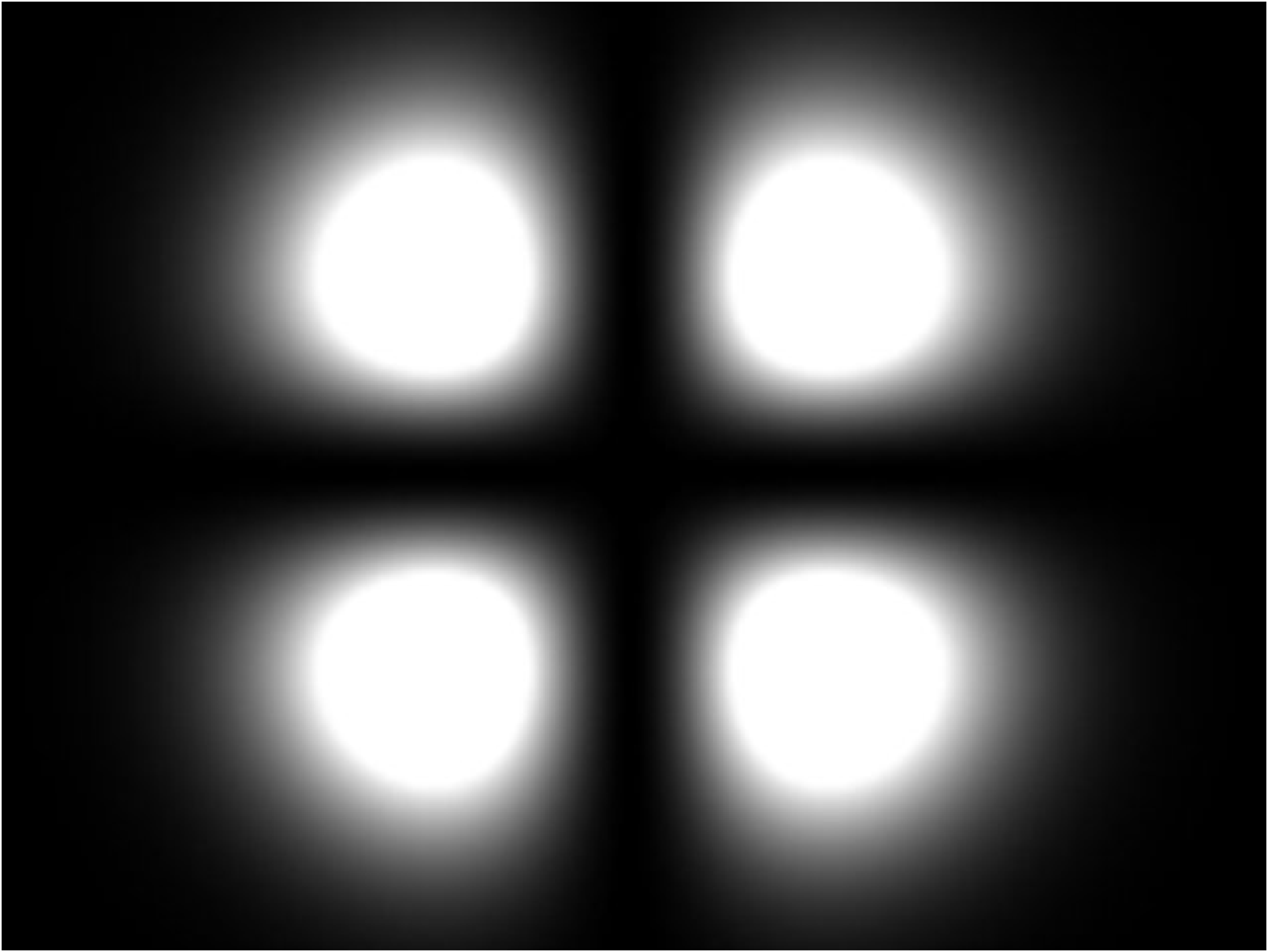}}&
{\includegraphics[width=000.075\textwidth]{11t.eps}}& 
{\includegraphics[width=000.075\textwidth]{Blank.eps}}& 
{\includegraphics[width=000.075\textwidth]{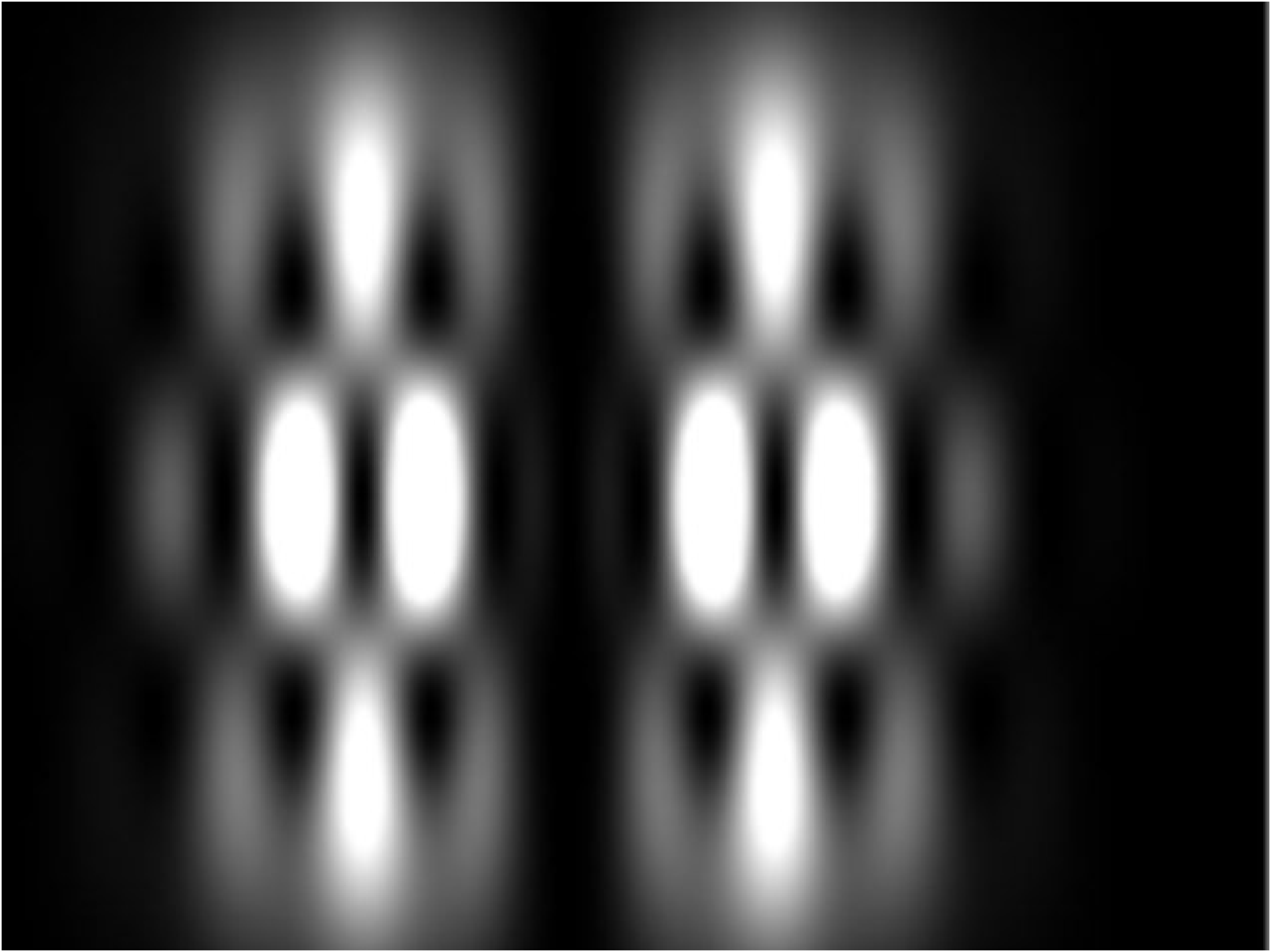}}\\ 

{\begin{array} {c} {HG_{15}} \\ { } \end{array}}& 
{\includegraphics[width=000.075\textwidth]{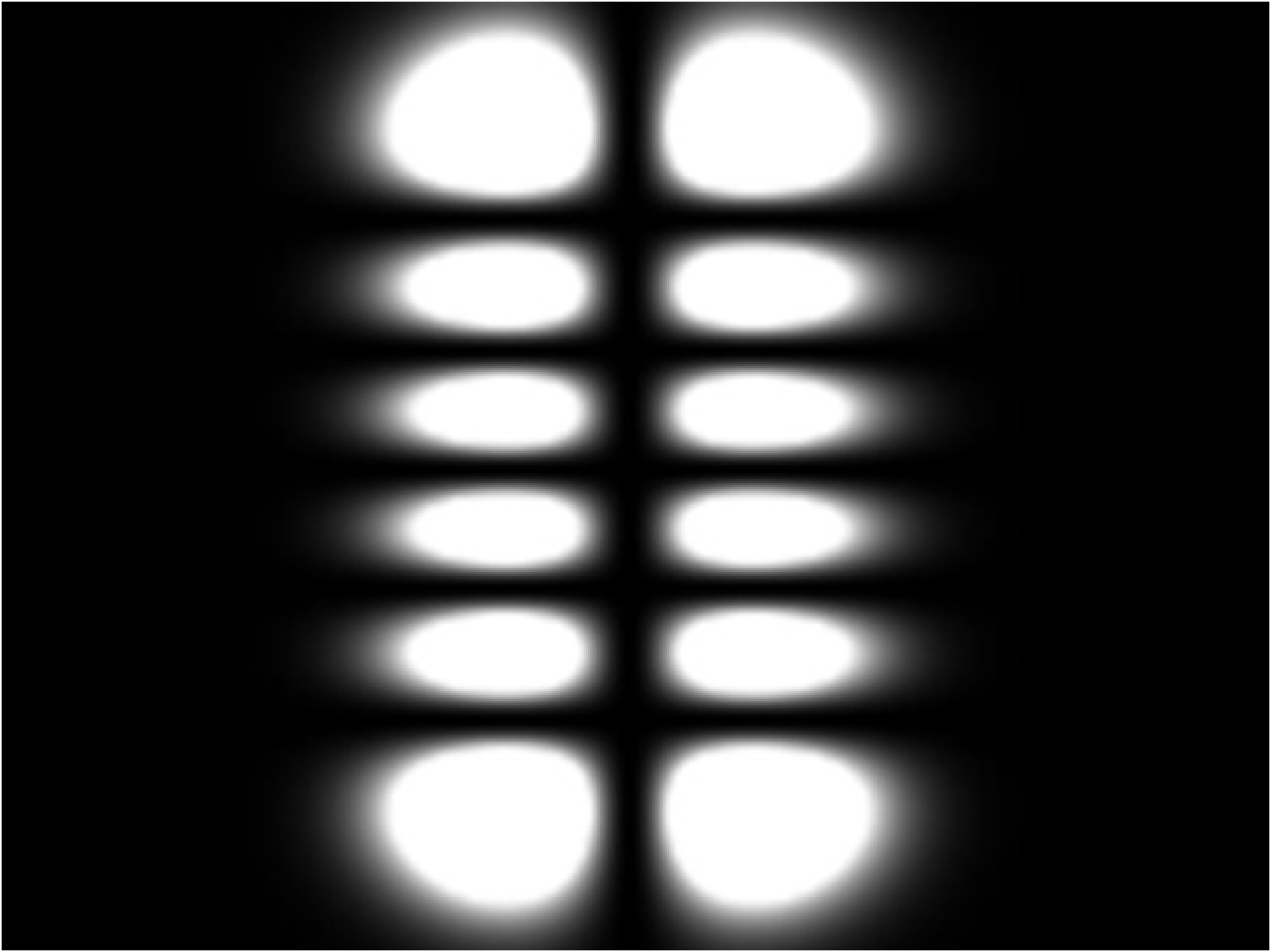}}& 
{\includegraphics[width=000.075\textwidth]{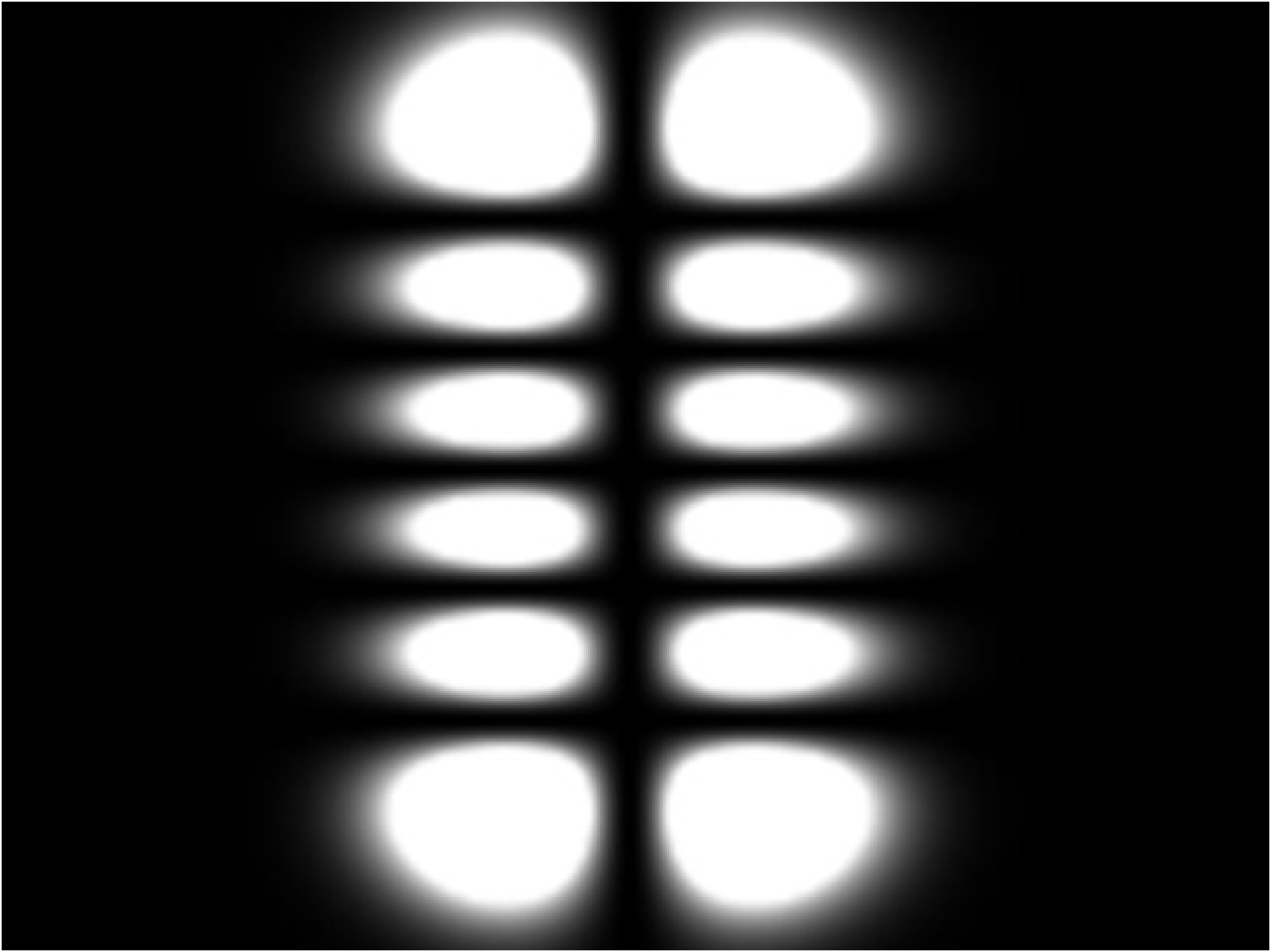}}& 
{\includegraphics[width=000.075\textwidth]{Blank.eps}}& 
{ }\\ 

{\begin{array} {c} {HG_{32}} \\ { } \end{array}}& 
{\includegraphics[width=000.075\textwidth]{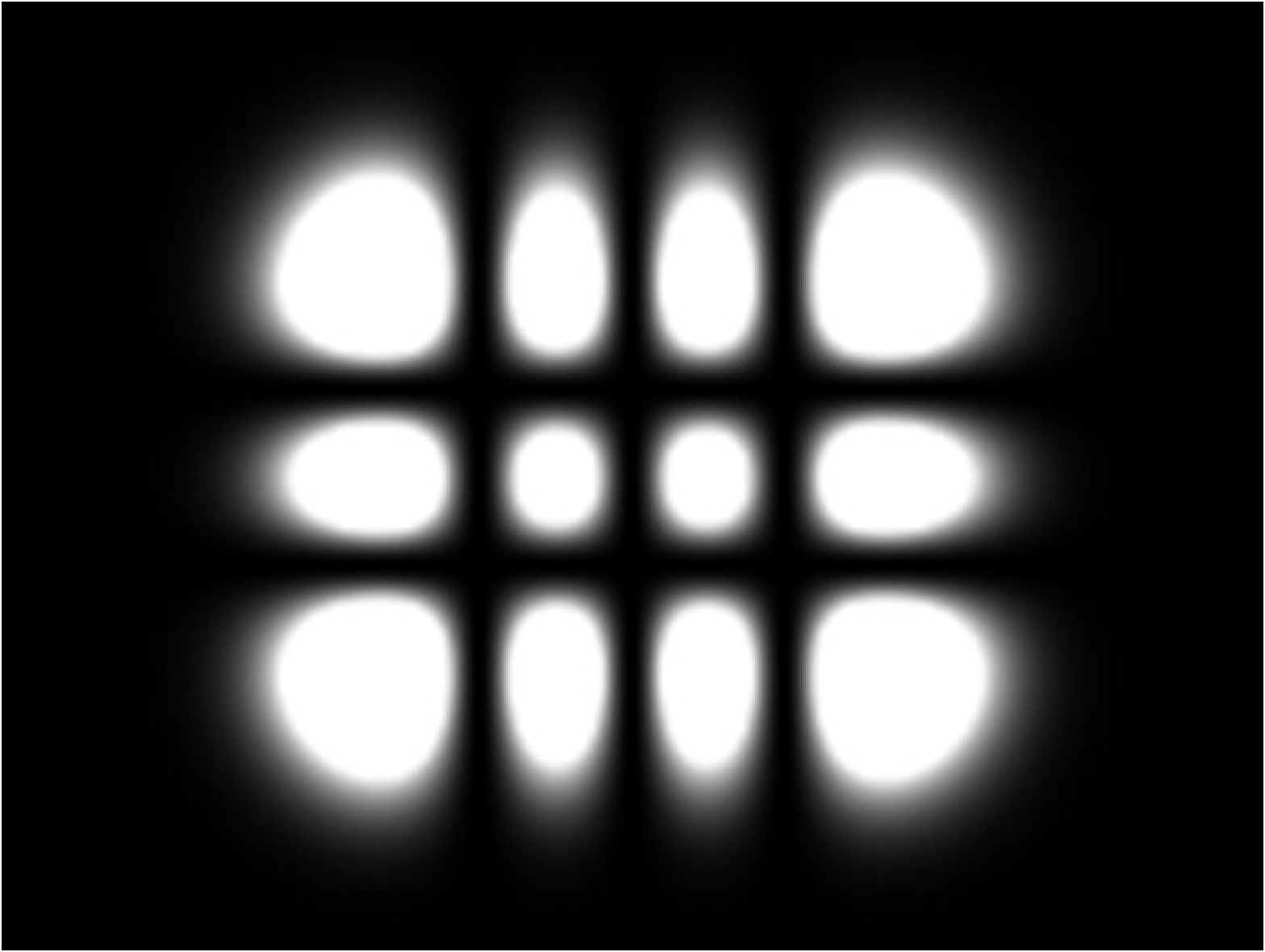}}&
{\includegraphics[width=000.075\textwidth]{Blank.eps}}&
{\includegraphics[width=000.075\textwidth]{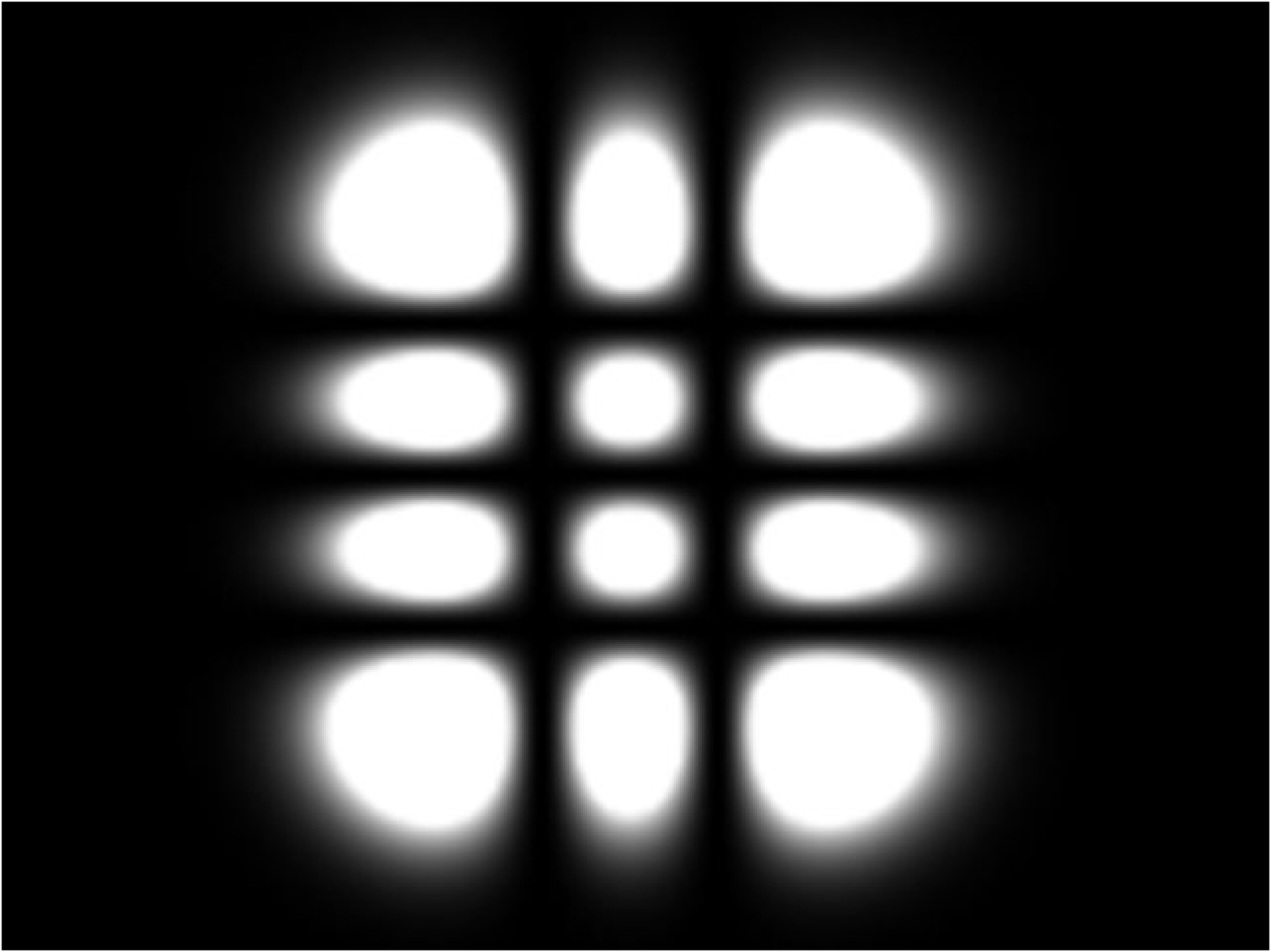}}&
{ } \\ 

{\begin{array}{c} {\textrm{\scriptsize{3 Mode}}} \\ {\textrm{\scriptsize{Fiber}}} \\ { } \\ { }\end{array}}&
{\includegraphics[width=000.075\textwidth]{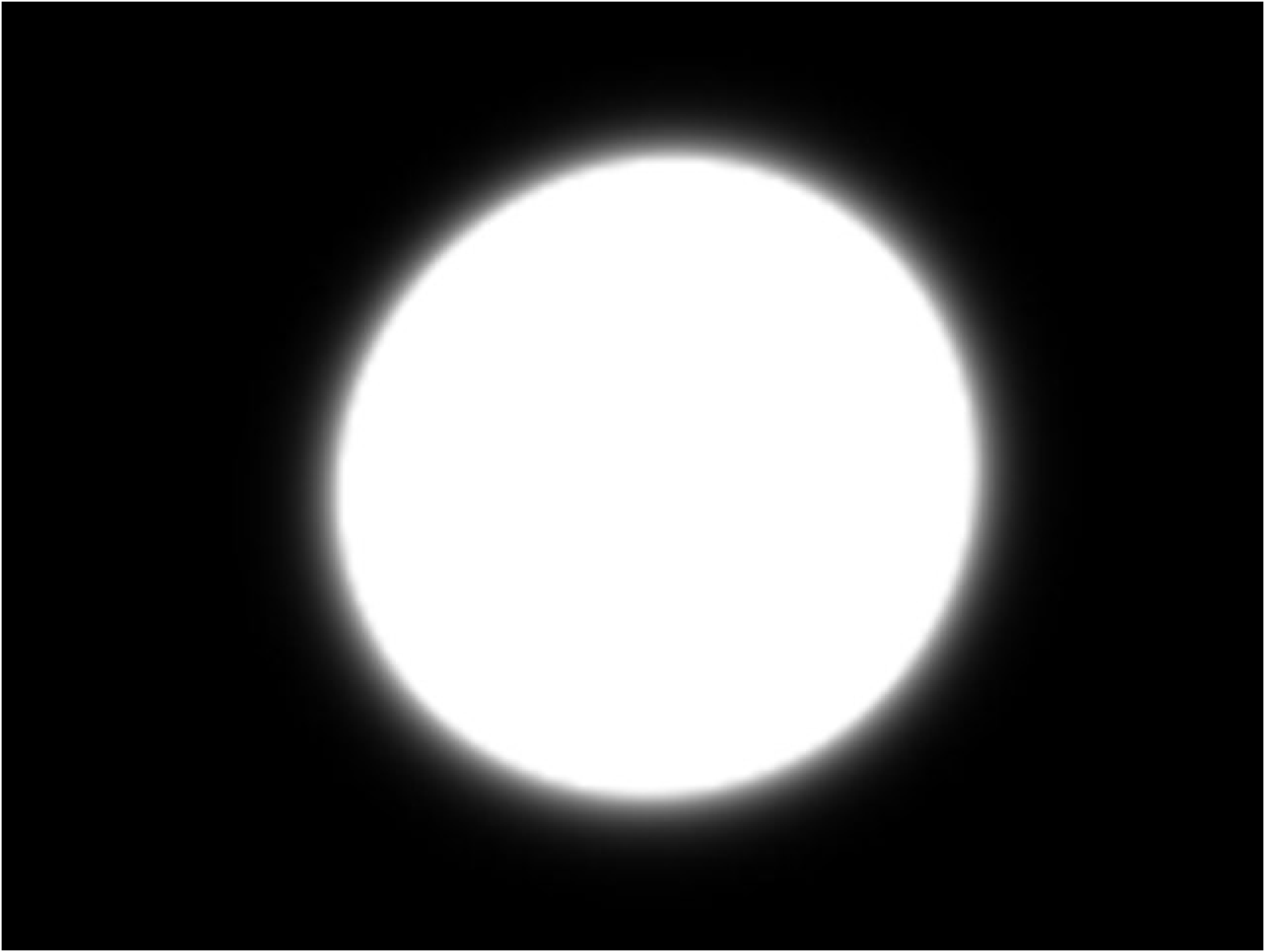}}&
{\includegraphics[width=000.075\textwidth]{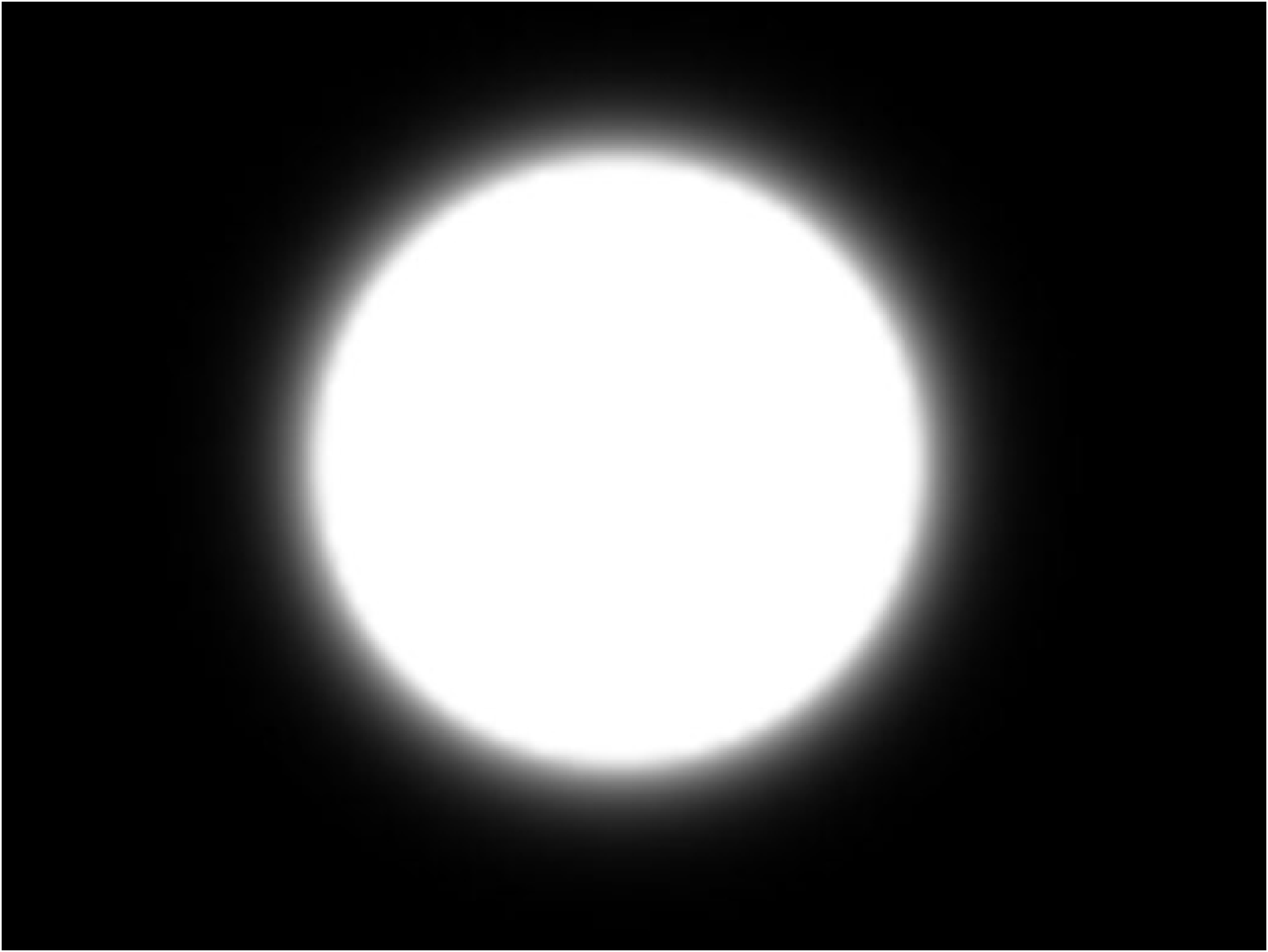}}&
{\includegraphics[width=000.075\textwidth]{HG45minust.eps}}&
{ } 

\end{array}} 

\end{array}$ 

\caption{\label{fig:Theory} Predicted output intensity profiles and interference patterns for a given input field.}

\end{figure}

 Thin crossed wires were inserted into the laser cavity in order to excite higher order HG modes by suppressing lower order transverse modes which lacked a nodal point near the location of the wire. An intracavity iris was then constricted in order to suppress the remaining excited modes that had a higher order then the desired mode. For a given input mode, we measured the output intensity profiles at both port A and port B of the interferometer using a CCD camera. In order to demonstrate the phase structure of the sorted HG modes, we employed a third beam splitter (BS3 in the figure) in order to pick off a portion of the input beam to use as a reference beam. We then interfered the reference beam for several input modes with their corresponding parity-sorted output modes by steering the input and output beam paths together so that they were co-propagating with a slight misalignment, and recorded the resulting intensity patterns. In a second experiment, the wire and iris were removed and the light was coupled into and out of a three-mode fiber (Thorlabs SM-780) which then acted as the input source. 

 We note here that due to the 50\% loss of the signal from port A from the presence of the beam splitter BS2, this setup was not capable of sorting single photons with high efficiency. However, by replacing beam splitter BS2 with a Faraday isolator and appropriately adjusting the compensators, 100\% efficiency can be obtained in principle as shown in Appendix \ref{app:Faraday}. 
 
 A representative sample of our results are presented in Fig. \ref{fig:Experiment}, which agree with the corresponding theoretical predictions in Fig. \ref{fig:Theory}. In these figures, the first column indicates the input mode, while the second gives the observed/predicted intensity distribution for that mode. Columns three and four give the observed/predicted distributions at output ports A and B, respectively. The final column gives the observed/predicted interference patterns between the input and output modes as discussed above, which exhibit a characteristic \textquotedblleft{forking}\textquotedblright $ $ of the vertical interference fringes which shows the phase structure of the modes \footnote{In order to interpret the final column of Figs. \ref{fig:Experiment} and \ref{fig:Theory}, recall that the output beam has been rotated 90\textdegree $ $ with respect to the input beam, regardless of output port. This has no effect upon the rotationally symmetric $HG_{00}$ mode, so the familiar interference fringes of a standard Mach-Zehnder interferometer are observed. However the 90\textdegree $ $ rotation does effect the first-order modes, so that the next two rows exhibit an interference pattern resulting form the superposition of the input HG mode and its 90\textdegree $ $ rotated counterpart. Note the characteristic \textquotedblleft{forking}\textquotedblright $ $ of the vertical fringes in both of these plots, which shows the nontrivial phase structure of these modes as they interfere. For the $HG_{45^{\circ}}$ mode, one would expect a similar \textquotedblleft{forked}\textquotedblright $ $ pattern, but rotated by 90\textdegree $ $. However, practical considerations required the presence of an extra mirror along the reference beam path in order to interfere the reference and output beams. Since an extra mirror reflection in the \textit{x}-\textit{y} plane transforms an $HG_{45^{\circ}}$ mode into its 90\textdegree $ $ rotated counterpart, the presence of the extra mirror canceled the effect of the out-of-plane rotation in this case. In this case therefore the resulting interference pattern resembled that of the $HG_{00}$ mode, which did not exhibit the phase structure of the mode. A similar issue occurs with the $HG_{11}$ mode, which is identical to its 90\textdegree $ $ rotated counterpart up to an overall phase. Therefore, in order to more clearly demonstrate the desired phase structure of the $HG_{45^{\circ}}$ and $HG_{11}$ modes, we steered the output beam so that its propagation axis was \textit{transversely shifted} with respect to the reference beam while still being (nearly) collinear with respect to it. For the case of the $HG_{45^{\circ}}$ mode, the transverse shift was directed both down and to the right, while for the $HG_{11}$ mode it was directed completely downwards. In this way, the interfering beams were only partially overlapping so that the resulting interference patterns, included in the fourth and fifth columns, clearly show the characteristic \textquotedblleft{forking}\textquotedblright $ $ effect in their interference patterns.}. Note that the the $HG_{00}$ and $HG_{11}$ modes exit port A, while the $HG_{10}$ and $HG_{01}$ modes exit port B regardless of their orientation as predicted in Figs. \ref{fig:2DParityTable} and \ref{fig:Theory}. Furthermore, the $HG_{15}$ mode (which has even parity since $n+m=1+5=6$) exits port A while the $HG_{32}$ mode (which has odd parity since $n+m=3+2=5$) exits port B. We note here for completeness that a $p^{\mathrm{th}}$ order $LG^{l}_{\pm p}$ mode is sorted by this interferometer according to whether $p$ is even or odd which parallels the current demonstration of $HG_{nm}$ mode sorting according to the even/odd parity of $n+m$. We have therefore shown with these results that stable, cascadable, single photon OAM sorting schemes corresponding to those discussed in \cite{Padgett, Wei} are indeed feasible with our interferometer. 
 
 Finally, we note that the bottom row of Fig. \ref{fig:Experiment} shows that the output field from the three-mode fiber is decomposed into the fundamental $HG_{00}$ mode, which exits out of port A, and an equal superposition of first order Laguerre-Gauss fiber eigenmodes, which interfere together to make a first order HG mode which exits out of port B (the theory plot in Fig. \ref{fig:Theory} is for a field composed of 85\% $HG_{00}$ and 15\% first order modes). This explicitly demonstrates the usefulness of our 2-D parity sorter as a \textquotedblleft{beam splitter}\textquotedblright $ $ for even and odd transverse spatial modes, while at the same time suggests that it can be employed as an alternative to relatively lossy holograms in spatial mode filtering, as discussed below.
 
\section{\label{sec:Applications}Applications\protect}
\subsection{\label{sec:Filter}First Order Mode Filter\protect}

\begin{figure}
\includegraphics[width=0.5\textwidth]{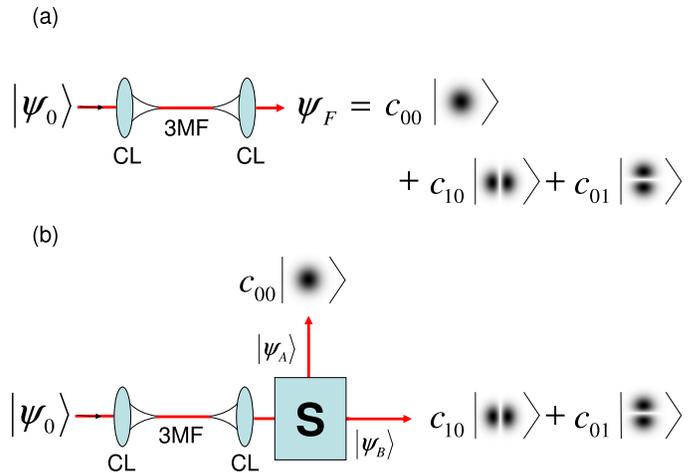}
\caption{\label{fig:Filter} A spatial filter for zero and first-order HG modes. (a) A three-mode optical fiber (3MF) with input and output coupling lenses (CL). The fiber, which has a parabolic refractive index profile, acts as a mode filter upon an arbitrary transverse input beam $\left|\Psi_{0}\right\rangle=\sum_{n,m=0}^{\infty}c_{nm}\left|HG_{nm}\right\rangle$ such that upon coupling into and out of the fiber the resulting beam state is of the form $\left|\psi_{F}\right\rangle=c_{00}\left|HG_{00}\right\rangle+c_{10}\left|HG_{10}\right\rangle+c_{01}\left|HG_{01}\right\rangle$. (b) Inserting a 2-D Parity Sorter (S) after the fiber separates the zero and first-order modes into ports A and B, respectively.}
\end{figure}

One particularly useful characteristic of our Sagnac interferometer is its ability to distinguish between zero-order and first-order transverse HG modes. When used in conjunction with a three-mode optical fiber, this ability allows the Sagnac to act as a spatial mode filter that passes only the first-order HG modes while rejecting all other mode orders. To see this, consider a linearly polarized monochromatic paraxial beam with electric field $E\left(x,y,z\right)e^{-i\omega_{p} t}$ coupled into an optical fiber with input and output coupling lenses as shown in Fig. \ref{fig:Filter} (a) (we align the fiber with the \textit{z}-axis and place the origin at the fiber input face). Such a beam comprises a general solution to the scalar paraxial wave equation, of which the paraxial HG modes form a complete basis \cite{Seigman}. Employing the notation $E\left(x,y,z\right)\equiv\left|\psi_{0}\right\rangle$, one may therefore write 

\begin{equation} \label{E}
\left|\psi_{0}\right\rangle=E_{0}\sum_{n,m=0}^{\infty}c_{nm}\left|HG_{nm}\right\rangle
\end{equation}

\noindent where $c_{nm}\equiv\left\langle \psi_{0}\right.\left|HG_{nm}\right\rangle$ is the overlap integral between the incident field and the $HG_{nm}$ mode, and the constant $E_{0}$ is the electric field magnitude. When the fiber and input coupling lens are properly aligned, the beam will be centered coaxially with the fiber, and the location of the beam waist will coincide with the plane of the fiber input face at $z=0$. Under these conditions, the paraxial HG modes evaluated at the fiber input plane have the simplified transverse form

\begin{align} \label{HG}
& \left|HG_{nm}\right\rangle \nonumber \\
&=A_{nm}\frac{1}{w_{0}}H_{n}\left(\sqrt{2}\frac{x}{w_{0}}\right)H_{m}\left(\sqrt{2}\frac{y}{w_{0}}\right)e^{-\frac{\left(x^{2}+y^{2}\right)}{w_{0}^{2}}}
\end{align}

\noindent where $A_{nm}\equiv\sqrt{\frac{2}{2^{n+m}\pi n!m!}}$, $H_{p}\left(u\right)$ in \eqref{HG} denotes a Hermite polynomial of order $p$, and $w_{0}$ is the beam waist radius (\eqref{HG} is obtained by taking the limit as \textit{z} tends toward zero of the associated paraxial expression of \cite{Seigman}).

 In comparison, the eigenmodes of an optical fiber with a parabolic refractive index profile $n\left(\rho\right)=n_{0}\left(1-\Delta\frac{\rho^{2}}{a^{2}}\right)$ obeying the weakly guided condition $\Delta\ll0$ are well approximated \cite{Liberman} by the linearly polarized Laguerre-Gauss modes $\left|LG_{q,m}\right\rangle$:

\begin{equation} \label{LG}
\left|LG_{q,m}\right\rangle =\left(\frac{\rho}{a}\right)^{\left|m\right|}L_{q}^{\left|m\right|}\left(V\frac{\rho^{2}}{a^{2}}\right)e^{-\frac{V}{2}\frac{\rho^{2}}{a^{2}}}e^{im\phi}
\end{equation}

\noindent where $a$ is the fiber radius, $V\equiv\frac{\omega_{p}n_{0}}{c}a\sqrt{2\Delta}$ is the normalized frequency, the $L_{q}^{\left|m\right|}$ are generalized Laguerre polynomials, and we have employed transverse cylindrical coordinates $\left(\rho,\phi\right)$. Furthermore, for sufficiently small $a$, the higher order modes will be cut off \cite{Snyder} so that the fiber will support only the three LG modes $\left|LG_{0,0}\right\rangle$, $\left|LG_{0,+1}\right\rangle$, and $\left|LG_{0,-1}\right\rangle$, or equivalently the three HG modes $\left|HG_{00}\right\rangle$, $\left|HG_{10}\right\rangle$, and $\left|HG_{01}\right\rangle$. Therefore, regardless of the form of the transverse input state $\left|\psi_{0}\right\rangle$ in equation \eqref{E}, a weakly guided parabolic fiber will in principle filter out the higher-order HG mode contributions while transmitting the fundamental and first-order modes with high efficiency such that the remaining output state $\psi_{F}$ will have the form

\begin{equation} \label{SF}
\left|\psi_{F}\right\rangle=c_{00}\left|HG_{00}\right\rangle+c_{10}\left|HG_{10}\right\rangle+c_{01}\left|HG_{01}\right\rangle
\end{equation}

\noindent provided that the mode matching condition $w_{0}=a\sqrt{\frac{2}{V}}$ is met. Given this fiber-filtered state, our Sagnac can then sort $\left|\psi_{F}\right\rangle$ such that the output modes at port A and port B are $\left|\psi_{A}\right\rangle=c_{00}\left|HG_{00}\right\rangle$ and $\left|\psi_{B}\right\rangle=c_{10}\left|HG_{10}\right\rangle+c_{01}\left|HG_{01}\right\rangle$, respectively, as shown in Fig. \ref{fig:Filter} (b). 

\subsection{\label{sec:QIP}Applications to Quantum Information Processing\protect} 

 Here we propose two applications of 2-D parity sorters to quantum information processing (QIP) with transverse spatial modes. Both make use of spontaneous parametric down conversion in addition to the first-order mode filter discussed above, as shown in Figs. \ref{fig:Applications} (a) and (b). In both cases, a crystal with a $\chi^{(2)}$ nonlinearity is cut and oriented for Type-II collinear phase matching and pumped with a sufficiently weak cw laser in a well-defined HG mode $HG_{nm}$, such that the resulting state is predominantly a superposition of the vacuum and a two-photon state $\left|\Psi\right\rangle=C_{1}\left|\left|vac\right\rangle\right\rangle+C_{2}\left|\left|\Psi_{2ph}\right\rangle\right\rangle$, where $C_{1} \gg C_{2}$ and the \textquotedblleft{double-ket}\textquotedblright $ $ symbol $\left|\left|\;\;\;\right\rangle\right\rangle$ denotes fock-space states of the electromagnetic field \cite{Hong}. 
 
 When sufficiently narrowband filters centered at half the pump frequency $\omega_{p}$ are employed, the co-propagating two-photon state $\left|\Psi_{2ph}\right\rangle$ is frequency degenerate, and is therefore not entangled in the spectral degree of freedom. However, photons have four degrees of freedom, and $\left|\Psi_{2ph}\right\rangle$ is entangled in each of the remaining three (polarization and the two transverse coordinates). In particular, if the appropriate birefringent phase compensators are used \cite{Kwiat}, then the biphoton mode function corresponding to the two-photon component $\left|\Psi_{2ph}\right\rangle$ of the state $\left|\Psi\right\rangle$ can be written as \cite{Kwiat, Walborn}

\begin{equation} \label{Factor}
\left|\psi_{2ph}\right\rangle=\left|\phi_{nm}\right\rangle \otimes\Big(\left|H\right\rangle\left|V\right\rangle+\left|V\right\rangle\left|H\right\rangle\Big)
\end{equation}
 
\noindent where $\left|H\right\rangle$ and $\left|V\right\rangle$ denote the horizontal and vertical single-photon polarizations, and $\left|\phi_{nm}\right\rangle$ denotes the spatial portion of the biphoton mode function. In \cite{Walborn}, $\left|\phi_{nm}\right\rangle$ was expanded in terms of the HG modes for an arbitrary HG pump beam $HG_{nm}$, and shown to be entangled: In terms of the HG mode functions, $\left|\phi_{nm}\right\rangle$ takes the form

\begin{equation} \label{Phi}
\left|\phi_{nm}\right\rangle=\sum_{j,k,s,t=0}^{\infty}c_{jkst}^{nm}\left|HG_{jk}\right\rangle\left|HG_{st}\right\rangle
\end{equation}
 
\noindent where the specific coefficients $c_{jkst}^{nm}$ depend on the order $nm$ of the pump beam via equation (18) of \cite{Walborn}. As shown there, the expansion contains an infinite number of nonzero terms, and converges slowly in general. However, many of the coefficients $c_{jkst}^{nm}$ are zero due to selection rules. We apply this result below in two special cases.

\subsubsection{\label{sec:Bell}Bell State Generation\protect} 

\begin{figure}[!t]
\includegraphics[width=0.5\textwidth]{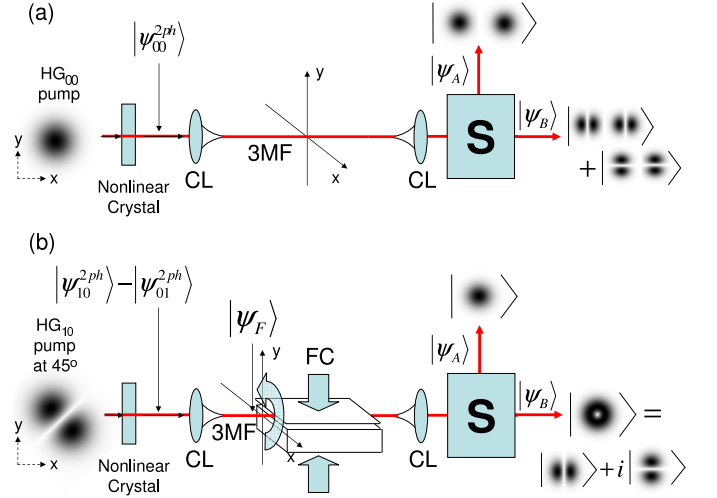}
\caption{\label{fig:Applications} Possible applications of a 2-D parity sorter to quantum information processing. (a) Proposed scheme to produce Bell states entangled in first-order transverse spatial modes. (b) Proposed scheme to produce heralded single photons in arbitrary first-order transverse spatial states.}
\end{figure}

 The first QIP application we discuss is the production of Bell states entangled in first-order transverse spatial modes, as shown in Fig. \ref{fig:Applications} (a). An HG Bell state generation experiment similar to this has been carried out by \cite{White}, which used holograms to sort and detect spatial modes. However, employing our Sagnac interferometer in the place of holograms will provide an improvement in efficiency over standard holographic techniques. In the current setup, the pump beam is in an $HG_{00}$ mode, and following \cite{Walborn} we find that the spatial part of the down converted biphoton mode function has the form

\begin{align} \label{HG00}
\left|\phi_{00}\right\rangle & =  c_{0}\left|HG_{00}\right\rangle\left|HG_{00}\right\rangle \nonumber \\ 
& +c_{1}\Big(\left|HG_{10}\right\rangle\left|HG_{10}\right\rangle+\left|HG_{01}\right\rangle\left|HG_{01}\right\rangle\Big)\nonumber \\ 
& +c_{2}\Big(\left|HG_{00}\right\rangle\left|HG_{02}\right\rangle+\left|HG_{02}\right\rangle\left|HG_{00}\right\rangle \Big. \nonumber \\ 
& +\Big. \left|HG_{00}\right\rangle\left|HG_{20}\right\rangle+\left|HG_{20}\right\rangle\left|HG_{00}\right\rangle\Big) \nonumber \\
& +...
\end{align}

\noindent where $c_{1}=\frac{1}{2}c_{0}\approx0.04$, and $c_{2}\approx-0.03$ under typical experimental conditions (pump beam width 0.1 mm and crystal length 1 mm), and all other coefficients of total order $j+k+s+t=2$ are zero. The omitted terms in \eqref{HG00} all involve biphoton mode functions of total order $j+k+s+t\geq 4$. However, since each individual photonic mode is filtered by the Fiber-Sagnac combination as discussed above, the mode function in \eqref{HG00} retains only its first-order component, so that the (renormalized) mode function exiting port B is given by \eqref{Factor} and \eqref{HG00} as

\begin{align} \label{Hyper}
\left|\psi_{B}\right\rangle=\frac{1}{\sqrt{2}}\Big(\left|HG_{10}\right\rangle\left|HG_{10}\right\rangle & + \left|HG_{01}\right\rangle\left|HG_{01}\right\rangle\Big) \nonumber \\ 
& \otimes\Big(\left|H\right\rangle\left|V\right\rangle+\left|V\right\rangle\left|H\right\rangle\Big)
\end{align}

\noindent as shown in the figure. Thus, the filtered biphoton mode function is maximally entangled in both polarization \textit{and} first-order spatial modes. This is a Bell state.

 We note here that the two photons comprising the Bell state are co-propagating; in order to turn this source into \textquotedblleft{useful}\textquotedblright $ $entanglement, one must separate these photons onto two separate paths without destroying the entanglement relationship. This can be done by inserting a 50:50 beam splitter after port B, albeit with a 50\% loss of the entanglement source. To obtain the HG-entangled Bell states without such a loss, one may instead use a polarizing beam splitter. The co-propagating photons are then path-separated with the cost of destroying the polarization entanglement, while the spatial mode entanglement remains.

\subsubsection{\label{sec:Herald}Heralded Photons in Arbitrary Spatial Modes\protect} 
 
 We now consider the second application of our mode sorter, shown in Fig. \ref{fig:Applications} (b). In this setup, the pump beam is a first-order $HG_{10}$ mode rotated through an angle of 45\textdegree $ $ in the transverse plane, denoted as $HG_{45^{\circ}}$.  Since $HG_{45^{\circ}}$ comprises an in-phase superposition of $HG_{10}$ and $HG_{01}$ modes, we express it in terms of these, and again employ equation (19) of \cite{Walborn}, finding that that down-converted biphoton mode function is
 
\begin{align} \label{HG45}
\left|\phi_{45^{\circ}}\right\rangle & \equiv \left|\phi_{10}\right\rangle+\left|\phi_{01}\right\rangle \nonumber \\
& =c_{1}\Big(\left|HG_{10}\right\rangle\left|HG_{00}\right\rangle+\left|HG_{00}\right\rangle\left|HG_{10}\right\rangle \Big. \nonumber \\ 
& \;\;\;\;+\Big.\left|HG_{01}\right\rangle\left|HG_{00}\right\rangle+\left|HG_{00}\right\rangle\left|HG_{01}\right\rangle\Big)+...
\end{align}
 
\noindent where all other coefficients of total order $j+k+s+t=1$ are zero. Similarly to the previous case, the omitted terms in \eqref{HG45} all involve biphoton mode functions of total order $j+k+s+t\geq 3$, and are therefore filtered by the fiber. Therefore, the action of the Sagnac on the present fiber-filtered biphoton mode function is to path-separate the photons in the fundamental mode $\left|00\right\rangle$ from those in the higher modes.  Thus, upon exiting the Sagnac the biphoton mode function is

\begin{equation} \label{Out}
\left|\psi\right\rangle\equiv\left|HG_{00}\right\rangle_{A}\left|HG_{45^{\circ}}\right\rangle_{B}\otimes\Big(\left|H\right\rangle_{A}\left|V\right\rangle_{B}+\left|V\right\rangle_{A}\left|H\right\rangle_{B}\Big)
\end{equation}

\noindent where $\left|HG_{45^{\circ}}\right\rangle\equiv\left|HG_{10}\right\rangle+\left|HG_{01}\right\rangle$ and A and B label the output ports, so that the detection of a $\left|HG_{00}\right\rangle$ photon in port A heralds a single polarization-entangled photon in the pure spatial mode $\left|HG_{45^{\circ}}\right\rangle$ in port B.

 In order to produce heralded single photons in \textit{arbitrary} spatial modes, one may apply a directional compression stress on the three-mode fiber with two plates as shown in Fig \ref{fig:Applications} (b). Such a compression breaks the cylindrical symmetry of the fiber medium which causes the non-cylindrically symmetric $HG_{10}$ and $HG_{01}$ modes to experience differing phase velocities as they propagate though the fiber, resulting in a nonzero relative phase at the fiber output that is controllable by the magnitude and direction of the applied stress \cite{McGloin}. That is, the compressor acts analogously to a fractional wave plate for polarization, with principal axes aligned parallel and perpendicular to the direction of compression, and with the degree of phase retardation proportional to the amount of stress. In the case illustrated, the biphoton mode function propagating in the fiber (before compression, but after being filtered) is $\left|\psi_{F}\right\rangle=\left|HG_{00}\right\rangle\left|HG_{45^{\circ}}\right\rangle+\left|HG_{45^{\circ}}\right\rangle\left|HG_{00}\right\rangle$, the direction of compression is along the \textit{y} axis, and the magnitude is such that so that the $HG_{10}$ and $HG_{01}$ mode components experience a relative phase factor of $e^{i\frac{\pi}{2}}=i$. Therefore, the spatial part of the single-photon mode function $\left|HG_{45^{\circ}}\right\rangle\equiv\left|HG_{10}\right\rangle+\left|HG_{01}\right\rangle$ becomes $\left|LG^{+1}_{0}\right\rangle\equiv\left|HG_{10}\right\rangle+i\left|HG_{01}\right\rangle$ after passing through the compressor, so that the heralded single-photon mode function at Sagnac output port B is both polarization entangled and in a well-defined first-order spatial mode:

\begin{equation} \label{HeraldedLG}
\left|\psi_{B}\right\rangle=\left|LG^{+1}_{0}\right\rangle\otimes\Big(\left|H\right\rangle\left|V\right\rangle+\left|V\right\rangle\left|H\right\rangle\Big)
\end{equation}

We note that to cover the entire Poincar\'{e} sphere of first-order transverse spatial modes \cite{van Enk}, two such successive compressors are needed with one applying twice the amount of pressure as the other, in analogy with the requirement of a half-wave plate followed by a quarter-wave plate in order to turn linearly polarized light into light of an arbitrary polarization state. Also, the fiber compressor as shown in Fig. \ref{fig:Applications} (b) can be used in conjunction with the setup in Fig. \ref{fig:Applications} (a) in order to produce, for example, LG-entangled bell states. Finally, we note that the above states can also be created by pumping the crystal with the appropriate higher order mode as opposed to using fiber compressors after the fact, since our Sorter can filter our any unwanted contributions to the biphoton mode function. However, in many cases, pumping the crystal with a Gaussian mode is preferable.

\section{\label{sec:Conclusions}Conclusions\protect}

 We have demonstrated experimentally a stable out-of-plane Sagnac interferometer which sorts photonic transverse spatial modes according to two-dimensional parity. We performed 2-D parity sorting on Hermite-Gauss modes of definite parity and on the mixed-parity output field from a three-mode fiber. This interferometer is the first phase-stable device capable of measuring the transverse spatial mode parity of single-photon states, and is in principle 100\% efficient. Because of this stability, many interferrometers can for the first time be practically cascaded in order to measure the orbital angular momentum of single photon states, without the use of holograms which are lossy in practice. The Sagnac can also be used to achieve spatial mode sorting of the output of a three-mode fiber into the fundamental Gaussian mode and the first-order mode family. This in turn makes feasible the generation of both HG-entangled Bell states and heralded single photons in arbitrary first-order spatial modes, covering the entire Poincar\'{e} sphere of first-order tranverse modes. In contrast to standard holographic techniques which path-separate spatial modes by converting them to the fundamental Gaussian, our Sagnac sorts modes while leaving their spatial mode functions unchanged, which may be desirable for certain applications. 
 
 The 2-D parity sorting interferometer belongs to a larger class of interferometers which differ according to the relative rotation $\Psi$ of their constituent propagating and counter-propagating beams \cite{Padgett}. For example, the 2-D sorter (with $\Psi=180$\textdegree $ $) can distinguish between the zero order ($HG_{00}$) and first order ($HG_{01}$ and $HG_{10}$) modes, but not between the zero order and second order ($HG_{02}$, $HG_{20}$, and $HG_{11}$) modes. Conversely, a Sagnac with $\Psi=90$\textdegree $ $ can distinguish between zero order and second order modes, but not between zero and first or third order modes. In general, the various interferometers discussed in this paper can be stably cascaded so as to sort photons belonging to any Poincar\'{e} sphere order, which enables the manipulation of multi-dimensional qudits encoded in orbital angular momentum. 
 
 Two-dimensional transverse spatial parity provides a novel type of qubit encoding that is based on continuous photonic degrees of freedom, in addition to the scheme based on one-dimensional parity previously realized by \cite{Yarnall}. Unfortunately, to our knowledge it is not possible to design a Sagnac interferometer that sorts according to 1-D parity, since the relative transverse rotation upon which our Sagnac depends treats both spatial variables on an equal footing. Therefore, it appears that encodings based on 2-D parity can be made more robust than those based on 1-D parity due to the phase-stable manipulations of qubits enabled by this work.
 
\begin{acknowledgments}
We wish to acknowledge Steven van Enk for fruitful discussions. This work was supported by the National Science Foundation, through its Physics at the Information Frontier (PIF) program, grant PHY-0554842.
\end{acknowledgments}

\appendix
\section{\label{app:OAM}Measuring the odd-valued OAM of Single Photons\protect}

\begin{figure}[!t]
\includegraphics[width=0.5\textwidth]{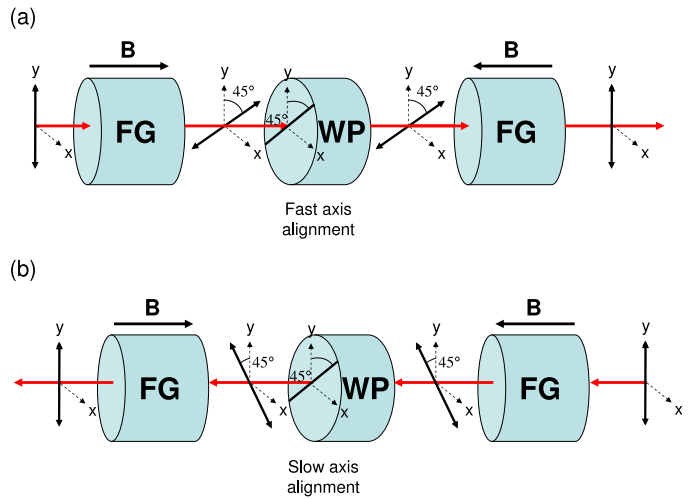}
\caption{\label{fig:Shifter} A device that imparts a relative phase shift between counter-propagating fields, consisting of a tiltable birefringent waveplate (WP) with identical lengths of Faraday glass (FG) on each side, distinguished only by the direction of the applied magnetic fields (B) permeating them. (a) A vertically polarized right-propagating photon becomes aligned with the fast axis of a birefringent waveplate (axis is denoted by the bold line), which induces a phase shift. (b) A vertically polarized left-propagating photon becomes aligned with the slow axis of the waveplate, thereby experiencing an unequal phase shift when compared to the forward-propagating beam.}
\end{figure}

 In this appendix, we propose an optical device that imparts a variable phase shift that is dependent upon whether the photon is propagating forwards or backwards through the device, as discussed in Section \ref{sec:Sagnac}. 

 The device consists of a tiltable birefringent waveplate surrounded on both sides by identical lengths of Faraday glass, distinguished only by the direction of the applied magnetic fields (B) permeating them as shown in Fig \ref{fig:Shifter}. The uniform magnetic fields cause the Faraday glass to act as a polarization rotator via the Faraday effect \cite{Saleh}, with the sense of rotation depending of the direction of the applied field. Upon entering the device from the left, a right-propagating vertically polarized photon undergoes a polarization rotation such that upon entering the waveplate the polarization is aligned with the waveplate's fast axis. The waveplate then imparts a phase shift to the photon before its polarization is returned to its original state, as shown in \ref{fig:Shifter} (a). Conversely, a vertically polarized left-propagating photon undergoes a transformation which leaves its polarization aligned with the waveplate's \textit{slow} axis. Therefore, the left-propagating photon experiences a different phase shift than the right-propagating photon before its polarization is returned to vertical, as shown in \ref{fig:Shifter} (b). The waveplate can be tilted about its fast or slow axes in order to vary this relative phase shift, similarly to the standard method of tilting a thin glass slide. By employing one of these phase shifting devices inside our Sagnac interferometer, one may effect variable phase shifts between the propagating and counter-propagating beams without changing the polarization state of the light.  We conclude that our Sagnac is capable of sorting odd OAM-valued photons from one another without the use of comparatively lossy holograms, and can therefore provide cascaded phase-stable sorting and measurement of single photons possessing arbitrary absolute OAM values, as discussed in Section \ref{sec:Sagnac}.

\section{\label{app:Faraday}Efficient sorting of single photon states\protect}

 In this appendix, we show that by replacing beam splitter BS2 in Fig. \ref{fig:Apparatus} with a Faraday isolator and appropriately adjusting the Berek compensators, 100\% sorting efficiency of single photon states can in principle be obtained, as stated in Section \ref{sec:Experiment}. 

\begin{figure}
\includegraphics[width=0.5\textwidth]{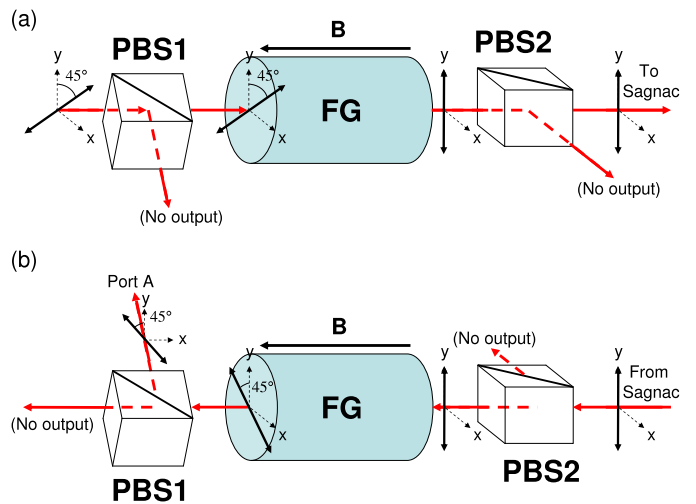}
\caption{\label{fig:Faraday} A Faraday isolator, consisting of a piece of Faraday glass (FG) with a polarizing beam splitter on each side (PBS1 and PBS2). (a) An input photon (propagating left to right) polarized at positive 45\textdegree $ $ angle with respect to the vertical is completely transmitted and becomes vertically polarized. (b) A back-propagating (from right to left) output photon from port A of the Sagnac with vertical polarization is completely transmitted through PBS2, but is subsequently completely deflected from PBS1.}
\end{figure}
 
 A Faraday isolator (shown in Fig. \ref{fig:Faraday}) consists of a piece of Faraday glass with a polarizing beam splitter (PBS) on each side. The application of a uniform magnetic field causes the Faraday glass to act as a polarization rotator via the Faraday effect \cite{Saleh}. The first PBS (PBS1) is oriented at a positive 45\textdegree $ $ angle with respect to the vertical, such that an incoming photon polarized along this same direction will be completely transmitted. Conversely the second PBS (PBS2) is oriented vertically (at a 45\textdegree $ $ angle with respect to the first), in order to completely transmit vertically polarized photons. Therefore, the polarization state of an incoming photon initially polarized at 45\textdegree $ $ with respect to the vertical will remain unchanged as the photon is completely transmitted though PBS1, but will then be transformed to vertical polarization so that the photon is also completely transmitted through PBS2, as shown in Fig. \ref{fig:Faraday} (a). However, an arbitrarily polarized photon propagating through the isolator in the opposite direction will have its horizontally polarized component deflected by PBS2, while its vertically polarized component will be transmitted and subsequently rotated by the Faraday glass to a \textit{negative} 45\textdegree $ $ angle with respect to the vertical, so that this remaining component is completely deflected by PBS1. 

 Applying this to our current problem, one finds that although our backwards-propagating beam from output port A will be completely deflected as desired upon employing a Faraday isolator, it would also in general be separated into two different paths by the two PBS's. Although one can easily recombine these beams using a half wave plate and an additional polarizing beam splitter, such a separation and recombination is undesirable as it introduces the same phase noise and drift that gives our setup an advantage over previous Mach-Zehnder-based 2-D sorter implementations. In our experiment, we can overcome this difficulty by adjusting the Berek compensators such that given a vertical input polarization, the output polarizations of both port A and port B are vertical. Therefore, with the compensators appropriately adjusted, we conclude that back-propagating photons from port A will be completely \textit{transmitted} by PBS2 and completely \textit{deflected} by PBS1, as shown in Fig. \ref{fig:Faraday} (b). In light of this, it is apparent that PBS2 of Fig. \ref{fig:Faraday} is actually unnecessary, however we have included it because of the practical consideration that Faraday isolators are readily commercially available in the form discussed above. We conclude that through the use of a Faraday isolator in conjunction with the Berek compensators, 100\% of the output field from port A can be deflected into a single beam path that is distinct from the input path, so that our 2-D parity sorting Sagnac is in principle 100\% efficient with respect to both output ports.

\end{document}